\documentclass[useAMS,usenatbib]{mnras}

\usepackage{url,times,graphicx,amsmath,amsfonts,amssymb,color,epsfig,epstopdf}

\usepackage[dvipsnames,svgnames,hyperref]{xcolor}
\usepackage{footnote}
\usepackage{comment}

\newcommand{\uc}{{'-uc'}}
\newcommand{\cone}{{`-c01'}}
\newcommand{\ctwo}{{`-c02'}}

\newcommand{\DLB}{\textsc{DLB07}}
\newcommand{\lgalaxy}{\textsc{Lgalaxies}}
\newcommand{\mice}{\textsc{Mice}}
\newcommand{\morgana}{\textsc{Morgana}}

\newcommand{\ysam}{\textsc{ySAM}}
\newcommand{\galform}{\textsc{Galform}}

\newcommand{\galformVGP}{\textsc{Galform-gp14}}

\newcommand{\sag}{\textsc{Sag}}
\newcommand{\sage}{\textsc{Sage}}

\newcommand{\galics}{\textsc{GalICS~2.0}}

\newcommand{\rockstar}{\textsc{Rockstar}}

\newcommand{\Mbnd}{{\ifmmode{M_{\rm bnd}}\else{$M_{\rm bnd}$}\fi}}
\newcommand{\Mfof}{{\ifmmode{M_{\rm fof}}\else{$M_{\rm fof}$}\fi}}
\newcommand{\Mcrit}{{\ifmmode{M_{\rm 200c}}\else{$M_{\rm 200c}$}\fi}}
\newcommand{\Mmean}{{\ifmmode{M_{\rm 200m}}\else{$M_{\rm 200m}$}\fi}}
\newcommand{\MBN}{{\ifmmode{M_{\rm BN98}}\else{$M_{\rm BN98}$}\fi}}

\newcommand{\Rcrit}{{\ifmmode{R_{\rm 200c}}\else{$R_{\rm 200c}$}\fi}}

\newcommand{\hMpc}{{\ifmmode{h^{-1}{\rm Mpc}}\else{$h^{-1}$Mpc}\fi}}
\newcommand{\hkpc}{{\ifmmode{h^{-1}{\rm kpc}}\else{$h^{-1}$kpc}\fi}}
\newcommand{\hMsun}{{\ifmmode{h^{-1}{\rm {M_{\odot}}}}\else{$h^{-1}{\rm{M_{\odot}}}$}\fi}}
\newcommand{\Mstar}{{\ifmmode{M_{*}}\else{$M_{*}$}\fi}}
\newcommand{\Mhalo}{{\ifmmode{M_{\rm halo}}\else{$M_{\rm halo}$}\fi}}
\newcommand{\Ngal}{{\ifmmode{N_{\rm gal}}\else{$N_{\rm gal}$}\fi}}
\newcommand{\Norph}{{\ifmmode{N_{\rm orphan}}\else{$N_{\rm orphan}$}\fi}}
\newcommand{\Nxorph}{{\ifmmode{N_{\rm non-orphan}}\else{$N_{\rm non-orphan}$}\fi}}
\newcommand{\Zsolar}{{\ifmmode{Z_{\odot}}\else{$Z_{\odot}$}\fi}}
\newcommand{\Msun}{{\ifmmode{{\rm {M_{\odot}}}}\else{${\rm{M_{\odot}}}$}\fi}}
\newcommand{\ltsima}{$\; \buildrel < \over \sim \;$}
\newcommand{\gtsima}{$\; \buildrel > \over \sim \;$}
\newcommand{\lsim}{\lower.5ex\hbox{\ltsima}}
\newcommand{\gsim}{\lower.5ex\hbox{\gtsima}}

\def\lesssim{\mathrel{\hbox{\rlap{\hbox{\lower4pt\hbox{$\sim$}}}\hbox{$<$}}}}
\def\gtrsim{\mathrel{\hbox{\rlap{\hbox{\lower4pt\hbox{$\sim$}}}\hbox{$>$}}}}

\newcommand{\Tab}[1]{Table~\ref{#1}}
\newcommand{\Sec}[1]{Section~\ref{#1}}
\newcommand{\App}[1]{Appendix~\ref{#1}}

\newcommand{\Fig}[1]{Fig.~\ref{#1}}
\newcommand{\beq}{\begin{equation}}
\newcommand{\eeq}{\end{equation}}
\def\beqa{\begin{eqnarray}}
\def\eeqa{\end{eqnarray}}
\def\hMpc{$h^{-1}\,{\rm Mpc}$}
\def\hkpc{$h^{-1}\,{\rm kpc}$}

\title[Cosmic CARNage]
{Cosmic CARNage I: on the calibration of galaxy formation models}
\author[Knebe et. al]
       {Alexander Knebe,$^{1,2}$\thanks{E-mail: alexander.knebe@uam.es}
       Frazer R. Pearce,$^{3}$				
       Violeta Gonzalez-Perez,$^{4,5}$			
       Peter A. Thomas,$^{6}$ 	 			
       \newauthor
       Andrew Benson,$^{7}$    				
       Rachel Asquith,$^{3}$                             
       Jeremy Blaizot,$^{8,9,10}$ 			
       Richard Bower,$^{5}$			        
       \newauthor
       Jorge Carretero,$^{11,12}$ 				
       Francisco J. Castander,$^{11}$	 		
       Andrea Cattaneo,$^{13,14}$			
       Sofia A. Cora,$^{15,16}$				
       \newauthor
       Darren J. Croton,$^{17}$			        
       Weiguang Cui,$^{1}$  			
       Daniel Cunnama,$^{18,19}$  			
       Julien E. Devriendt,$^{20}$			
       \newauthor
       Pascal J. Elahi,$^{21}$				
       Andreea Font,$^{22}$	  		        
       Fabio Fontanot,$^{23}$			
       Ignacio D. Gargiulo,$^{15,16}$	 		
       \newauthor
       John Helly,$^{5}$			
       Bruno Henriques,$^{24,25}$		        
       Jaehyun Lee,$^{26}$				
       Gary A. Mamon,$^{27}$			
       Julian Onions,$^{3}$                                    
       \newauthor
       Nelson D. Padilla,$^{28,29}$ 				
       Chris Power,$^{21}$				
       Arnau Pujol,$^{11,30,31}$				
       Andr\'es N. Ruiz,$^{32,33}$                 
       \newauthor
       Chaichalit Srisawat,$^{6}$ 	                
       Adam R. H. Stevens,$^{21,17}$                       
       Edouard Tollet,$^{34,35}$				
       \newauthor
       Cristian A. Vega-Mart\'inez,$^{15}$ 	
       Sukyoung K. Yi$^{36}$			        
\\
\\
$^{1}$Departamento de F\'isica Te\'{o}rica, M\'{o}dulo 15, Facultad de Ciencias, Universidad Aut\'{o}noma de Madrid, 28049 Madrid, Spain\\
$^{2}$Astro-UAM, UAM, Unidad Asociada CSIC\\
$^{3}$School of Physics \& Astronomy, University of Nottingham, Nottingham NG7 2RD, UK\\
$^{4}$Institute of Cosmology and Gravitation, University of Portsmouth, Portsmouth PO1 3FX, UK\\
$^{5}$Institute for Computational Cosmology, Department of Physics, University of Durham, South Road, Durham, DH1 3LE, UK\\
$^{6}$Department of Physics \& Astronomy, University of Sussex, Brighton, BN1 9QH, UK\\
$^{7}$Carnegie Observatories, 813 Santa Barbara Street, Pasadena, CA 91101, USA\\
$^{8}$Universit\`e de Lyon, Lyon, F-69003, France\\
$^{9}$Universit\`e Lyon 1, Observatoire de Lyon, 9 avenue Charles Andr\`e, Saint-Genis Laval, F-69230, France\\
$^{10}$CNRS, UMR 5574, Centre de Recherche Astrophysique de Lyon ; Ecole Normale Sup\`erieure de Lyon, Lyon, F-69007, France\\
$^{11}$Institut de Ci\`encies de l'Espai, IEEC-CSIC, Campus UAB, 08193 Bellaterra, Barcelona, Spain\\
$^{12}$Port d{'}Informaci\'{o} Cient\'{i}fica (PIC) Edifici D, Universitat Aut\`{o}noma de Barcelona (UAB), E-08193 Bellaterra (Barcelona), Spain.\\
$^{13}$GEPI, Observatoire de Paris, CNRS, 61, Avenue de l'Observatoire 75014, Paris  France\\
$^{14}$Institut d'Astrophysique de Paris, 98bis Boulevard Arago, 75014, Paris, France\\
$^{15}$Instituto de Astrof\'isica de La Plata (CCT La Plata, CONICET, UNLP), Paseo del Bosque s/n, B1900FWA, La Plata, Argentina.\\
$^{16}$Facultad de Ciencias Astron\'omicas y Geof\'{\i}sicas, Universidad Nacional de La Plata, Paseo del Bosque s/n, B1900FWA, La Plata, Argentina\\
$^{17}$Centre for Astrophysics and Supercomputing, Swinburne University of Technology, Hawthorn, Victoria 3122, Australia\\
$^{18}$South African Astronomical Observatory, PO Box 9, Observatory, Cape Town 7935, South Africa\\
$^{19}$Department of Physics and Astronomy, University of the Western Cape, Cape Town 7535, South Africa\\
$^{20}$Astrophysics, University of Oxford, Denys Wilkinson Building, Keble Road, Oxford, OX1\,3RH, UK\\
$^{21}$International Centre for Radio Astronomy Research, University of Western Australia, 35 Stirling Highway, Crawley, Western Australia 6009, Australia\\
$^{22}$Astrophysics Research Institute, Liverpool John Moores University,  IC2, Liverpool Science Park, 146 Brownlow Hill,  Liverpool L3 5RF, UK\\
$^{23}$INAF - Astronomical Observatory of Trieste, via Tiepolo 11, I-34143 Trieste, Italy\\
$^{24}$Institute for Astronomy, ETH Zurich, CH-8093 Zurich, Switzerland\\
$^{25}$Max-Planck-Institut f\"ur Astrophysik, Karl-Schwarzschild-Str. 1, 85741 Garching b. M\"unchen, Germany\\
$^{26}$Korea Institute for Advanced Study, 85 Hoegi-ro, Dongdaemun-gu, Seoul 02455, Republic of Korea\\
$^{27}$Institut d'Astrophysique de Paris (UMR 7095: CNRS \& UPMC), 98 bis Bd Arago, F-75014 Paris, France\\
$^{28}$Instituto de Astrofisica, Universidad Catolica de Chile, Santiago, Chile\\
$^{29}$Centro de Astro-Ingenieria, Universidad Catolica de Chile, Santiago, Chile\\
$^{30}$DEDIP/DAP, IRFU, CEA, Universit\'e Paris-Saclay, F-91191 Gif-sur-Yvette, France\\
$^{31}$Universit\'e Paris Diderot, AIM, Sorbonne Paris Cit\'e, CEA, CNRS, F-91191 Gif-sur-Yvette, France\\
$^{32}$Instituto de Astronom\'ia Te\'orica y Experimental (CONICET-UNC), Laprida 854, X5000BGR, C\'ordoba, Argentina\\
$^{33}$Observatorio Astron\'omico de C\'ordoba, Universidad Nacional de C\'ordoba, Laprida 854, X5000BGR, C\'ordoba, Argentina\\
$^{34}$Observatoire de Paris, GEPI, CNRS \& PSL Research University, 61 Avenue de l'observatoire, F-75014 Paris, France\\
$^{35}$Universit{\'e} Paris Diderot, Sorbonne Paris Cit{\'e}, F-75013 Paris, France\\
$^{36}$Department of Astronomy and Yonsei University Observatory, Yonsei University, Seoul 03722, Republic of Korea\\
 }

\begin{document}

\date{Accepted XXXX . Received XXXX; in original form XXXX}

\pagerange{\pageref{firstpage}--\pageref{lastpage}} \pubyear{2010}

\maketitle

\label{firstpage}

\clearpage

\begin{abstract}

We present a comparison of nine galaxy formation models, eight semi-analytical and one halo occupation distribution model, run on the same underlying cold dark matter simulation (cosmological box of co-moving width $125$\hMpc, with a dark-matter particle mass of $1.24\times10^9$\hMsun) and the same merger trees. While their free parameters have been calibrated to the same observational data sets using two approaches, they nevertheless retain some `memory' of any previous calibration that served as the starting point (especially for the manually-tuned models). For the first calibration, models reproduce the observed $z\!=\!0$ galaxy stellar mass function (SMF) within $3$-$\sigma$. The second calibration extended the observational data to include the $z\!=\!2$ SMF alongside the $z\sim 0$ star formation rate function, cold gas mass and the black hole-bulge mass relation. Encapsulating the observed evolution of the SMF from $z\!=\!2$  to $z\!=\!0$ is found to be very hard within the context of the physics currently included in the models. We finally use our calibrated models to study the evolution of the stellar-to-halo mass (SHM) ratio. For all models we find that the peak value of the SHM relation decreases with redshift. However, the trends seen for the evolution of the peak position as well as the mean scatter in the SHM relation are rather weak and strongly model dependent. Both the calibration data sets and model results are publicly available.

\end{abstract}
\noindent

\section{Introduction} \label{sec:introduction}
Galaxy formation is one of the most complex phenomena in astrophysics as it involves physical processes that operate and interact on scales from the large-scale structure of the Universe (i.e. several Gpcs) down to the sizes of black holes (i.e. sub-pc scales). This enormous dynamic range in scale prevents us from modelling galaxies `ab initio' and therefore any model of galaxy formation depends upon a number of recipes that encode all the physical processes we believe are relevant at all those different scales \citep[see, e.g.][for recent reviews]{baugh_review_2006,Frenk12,Silk12,Somerville15}. These recipes are not precisely known but are each regulated by parameters that are chosen to satisfy observational constraints. This is primarily accomplished by means of one-point functions -- such as stellar mass or luminosity functions, the black-hole bulge-mass relation, the star formation rate density, etc. \citep[e.g.][]{Kauffmann93,Somerville99,Croton06,Bower06,Monaco07,Guo13}, although the first steps have been taken in the direction of extending this to two-point functions \citep[e.g. the two-point correlation function of galaxies,][]{vanDaalen16}.

To calibrate the parameters of a galaxy formation model one picks an observational data set and adjusts the free parameters (noting that some parameters might be fixed during that process) until a sufficient match is obtained \citep[e.g.][]{henriques_mcmc_2009}. Here, `sufficient match' depends on several factors, including the objectives of the science project the particular galaxy formation models are being developed for. Some models are designed to reproduce certain observations better than others -- at the expense of matching the latter. And as we will see later, simultaneously matching multiple observations adds additional degrees of freedom in how to weight the various calibration data sets. In that regard an idealised project comparing galaxy formation codes would use the same automated tuning method on all the models, as well as defining both a clear weighting scheme for the different observations and criteria for calibration. While it would be interesting to adopt such an approach of attaching all the models to the same automated tuning engine, we leave this for a future study. We note that several of the methods incorporated here do not presently contain a fully automated calibration procedure and hence insisting on this approach would have severely limited the scope of this project.

Any observational data input as a calibration constraint cannot be viewed as an output prediction of the model. But in that regards it is important to note that having a model that self-consistently matches the calibration data is a non-trivial and interesting exercise; it shows that there is a plausible physical model that is consistent with the observed Universe -- at least as described by the calibration data. However, properties independent of those used for model calibration can be considered genuine predictions. Of course it is usually the case that observational properties depend somewhat on one another \citep[e.g., the `fundamental metallicity relation'][]{Lara-Lopez10,Mannucci10,Salim15,Lagos16}. The extent of this intrinsic overlap needs to be judged when considering the strength of the prediction. Such an approach separates the resulting galaxy properties into two broad categories: the `prescriptions' and the `predictions'. The extent to which a model `prescribes' depends largely upon what it is intended to be used for. A well-calibrated highly prescribed model will be guaranteed to match a wide range of observations {\it by construction} and may be highly desirable for testing observational pipelines. Alternatively, a model with few prescriptions might be more suitable for examining where the physics of galaxy formation breaks down as there will be a wide range of available predictions.

In a previous study \citep{Knebe15} we applied 14 different galaxy formation models to the same underlying cosmological simulation and the merger trees derived from it. Those models were used with their published parameters and not re-calibrated to the new simulation data. We have seen that using models \textit{as is} introduces model-to-model variations larger than reported in previous comparisons \citep[e.g.][]{Fontanot09,Lu14}. We attributed this to the missing re-calibration to the new simulation and merger trees used for that particular study. In the present work we now both re-calibrate the models and introduce a unified set of observations to produce a common calibration. We investigate how well multiple simultaneously applied constraints are reproduced by the models (i.e. the `prescriptions') and also study how calibrating to different data sets affects some properties that were not used during the calibration).

We stress that the aim of this work is not to investigate the calibration process of galaxy formation models which has been subject of previous works \citep[e.g.][]{Benson14,Rodrigues17}. Our prime objective is to obtain models that can be used to study average, statistical properties of galaxies representative of the observable population. Further we want to understand what is needed to achieve that goal. The common ground for all our models is the same simulation, merger tree, and the observational constraints used for calibration. We further fixed a few more ingredients that enter into each of the models such as the initial mass function of stars, stellar yields (how chemical elements are produced in stars), and the recycled fraction (the fraction of gas available for new star formation). We are therefore left with a selection of galaxy formation models that use similar assumptions, but are different in design and how to model galaxy formation. We are additionally including a halo occupation distribution (HOD) model \citep[\mice,][]{Carretero14} in the comparison: as such models relate numerical data (for a given redshift) to observations in a statistical manner, they provide -- by construction -- an accurate reproduction of the galactic content of haloes. While they do not provide galaxy populations that self-consistently evolve in time, they nevertheless have great value when it comes to interpreting data from on-going and future galaxy redshift surveys \citep[especially for clustering studies,][]{Pujol17}. In fact, the \mice\ model has been used to generate the flagship galaxy mock catalogue for the Euclid satellite mission (which is available at the {\sc CosmoHub}\footnote{\url{https://cosmohub.pic.es}} database).

We have made all the data for this project publicly available. A link for the observational data used for the common calibration (i.e. the so-called `CARNage calibration set')  can be found in \App{app:observations}), and the resulting galaxy catalogues (ca. 40GB of data) are stored on a data server to which access will be granted upon request to the leading author.

\section{The Simulation Data} \label{sec:simulation}
The halo catalogues used for this paper are extracted from 125 snapshots\footnote{It has been shown that the number of snapshots as chosen here is sufficient to achieve convergence to within 5 per cent for galaxy masses \citep{Benson12b}.} of a cosmological dark-matter-only simulation undertaken
using the \textsc{Gadget-3} $N$-body code \citep{Springel05} with
initial conditions drawn from the Planck cosmology
\citep[][$\Omega_{\rm m}=0.307$, $\Omega_\Lambda=0.693$, $\Omega_{\rm
b}=0.048$, $\sigma_8=0.829$, $h=0.677$, $n_s=0.96$]{Planck2013}. We
use $512^3$ particles in a box of co-moving width $125$\hMpc, with a
dark-matter particle mass of $1.24\times10^9$\hMsun. Haloes were
identified with \rockstar\ \citep{Behroozi12} and merger trees
generated with the \textsc{ConsistentTrees} code
\citep{Behroozi12b}. 
Even though both halo finder and tree builder
have changed with respect to \citet{Knebe15}, the files passed to the
modellers were in the same format. Essentially, these changes have
improved the underlying simulation framework upon which this work is
based: the box is bigger, there are more snapshots, the halo
catalogues are more complete at early times and the tree construction
includes patching when objects briefly disappear between snapshots. 

\begin{savenotes}
\begin{table*}
  \caption{Participating galaxy formation models. We list here the reference where the model used in this work has been introduced and where the original calibration details can be found. We further provide columns summarizing the calibration approach (MCMC: Monte Carlo Markov Chain, PSO: Particle Swarm Optimization) and comments that are of relevance for the data sets used in this work.}
\label{tab:models}
\begin{center}
\begin{tabular}{llll}
\hline
code name	& reference				&  calibration approach		& comments  \\
\hline
 \DLB 		& \citet{delucia_sam_2007}	& manual					& SMF  $z\!=\!2$ has not been used for \ctwo\\
 \galics		& \citet{Knebe15}	        		& manual					& \cone\ and \ctwo\ data sets are identical \\
 \galformVGP	& \citet{gp14}				& manual					& --- \\
 \lgalaxy		& \citet{Henriques13}		& automated: MCMC			& for \cone\ also SMF $z\!=\!2$ has been used 	\\
 \mice 		& \citet{Carretero14}			& manual					& neither CGMF nor BHBM for \ctwo	\\
 \morgana 	& \citet{Monaco07}			& manual					& SMF  $z\!=\!2$ has not been used for \ctwo\\
 \sag			& \citet{gargiulo_2014}, Cora et al. (in prep.)		& automated: PSO			& --- \\
 \sage 		& \citet{Croton16}			& manual 					& \cone\ and \ctwo\ data sets are identical\\
 \ysam 		& \citet{lee13}				& manual					& ---\\
\hline
\end{tabular}
\end{center}
\end{table*}
\end{savenotes}

\section{Galaxy Catalogues} \label{sec:catalogues}
The nine participating galaxy formation models are listed in \Tab{tab:models}. While we include a reference to the paper where the model and its parameters are presented in detail, we also refer the reader to the Appendix of \citet{Knebe15} where all the models have been summarized in a concise and unified manner. However, we need to mention that the \sag\ model used here differs from the previous version and corresponds to the one presented in Cora et al. (in prep.): the model includes a revised supernova feedback scheme and a detailed modelling of environmental effects coupled with an improved orbital evolution of orphan galaxies. Further, the \galics\ model used here is exactly the one described in the Appendix of \citet{Knebe15} and \textit{not} the one presented in \citet{Cattaneo17}.

The third column in \Tab{tab:models} provides information about whether the model has been calibrated manually or using some automated technique (to be discussed in more detail below). The last column in \Tab{tab:models} provides some remarks about particulars affecting the common calibration strategy.

\begin{savenotes}
\begin{table*}
  \caption{Observational data sets used for the common calibration approach: stellar mass function (SMF), star formation rate function (SFRF), cold gas mass fraction (CGMF), and black-hole bulge-mass relation (BHBM). The actual data used for this paper can be downloaded using the hyperlink provided in the \App{app:observations}. The last column indicates for which calibration set the data was used.}
\label{tab:constraints}
\begin{center}
\begin{tabular}{lllr}
\hline
observation & redshift	& reference 											& calibration\\
\hline
SMF		& $z\!=\!0$		& \citet{baldry_etal12,li_distribution_2009,Baldry08}				& c01 + c02 \\
SMF		& $z\!=\!2$		& \citet{Tomczak14,Muzzin13,ilbert_etal13,Dominguez-Sanchez11}	& c02 \\
SFRF	& $z\!=\!0.15$		& \citet{Gruppioni15}										& c02 \\
CGMF	& $z\!=\!0$		& \citet{Boselli14}										& c02 \\
BHBM	& $z\!=\!0$		& \citet{McConnell13,Kormendy13}							& c02 \\
\hline
\end{tabular}
\end{center}
\end{table*}
\end{savenotes}

Each model has been applied to the simulation data generating three distinct galaxy catalogues:

\begin{itemize}
\item {\textbf{un-calibrated (uc)}:}
As has been done in \citet{Knebe15} each model was applied without any re-calibration to the new simulation, merger trees and the common assumptions detailed in \Sec{sec:common}.

\item {\textbf{calibration \#1 (c01, SMF at $z\!=\!0$)}:}
The models were calibrated to the provided SMF at $z\!=\!0$ in the range around the SMF knee, $9.95\leq {\rm log}_{10}(M_*/M_{\odot})\leq 11.15$. We will refer to this as calibration \cone.

\item {\textbf{calibration \#2 (c02, SMF at $z\!=\!0 \ {\rm \&}\ 2$ + extra constraints)}:}
In addition to the constraint used for calibration~\#1, the black hole--bulge mass and cold gas mass at $z\!=\!0$, the star formation rate function at $z\!=\!0.15$, and the stellar mass function at $z\!=\!2$ have been used. We will from now on refer to this either as calibration \ctwo\ or simply the `CARNage calibration'. This observational data set is motivated and described in detail in \Sec{sec:observations}.
\end{itemize}

Note that the un-calibrated model will only serve as a connecting point to our previous comparison \citep{Knebe15} and will not enter the main part of this work. And even though results from the two calibration approaches are not mixed together in a single plot, we also chose to use different linestyles for these two  catalogues: results from \cone\ will always be presented as dashed lines whereas the results from \ctwo\  are shown as solid lines. This facilitates comparison and allows for a quicker identification of calibrations in the plots.

\subsection{Common modelling}\label{sec:common}
In this work we have populated one simulation with eight different semi-analytical models of galaxy formation and evolution and one halo occupation model. In order to further align the various galaxy formation models they all assumed a Chabrier initial mass function (IMF), a metallicity yield of 0.02 and a recycled fraction of 0.43.\footnote{Since the \sag\ code does not use an instantaneous recycling approximation but relies on a set of tables with yields and ages for stars in different mass ranges, the yield and recycled fraction are not fixed to these values.} The first uncalibrated (\uc) stage has populated the simulation described in \Sec{sec:simulation} using the models with parameters as previously published (see \Tab{tab:models} for the list of references), but with a Planck cosmology and the assumptions just mentioned with regards to IMF, yield and recycle fraction. The results from this uncalibrated comparison are presented in \App{app:uc} which demonstrates that changing the simulation and merger trees has not changed our findings in \citep{Knebe15} and that the same results as those obtained previously are recovered, respectively.

\subsection{Calibration}
Models of galaxy formation and evolution provide a research tool that can be used to explore a vast range of dynamical scales, from stellar nurseries to the large scale structures seen in the observed Universe. Covering such an enourmous dynamical range in scales implies that approximations are needed in order to reduce the computational cost to a feasible level. Moreover, not all the relevant physical processes are well understood and constrained observationally. Thus, any model of galaxy formation has free parameters or parameters that can vary within a reasonable range constrained by direct observations. These free parameters  summarize our ignorance with respect to the physical processes pertinent for the formation of galaxies. They are chosen by comparing certain model properties to observations. This choice depends on the physical processes and cosmic times one is interested in. We refer to calibration as the process of choosing these free parameters.

For this work two sets of observational data were given, one for \cone\ and one for \ctwo, but the specific calibration approach was left to each group of modellers, so it could be close to those used for their published models. 

\subsubsection{Calibration approach}
The calibration approach can be split into two distinct categories: manual and automated. In \Tab{tab:models} we provide this information for each of the models and describe these two classes here in more detail:
\begin{itemize}
   \item {\textbf{Manual calibration}:} This calibration approach explores the parameter space in a manual way, basing the choices of parameters to be varied on previous knowledge \citep[e.g.][]{Lacey16}. When this approach is used, only a handful of the total free model parameters are modified. The others are inherited from previous work and thus, they have intrinsically been chosen with respect to a certain set of observations. For instance, the SMF at $z=0$ only constrains physical processes related to the star formation and feedback in galaxies, but not other processes such as the growth of black holes or the growth of bulges \citep{Rodrigues17}. However, in \cone, the models that were calibrated manually are leaving the parameters controlling these later processes as they were in previous incarnations of the models and thus, indirectly inheriting previous knowledge.

   \item {\textbf{Automated calibration}:} This calibration approach explores either the whole parameter space or a subset in a numerical manner. For this work there are models either using Monte Carlo Markov Chains \citep[MCMC,][used by \lgalaxy]{henriques_mcmc_2009} or a particle swarm optimisation \citep[PSO,][used by \sag]{ruiz2014}. When models are calibrated using a numerical exploration of the parameter space, there is a choice on which parameters are set free. If some parameters are set free but no adequate observable is used during the calibration, the resulting best set of parameters would not be a robust choice as they were basically unconstrained by the choice of the observational data set. While PSO focuses on quickly finding a best fit model given the observational contraints and chosen priors, the MCMC approach allows to have a full understanding of the statistical significance of a given set of parameters. However, in this case special care needs to be taken on how the observational errors are considered \citep[e.g.][]{Benson14}. Beyond these two numerical approaches, semi-analytical models have also been calibrated using emulators \citep[][]{Bower10}, an approach not used by any of the models presented here. 
\end{itemize}

As detailed above, the calibration process often entails that the models retain some `memory' of any previous calibration; this is especially true for the models that are tuned manually. This may particularly affect the `prescriptions' and `predictions' of the models. If a model usually uses, for instance, the evolution of the cosmic star formation rate density or the stellar-to-halo mass relation as one of its constraints for the parameters, this might still be reflected in the catalogues presented here. Practically what this means is that in general each model did not start each calibration process with an entirely clean slate but rather began with a set of parameters previously understood to produce a not completely unreasonable result. 

\subsubsection{Calibration steps and parameter changes}
The calibration~\#1 catalogue (\cone) starts from the parameters used for \uc\ (i.e. the original parameters presented in the model's reference publication) and uses the observed SMF at redshift $z\!=\!0$ as the only constraint. The calibration~\#2 catalogue (\ctwo) starts from the parameter values found for \cone, but now adds the provided SFRF, CGMF, and BHBM relations at redshift $z\sim 0$ as well as the SMF at $z\!=\!2$ to the constraints (see \Sec{sec:observations} for further details). The intention here was to add the minimum number of additional calibration datasets which simultaneously constrained the key physical processes required for a model of galaxy formation. In order to limit the number of different calibrations for each model we decided to simultaneously apply all four of the constraints in \ctwo. At this stage each modeller was given the freedom to assign weight to each of the additional constraints as they saw fit. The idea here was not to expect a perfect fit to all constraints but rather to provide a `best case' solution to the entirety of the constraints. 

Here there is a brief discussion of the changes seen in the model parameters when going from \uc\ to \cone\ and eventually \ctwo. Most of the models applied changes to the same parameters when calibrating to \cone\ and \ctwo. The only exception to this was \galformVGP\ where active galactive nuclei (AGN) feedback was changed for \cone\ but not for \ctwo\ and the level of stellar feedback only changed for \ctwo. In general the parameter changes required primarily focused on changing the feedback (either stellar or AGN) and/or the treatment of mergers. The latter impacts upon bulge, disc, and black hole growth which are amongst the parameters modified by most (but not all) modellers. However, the only changes for \galics\ were the limiting mass above which the accretion of gas onto galaxies is shut down \citep[the $M_{\rm shock\,max}$ parameter in Appendix 2 of][]{Knebe15} and a lowering of all outflow rates by a factor $\sim 1.3$. The HOD model \mice\ starts from the galaxy luminosity function and then converts it to stellar masses: for \cone\ this conversion has been updated to obtain a better fit to the provided SMF at $z=0$. For \ctwo\ this has been implemented even more self-consistently for all redshifts while also changing the calibration of the star formation rate. We close by mentioning that some of the models only changed a few parameters for the calibration sequence (e.g. \ysam\ only adjusted four parameters) whereas models applying an automated calibration process feature changes in substantially more parameters (e.g. for \lgalaxy\ seven parameters were varied).

\subsection{The `CARNage calibration' data set} \label{sec:observations}
The observational data sets used for the calibration are detailed in \App{app:observations} where we also provide a link for download. They are designed to constrain different aspects of galaxy formation and evolution, yet observationally are well established. Galaxy formation models aim at encapsulating the main physical processes governing galaxy formation and evolution in a set of coupled parameterised differential equations. These parameters are not arbitrary but set the efficiency of the various physical processes being modelled -- and they have to be tuned using observational data. All models are hence calibrated against a key set of observables by which different physics in the models are fixed -- depending on the actual observations used. The observational data sets used for the `CARNage calibration' where chosen in a way to be as complementary as possible (references are given in \Tab{tab:constraints}):

\paragraph*{Stellar Mass Function (SMF):} The SMF at $z\!=\!0$ is a fundamental property that can constraint alone much of the stellar formation and feedback processes that shape the formation and evolution of galaxies. In addition to the observed stellar mass function at $z\!=\!0$ (that formed the basis of calibration \cone) we added the observed stellar mass function at $z\!=\!2$. While this is still somewhat observationally uncertain it provides a constraint on the evolution of the stellar mass assembly. For the process of calibration we further agreed that when tuning to the SMF, the model stellar masses should be convolved with a $0.08(1+z)$ dex scatter to account for at least part of the observational errors in measuring stellar masses. Note, while each code wrote to the resulting output files the internally calculated stellar masses for each galaxy, these masses were subjected to this scatter only during the calibration process. And in order to mimic this effect when again comparing the model stellar masses to observations, we modify them in the same way while reading them in from the galaxy catalogues. I.e. our analysis pipeline convolved the model data with aforementioned scatter before comparing to observations.

\paragraph*{Star Formation Rate Function (SFRF):} While the SMF constrains the amount of mass in stars, the star formation rate function (SFRF) quantifies the change in the SMF as a function of time. It provides information on the efficiency of transforming (cold) gas into stars (and the fraction of stars whose mass is lost back as gas) in a given galaxy and hence drives any shape changes or evolution of the SMF. In the models, the cold gas fraction at low redshift is thus determined both by the star-formation law and by the effectiveness of quenching processes, such as AGN feedback and stripping, that remove cold gas from galaxies.
\paragraph*{Cold Gas Mass Fraction (CGMF):} Stars can only form when sufficient cold gas is present in a galaxy. Therefore, an important tracer for star formation is the fraction of cold gas to stellar mass. 
\paragraph*{Black Hole--Bulge Mass Relation (BHBM):} The observed relation between the mass of a galaxy's central black hole and the mass (or velocity dispersion) of a galaxy's bulge suggests that the central black hole plays a key role in galaxy evolution. The black-hole bulge-mass (BHBM) relation is used to constrain black hole masses as there is otherwise a large degeneracy between the black hole mass and the AGN feedback efficiency in the models \citep{Henriques09,Croton06,bower_agn_2006,Mutch13a,Croton16}: black hole growth plays a critical role in galaxy evolution \citep[e.g.][for recent observational results from the MaNGA survey]{Cheung16}. \\

We refrained from using galaxy luminosities, even though they are directly observable, because their calculation requires an additional layer of modelling which adds extra complexity to the comparison -- something to be avoided for this paper and left for a future study. 


\subsection{Common analysis}\label{sec:common_analysis}
All provided galaxy catalogues have been post-processed with a common analysis pipeline that is also made available alongside the numerical and observational data sets. While catalogues contain galaxies with much smaller masses, we limit all of the comparisons presented here to galaxies with stellar masses $M_{*}>10^{9}$\hMsun\ -- a mass threshold appropriate for simulations with a resolution in dark matter mass comparable to the Millennium simulation \citep[see][]{Guo11}. 

We further remark that the points for the models have been obtained by binning the data using logarithmically spaced bins on the $x$-axis and then calculating the median in that bin for both the $x$- and $y$-value.

\section{Calibration with the local SMF} \label{sec:calibration01}
Calibration \#1 presented in this section only uses the SMF at redshift $z\!=\!0$ as a constraint and we will explore its {\it prescription} in \Sec{subsec:calibration-c01}. We continue in \Sec{subsec:beyond-c01} to examine these \cone\ models compared to the observations used later on for calibration \ctwo. 

\subsection{The Calibration} \label{subsec:calibration-c01}
 \begin{figure}
   \includegraphics[width=\columnwidth]{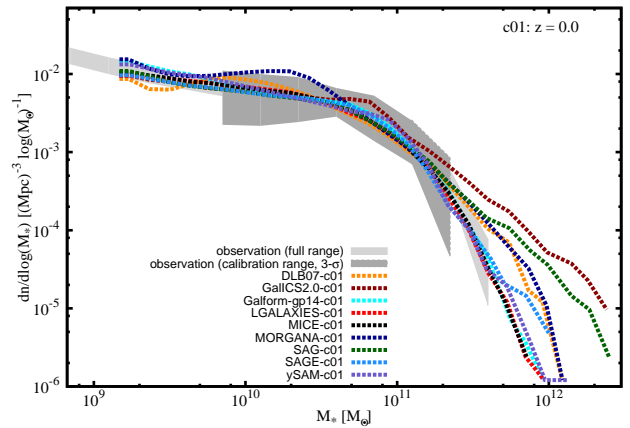}
   \caption{Stellar mass function at redshift $z\!=\!0$ for calibration \#1 (c01) compared to the observational data as described in detail in \App{app:SMF}. The light-shaded region shows the compilation and its $1$-$\sigma$ errors whereas the dark shaded region shows the mass range used to calibrate the models and its $3$-$\sigma$ errors.}
 \label{fig:SMFz0-c01}
 \end{figure}
 
 \begin{table}
  \caption{$\chi^2$-values for models with respects to observational data. The first column after the model name is for the SMF at $z\!=\!0$ in the calibration \cone\ data set; the following two columns are for the SMF at $z\!=\!0$ and $z\!=\!2$, respectively, in the calibration \ctwo\ data set. Note, only the range used during model calibration has entered into the $\chi^2$-calculation.}
\label{tab:SMFchi2}
\begin{center}
\begin{tabular}{ll|ll}
\hline
model & $\chi^2_{z\!=\!0,\rm c01}$ & $\chi^2_{z\!=\!0,\rm c02}$ & $\chi^2_{z\!=\!2,\rm c02}$  \\
\hline
\DLB 			& 3.0		& 30.	0	& 41.0\\ 
\galics 			& 0.8		& 0.8		& 19.0\\ 
\galformVGP 		& 0.9		& 3.4		& 14.0\\ 
\lgalaxy 			& 0.7		& 1.7		& 2.0\\ 
\mice	 		& 0.5		& 0.9		& 0.91\\ 
\morgana 			& 11.7	& 9.2		& 83.0\\ 
\sag	 			& 0.5		& 2.1		& 0.19\\ 
\sage	 		& 0.2		& 0.2		& 15.0\\ 
\ysam	 		& 1.4		& 5.7		& 44.0\\ 
\hline
\end{tabular}
\end{center}
\end{table}

For calibration \#1 the individual model parameters have been tuned to the same (compiled) stellar mass function (in the mass range $9.95\leq {\rm log}_{10}(M_*/M_{\odot})\leq 11.15$).\footnote{Remember, this mass range -- bracketing the knee of the SMF -- was agreed upon by all modellers during the Cosmic CARNage workshop.} We compare the \cone\ models against the used observational SMF with the latter shown in \Fig{fig:SMFz0-c01} as a light shaded region (see \App{app:SMF} for more details about this compilation of observations); we additionally indicate the mass range used for the model calibration as a dark shaded region (3-$\sigma$ error bars). Most of our galaxy formation models lie well within that `3-$\sigma$ area' as indicated by the low values given in \Tab{tab:SMFchi2} where we list the respective $\chi^2$-values (alongside the corresponding values for the other calibration and redshifts to be discussed later). Comparing this to the equivalent Fig.~2 of \citet{Knebe15} (as well as the uncalibrated analog in \App{app:uc}) we see a clear tightening when using the SMF at $z\!=\!0$ as a common constraint. The most apparent deviations from the observations and amongst the models is at the high-mass end where there are only a few objects with little constraining power: the data in that range is dominated by systematic errors which is why this range has been excluded from the calibration set.

\subsection{Beyond the calibration} \label{subsec:beyond-c01}
Even though the models have not been (actively) calibrated against anything else but the stellar mass function at redshift $z\!=\!0$, we will now study the model data for the reminder of the observational `CARNage calibration' set. This allows us to investigate how the residual scatter seen for the SMF in \Fig{fig:SMFz0-c01} propagates into other galaxy properties and their respective correlations. For most of the subsequent plots  in this sub-section (except the BHBM relation) the `CARNage set' will be represented as a shaded region indicating 1-$\sigma$ error bars.

\subsubsection{Star Formation Rate Function} \label{subsec:SFRFc01}
 \begin{figure}
   \includegraphics[width=\columnwidth]{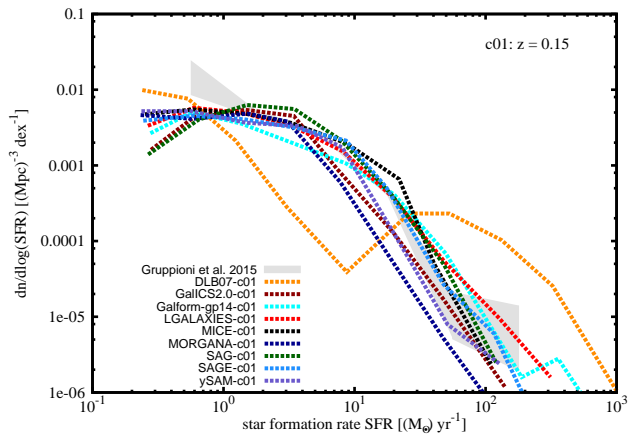}
   \caption{Star formation rate function at redshift $z\!=\!0.15$ for calibration \#1 (c01). The shaded regions shows the observations and its $1$-$\sigma$ errors.}
 \label{fig:SFRF-c01}
 \end{figure}
In \Fig{fig:SFRF-c01} we present the SFR distribution function, i.e. the number density of galaxies in a certain SFR interval compared to the observations of \citet{Gruppioni15}. Within the range of models considered here, most of them already lie close to the observations and they follow the same general trend even {\it before} this is used as a constraint. i.e. the form of the SFRF is largely already imposed by the requirement of matching the SMF at $z\!=\!0$. However, while SFR and stellar mass are certainly connected, this relation has to be viewed with care because of the recycle fraction of exploding stars: the integral over the SFR is not necessarily the total stellar mass. And we have found (though not explicitly shown here) that requesting the \DLB\ model (and to a lesser extent \morgana, too) to fit the provided SMF at $z\!=\!0$ degrades the quality of the matching to the SFRF: when using the uncalibrated \uc\ data for \DLB\ (and \morgana) we find that the SFRF for the two models lies within the observed 1-$\sigma$ range (not explicitly shown here).

\subsubsection{Cold Gas Mass Fraction}
 \begin{figure}
   \includegraphics[width=\columnwidth]{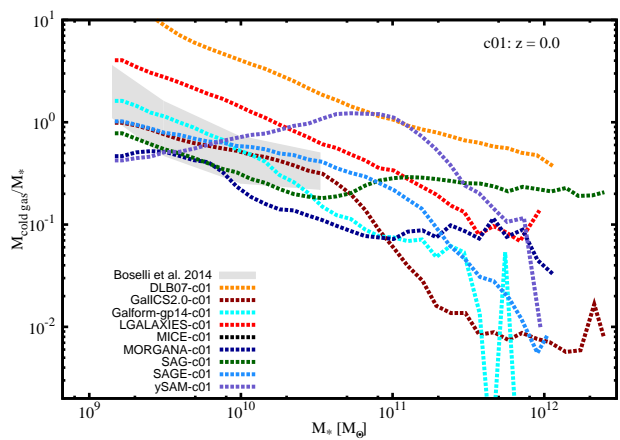}
   \caption{Cold gas fraction as a function of stellar mass at redshift $z\!=\!0$ for calibration \#1 (c01). The shaded regions shows the observations and its $1$-$\sigma$ errors.}
 \label{fig:CGMF-c01}
 \end{figure}
The cold gas fraction in galaxies as a function of stellar mass is shown in \Fig{fig:CGMF-c01}. In this figure, model galaxies are compared to the observational data from \citet{Boselli14}. Almost all the models (bar \ysam) reproduce the declining trend seen in the observations. However, leaving this property unconstrained leads to a substantial model-to-model variation amplitude-wise. We can already deduce some interesting conclusions from this figure: for instance, if unconstrained by the CGMF both \DLB\ and \lgalaxy\ would prefer higher cold gas mass fractions than observed.  

\subsubsection{Black Hole--Bulge Mass Relation}
\begin{figure}
   \includegraphics[width=\columnwidth]{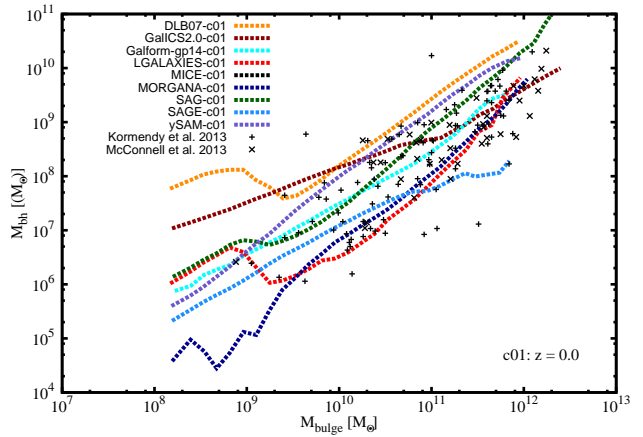}
   \caption{Black hole vs. bulge mass at redshift $z\!=\!0$ for calibration \#1 (c01) alongside the observational data.}
 \label{fig:BHBM-c01}
 \end{figure}
In \Fig{fig:BHBM-c01} we examine how the BHBM relation is reproduced by the models without constraining to it -- in comparison to the observational data of \citet{McConnell13,Kormendy13}. While one might conclude that such an agreement is related to the requirement of fitting the SMF at $z\!=\!0$, we confirm (though not explicitly shown here) that using the uncalibrated \uc\ data set gives a plot very similar to \Fig{fig:BHBM-c01}. This could be interpreted as the BHBM being largely insensitive to the parameters governing the SMF.

\subsubsection{Stellar Mass Function at $z\!=\!2$}
\begin{figure}
   \includegraphics[width=\columnwidth]{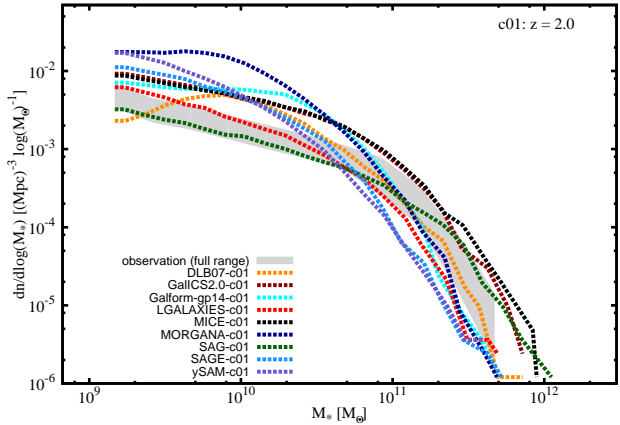}
   \caption{Stellar mass function at redshift $z\!=\!2$ for calibration \#1 (c01). The shaded regions shows the observations and its $1$-$\sigma$ errors.}
 \label{fig:SMFz2-c01}
 \end{figure}

All the previous observational data sets were restricted to (or close to) redshift $z\!=\!0$. We now extend our investigations to higher redshifts by considering the stellar mass function at $z\!=\!2$. We can observe in \Fig{fig:SMFz2-c01} that this poses a challenge for the majority (if not all) of the models. The scatter is considerably larger than for redshift $z\!=\!0$. We reconfirm that reproducing high-redshift observations is a challenge for most models: only \sag\ \& \lgalaxy\ lie within the $1$-$\sigma$ error range and in both cases the physics in the model was tuned to reproduce the stellar mass function evolution \citep{Henriques15}.

We see that all the other models lie above the observations at small-to-intermediate masses, indicating that when unconstrained by this observation they predict the presence of a large number of small objects that are not observed.

\subsection{Discussion}
In this work we required all models to use the same observational data during parameter calibration and, for this first approach, only use the stellar mass function at redshift $z\!=\!0$. We find that the scatter seen in Fig.2 of \citet{Knebe15} (and also \Fig{fig:SMFz0-uc} here) substantially tightens and now lies within the 3-$\sigma$ error bars of the observations -- at least for the mass range considered during the calibration. As another example, the model star formation rate function and the cold gas fraction follow the observational trends reasonably well, albeit still showing pronounced model-to-model variations. We remark that while for some of the models the change from the uncalibrated data set to calibration \cone\ clearly improved the match to the SMF at $z\!=\!0$, this was accompanied by a degrading of the match to other observational data. This is particularly prominent for the CGF where the uncalibrated data set (not shown here) shows far less model-to-model variation than seen in \Fig{fig:CGMF-c01}.

We have further found that the model SMF at redshift $z\!=\!2$ exhibits scatter to the same degree as found for models when not re-calibrated \citep[cf. Fig.2 in][again]{Knebe15}. We reconfirm the well-known problem that galaxy formation models readily overproduce low-mass galaxies at high redshift \citep[e.g.][]{Fontanot09,Weinmann12,Somerville15,Hirschmann16}.

\section{Calibration with the CARNage data set} \label{sec:calibration02}
The `CARNage calibration' data set has been introduced and motivated already in \Sec{sec:observations} and its details (including a link to a public database) are given in \App{app:observations}. All models have now either manually or automatically tuned their parameters with that particular set simultaneously. However, the modellers were given the freedom to assign weights to each observation individually: different models might be designed to perform better for some predictions/prescriptions and hence put more emphasis on reproducing, for instance, the black hole--bulge mass relation as opposed to the cold gas fraction. In passing we note that there is no difference between the \cone\ and \ctwo\ galaxy catalogues for \galics\ and \sage: their respective catalogues are based upon the same set of calibration parameters.\footnote{While all models went through the \ctwo-calibration process, both \galics\ and \sage\ eventually realized that the best parameters actually agree with the ones already obtained during the \cone\ calibration.}

\subsection{The Calibration}

\subsubsection{Stellar Mass Function at $z\!=\!0 \ {\rm \&}\ 2$}
 \begin{figure}
   \includegraphics[width=\columnwidth]{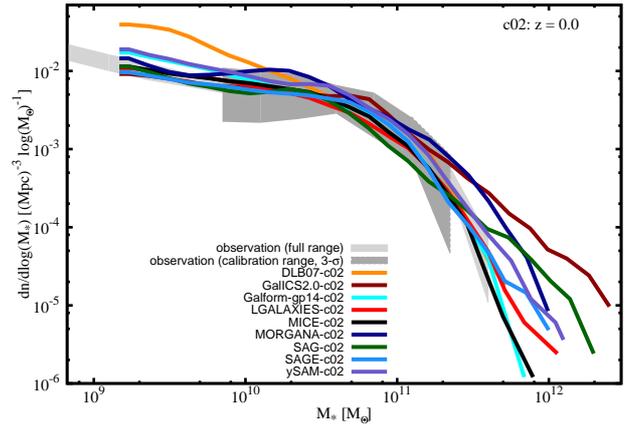}
   \includegraphics[width=\columnwidth]{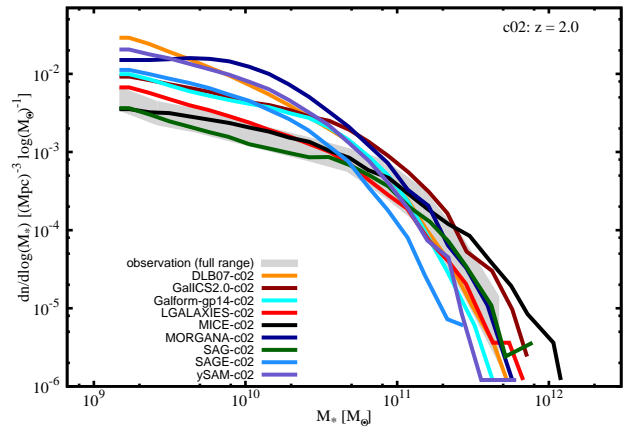}
   \caption{Stellar mass function at redshift $z\!=\!0$ (top panel) and $z\!=\!2$ (bottom panel) for calibration \#2 (c02).}
 \label{fig:SMFz02-c02}
 \end{figure}

In \Fig{fig:SMFz02-c02} we show the stellar mass function as given by the models for both redshift $z\!=\!0$ (upper panel) and $z\!=\!2$ (lower panel) in comparison to the observational data. These plots are again accompanied by the respective $\chi^2$-values listed in \Tab{tab:SMFchi2}. We find that adding the new constraints (including one at higher redshift) reduces the agreement at redshift $z\!=\!0$ for most models with the scatter between the models clearly increased. This scatter now spans the 3-$\sigma$ band at redshift $z\!=\!0$ (for the considered mass range, see \Sec{subsec:calibration-c01}) and they still show substantial variation at higher redshift: at $z\!=\!2$ only the \lgalaxy, \sag\ and \mice\ models lie close to the observational band across all stellar masses.

Even when constraining the SMF at $z\!=\!2$ most models clearly overproduce galaxies at the low-mass end at $z\!=\!2$, as already noted before. Suffice it to say that the $z\!=\!2$ SMF provides significant tension and it is already well known that it is difficult to concurrently obtain good fits to the SMF at both redshifts \citep[e.g.][]{Fontanot09,Weinmann12,Somerville15,Hirschmann16}.  A more in-depth study of the physics of simultaneously matching the redshift $z\!=\!0$ and $z\!=\!2$ SMF will be presented in a companion paper (paper II, Asquith et al. 2017). That work reasserts that tension exists to some extent in all the semi-analytic models of galaxy formation studied here. In that paper we investigate the evolution of the stellar mass function with redshift for all galaxies (passive and star forming) up to $z\!=\!3$ and find that all the models, despite the wide range of physical processes implemented, produce too many small galaxies at high redshift. These excess galaxies appear to be mainly star-forming and are not  present in the latest observations \citep{Mortlock15,Muzzin13}.

But we also noted before (see \Sec{subsec:SFRFc01}) that there is an interplay between matching the SMF and SFRF, especially for the \DLB\ and \morgana\ models. And the increased model-to-model variation for the SMF seen here might also be attributed to an improved matching of the SFRF as presented in the following sub-section.

\subsubsection{Star Formation Rate Function}

 \begin{figure}
   \includegraphics[width=\columnwidth]{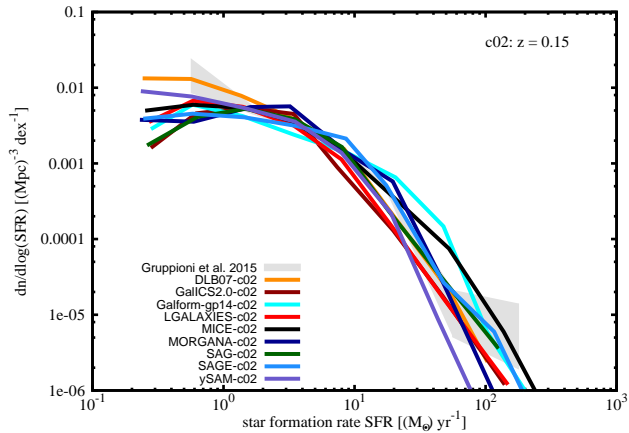}
   \caption{Star formation rate function at redshift $z\!=\!0.15$ for calibration \#2 (c02).}
 \label{fig:SFRF-c02}
 \end{figure}

In \Fig{fig:SFRF-c02} we can appreciate that when adding the SFRF as a constraint (along with the four additional constraints used in stage \ctwo) the scatter seen before in \Fig{fig:SFRF-c01} noticeably tightens. 

\subsubsection{Cold Gas Mass Fraction}
 \begin{figure}
   \includegraphics[width=\columnwidth]{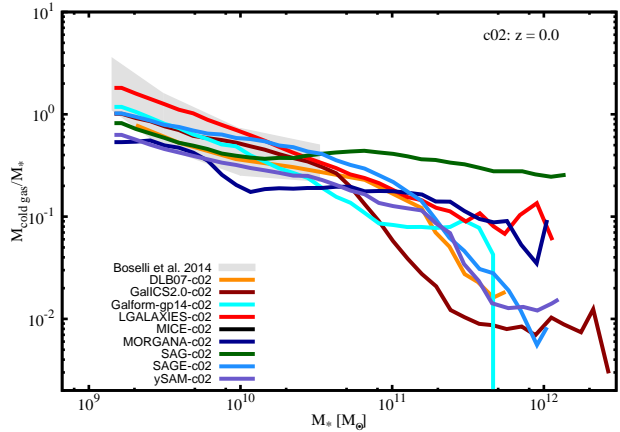}
   \caption{Cold gas fraction as a function of stellar mass at redshift $z\!=\!0$ for calibration \#2 (c02).}
 \label{fig:CGMF-c02}
 \end{figure}

We have seen before that leaving the cold gas fraction unconstrained leads to a substantial model-to-model variation in amplitude. This is somewhat alleviated by using it as a calibration constraint as can be verified in \Fig{fig:CGMF-c02}: all models lie within the 2-$\sigma$ range of observations. The most prominent change happens for \ysam, which had a generally rising trend when the observed CGMF was not used as a constraint.

\subsubsection{Black Hole--Bulge Mass Relation}
\begin{figure}
   \includegraphics[width=\columnwidth]{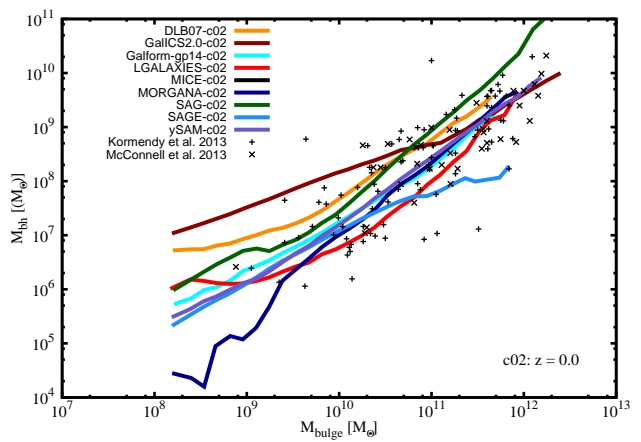}
   \caption{Black hole vs. bulge mass at redshift $z\!=\!0$ for calibration \#2 (c02).}
 \label{fig:BHBM-c02}
 \end{figure}

For the BHBM relation -- as presented in \Fig{fig:BHBM-c02} -- we find that all models lie within the observations with a slight tightening of the range when this relation is added as a constraint to the models. Note that in practice several models already included the BHBM relation as one of their usual constraints (and hence keeping its `memory'); we therefore did not expect a considerable change when adding it as a constraint for data set \ctwo.

The rather large black hole masses at the low-$M_{\rm bulge}$ end for \galics\ -- in comparison to the other models and even when adding this relation as a constraint -- are related to the treatment of major mergers, which instantaneously converts $1$ per cent of the gas into a central black hole (BH), while most of the remaining gas is ejected to the high mass-loading factors of low-mass galaxies and hence not available for star formation. In mergers with $M_{\rm gas}\gg M_*$, this assumption can lead to remnants with $M_{\rm bh}/M_{\rm bulge}\gg 0.01$.

\subsection{Beyond Calibration}
While all distribution functions and correlations (apart from the SMF at $z\!=\!0$) studied for data set \cone\ in \Sec{subsec:beyond-c01} could be considered predictions, the \ctwo\ data set has been constrained by the local SFRF and SMF as well as the redshift $z\!=\!2$ SMF. But will this be sufficient to `predict', for instance, the so-called {`Madau-Lilly'} plot \citep{Madau14,Madau96,Lilly96}, i.e. the evolution of the cosmic star formation rate density (cSFRD). We note that this plot is closely related to the the SMF as well as the SFRF: for instance, the integral over all masses of the SMF at a fixed redshift corresponds to the integral of the cSFRD up to that redshift; further, the integral over all SFR values in the SFRF gives the point in the cSFRD at the corresponding redshift. We have previously seen that matching higher redshift observations is far from trivial. This discrepancy is well known \citep[e.g.][]{Fontanot09,Weinmann12,Somerville15,Hirschmann16} and is somewhat driven by the fact that the integral under the observational curve is inconsistent with the observed stellar density today, a requirement that is enforced in the models \citep[][]{Nagamine04, Dave09, Wilkins08}, but influenced and modified by the recycled fraction as mentioned before.

\begin{figure}
   \includegraphics[width=\columnwidth]{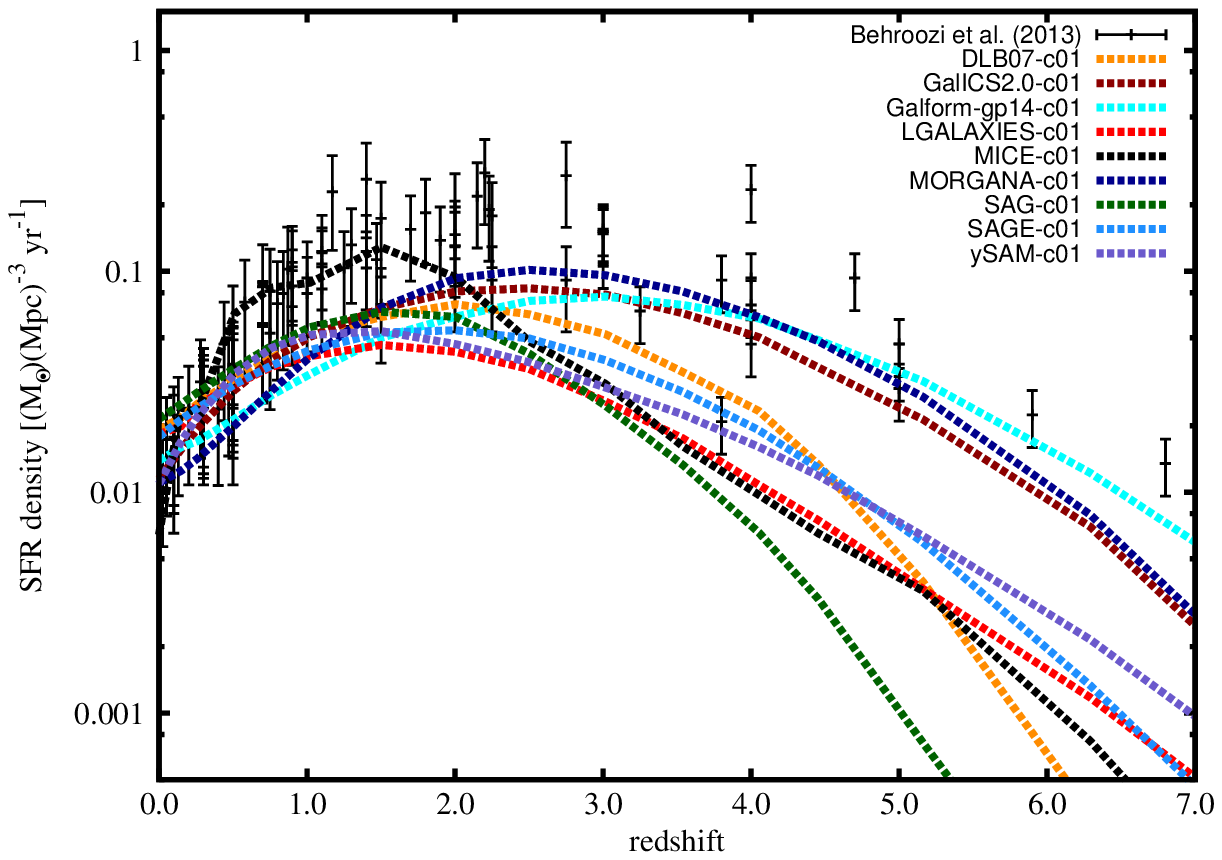}
   \includegraphics[width=\columnwidth]{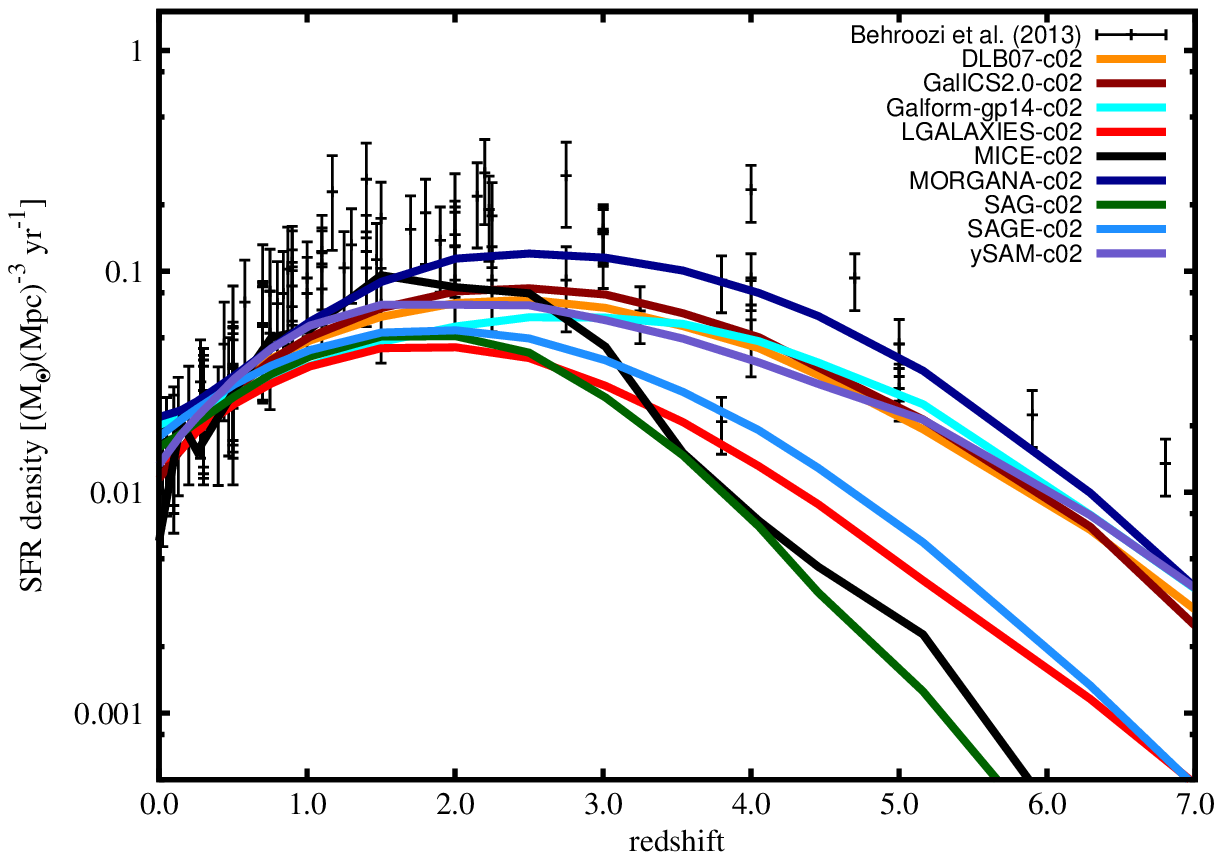}
   \caption{Evolution of the star formation rate density.}
 \label{fig:SFRzred}
 \end{figure}

\Fig{fig:SFRzred} shows the results for the evolution of the cosmic star formation rate density for each model, with observational data taken from \citet{Behroozi13}. We find that all the models reproduce the form of the star formation rate density evolution with a pronounced peak and a significant decrease at late times of approximately the observed amplitude. While the model-to-model variations appear to be the same for both calibrations \cone\ and \ctwo, we note that individual models substantially change their behaviour from one to the other. For instance, the \DLB\ model has a higher star formation rate at early times in \ctwo. \sag\ and \lgalaxy\ are towards the bottom end of the star formation rates at early times as is the HOD model, \mice. For \lgalaxy\ this is related to a lack of resolution in the provided $N$-body simulation used here: in \citet{Henriques15} it has been shown that with the addition of the Millenium-II simulation with an increased mass resolution, the model matches the observations at high-$z$ (but still falls well below at $z\!=\!1$-$2$). For \mice\ this is due to not having calibrated at those high redshifts; it only applied evolutionary correction up to $z\sim 3$. However, these are the three models that provide the best match to the SMF at $z\!=\!2$. We close by mentioning that both the \sage\ and \ysam\ models usually use the cSFRD as a constraint during their calibration; however, they utilize observational data presented in \citet{Somerville01} (\sage) and \citet{Panter07} (\ysam), respectively. 

We end this sub-section with a word of caution: we applied a general lower limit for galaxies entering our plots, i.e. $M_{*}>10^{9}$\hMsun. But this will bias the results presented in \Fig{fig:SFRzred} as it leaves out star formation taking place in smaller objects which is even more relevant at early times and high redshifts, respectively. In order to investigate the size of this effect we have performed two different tests. First, we have lowered the mass threshold in several steps from $M_{*,\rm cut}=10^{9}$\hMsun\ down to $M_{*,\rm cut}=10^{6}$\hMsun\ always using the galaxies as provided in the respective catalogue. We confirm that this does not alter the behaviour of the models for redshifts $z<2$, but increase the SFRD at higher redshifts bringing them into closer agreement with the observations. Second, we performed a more elaborate test to investigate resolution effects entering this plot: instead of simply adding the galaxies below the resolution limit, at all redshifts we fit the SMF to a Schechter function of the form $dn/d\log{M_{*}}=n (M_{*}/M_0)^p \exp{(-M_{*}/M_{0})}$ (with free parameters $n,p,M_{0}$) and the relation between SFR and stellar $M_{*}$ to a power-law ${\rm SFR}=A M_{*}^q$ (with free parameters $A,q$). These best-fit functions are then used to add the contribution from galaxies with $M_{*}<10^{9}$\hMsun\ to the cSFRD given as $\int_{M_{*}=10^{6}\hMsun}^{M_{*}=10^{9}\hMsun} {\rm SFR}(M_{*}) dn/dM_{*} \ dM_{*}$. We confirm again that the conclusions drawn from \Fig{fig:SFRzred} for the comparison between models remain unchanged when post-correcting in this way, but the curves are shifted upwards bringing them into better agreement with the observations.
However, as both these methods have significant uncertainties in the correction to be applied, we decided to simply describe their effects rather than incorporating them in the figure. 

\subsection{Discussion}
We have seen in the previous section that constraining the SMF at redshift $z\!=\!0$ alone will already be sufficient to give the right trends in correlations and shapes of distribution functions, but with noticeable model-to-model amplitude variation as well as large offsets to the observations for some models -- especially when higher redshifts are considered. In particular, models have difficulty producing the correct number density of low-mass galaxies at higher redshift \citep{Fontanot09,Weinmann12,Somerville15,Hirschmann16} -- a problem that has been addressed by adding some form of preventive/ejective stellar feedback \citep{White15,Hirschmann16} and/or modulation of reaccreation \citep{Henriques13} to the models in order to recover the evolution of the SMF.

When adding additional (orthogonal) constraints, the scatter generally tightens and models move towards closer agreement with each other. This is most prominent for the CGMF, i.e. the model-to-model variation substantially reduces when comparing \Fig{fig:CGMF-c01} (\cone\ calibration that only uses the stellar mass function at redshift $z=0$) to \Fig{fig:CGMF-c02} (\ctwo\ calibration that now includes additional observational data at different redshifts as described in \Sec{sec:catalogues}). Further, the SFRF scatter also decreases going from \cone\ to \ctwo. We can see that there is one model in particular, \DLB, which is a clear outlier in both the CGF and SFRF for \cone, and shows dramatic improvement going to \ctwo. The other models that have a notable change in CGF have an accompanying change in SFRF -- typically gas fractions drop and SFR goes up. However, for the SMF at redshift $z=0$ we note a marginal increase in the scatter as it is no longer the sole constraint.

When moving to a non-calibrated (yet related) property like the evolution of the star formation rate density, we find that switching from \cone\ to \ctwo\ will not tighten the scatter across models. Rather it impacts upon certain models more than others, e.g. \DLB\ sees an increase in amplitude at higher redshift; as the \DLB\ model did not use the SMF at $z\!=\!2$ when calibrating, this improvement is mainly due to the use of SFRF at $z\!=\!2$ as an additional constraint.

We will return to the evolution of the SMF in a companion paper (paper II, Asquith et al., in prep.) where we more closely investigate the evolution of the stellar mass function with redshift for the same models presented here, and we separate the galaxies into `passive' and `star-forming' classes.


\section{Stellar-to-halo mass ratio} \label{sec:SHM}
All the galaxy formation models presented and studied here populate given dark-matter haloes with galaxies whose properties depend on the particulars of the formation history of the halo they are placed in. Subsequent galaxy evolution then shapes the galaxy stellar mass function leading to the well-known shape that can be roughly described as two power-laws: at the low-mass end supernova feedback suppresses star formation, whereas various feedback mechanisms due to the accretion of gas onto a BH are responsible for a suppression of star formation at the high-mass end \citep[see][for a succinct review]{Silk12}.

None of the galaxy properties in the previous Sections have been related to the halo the galaxy resides in. Here we provide a link between the two by investigating the ratio between stellar and halo mass (SHM) as a function of halo mass. This ratio -- normalized by the cosmic baryon fraction -- can also be interpreted as an `efficiency of star formation' that depends on halo mass, i.e. how many of the maximally available baryons have been converted into stars. Its correlation with halo mass shows a distinct peak whose position coincides with the knee of the SMF. The temporal evolution of the SHM relation has caught a lot of attention recently: there appears to be no consensus yet whether it is evolving with redshift or not. Some authors claim that the peak position evolves (rises) with increasing redshift \citep{Moster13,Behroozi13,Leauthaud12,Matthee17} as opposed to works indicating no such evolution \citep{Hudson15,McCracken15}. Likewise, the same works indicate that the peak value of the SHM relation either evolves \citep{Moster13,Hudson15,Matthee17} or remains constant \citep{Behroozi13,Leauthaud12} with redshift. Results stemming from SAMs and hydrodynamical simulations are likely sensitive to the particulars of the modelling \citep{Mitchell16}. As mentioned before, both stellar and AGN feedback leave their imprint in the SHM relation, but the same also holds for disk instabilities and mergers: a halo with its galaxy falling into another larger halo will see the halo and stellar mass added to the host halo in case of merging -- irrespective of the star formation efficiency of the host.

Here we are addressing this point with the catalogues from our galaxy formation models, but limiting the analysis to central galaxies only. Orphan galaxies -- by definition -- do not have a dark matter halo \citep[for a detailed discussion and definition of 'orphan' galaxies and halo mass, respectively, please refer to][]{Knebe15}; and as subhaloes lose mass while orbiting about their host, their (satellite) galaxy will show a more complex SHM relation and are therefore also excluded from the analysis presented here.
Further, while other work has shown that the SHM relation is different for passive and star forming galaxies \citep[e.g.][]{Mitchell14}, we leave such a classification for a future investigation.

We further include the scatter in the SHM relation in our study here: while halo and semi-analytical models are based upon the assumption that galaxy evolution is directly related to halo growth, halo mass alone is not sufficient to explain the stellar mass of galaxies. This then naturally leads to a scatter in the SHM relation and, for instance, \citet{Matthee17} found in the EAGLE simulation that this scatter increases with redshift, but also decreases with halo mass. \citet{Wang13} even claim that how galaxies populate the scatter in the SHM relation plays an important role in determining the correlation functions of galaxies.

For the comparison presented here it is worth remembering again that all galaxy formation models used the same halo catalogues and hence the same halo masses. Therefore, all differences seen here in the stellar-to-halo mass relation can purely be ascribed to the variations in the modelling of the stellar component of galaxies.

\subsection{SHM relation} \label{sec:SHMrelation}
 \begin{figure}
   \includegraphics[width=\columnwidth]{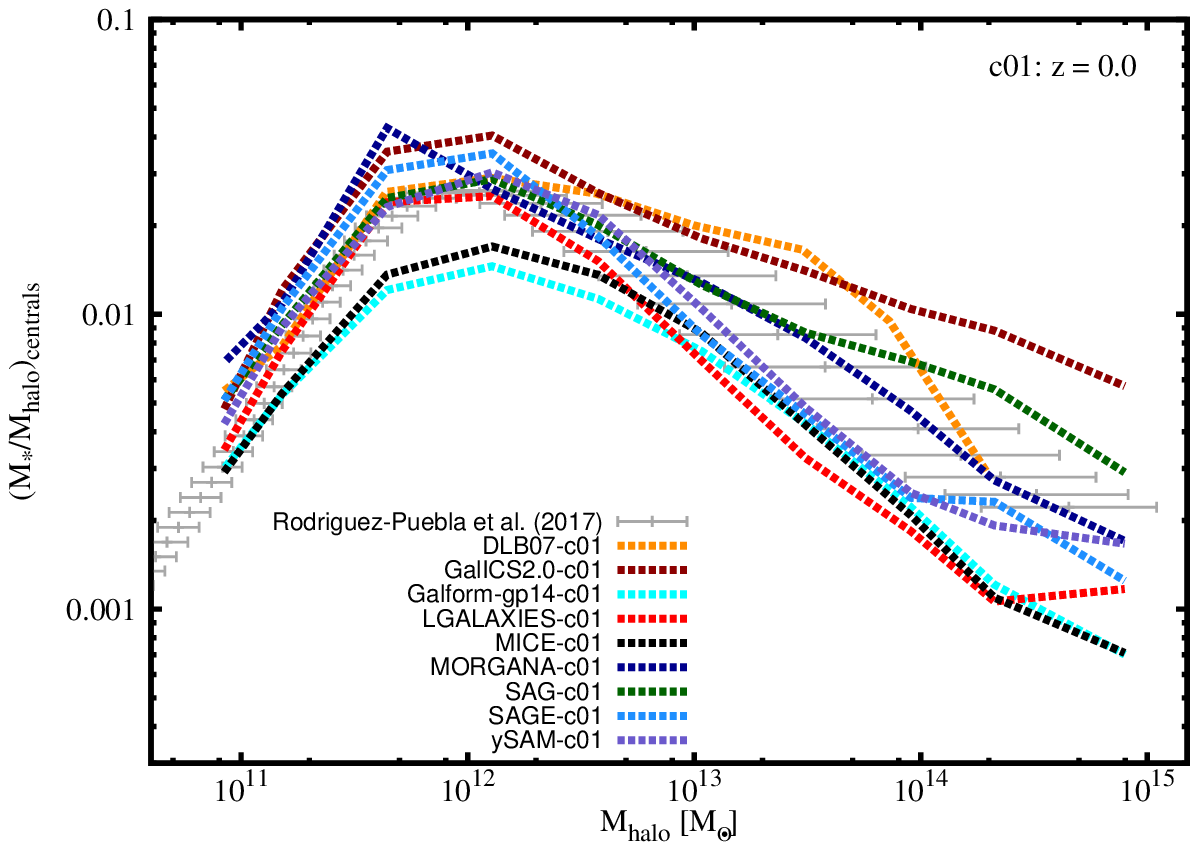}
   \includegraphics[width=\columnwidth]{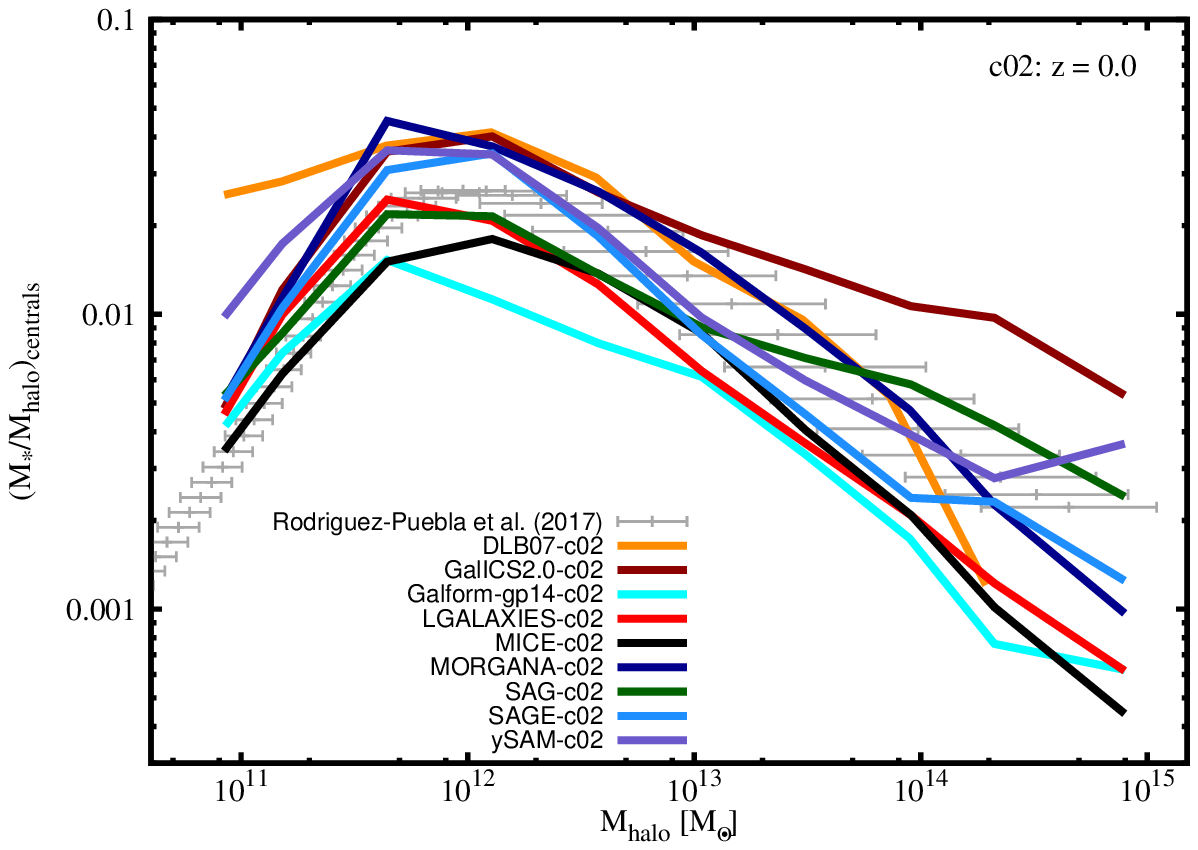}
   \caption{Stellar to halo mass ratio as a function of halo mass for central galaxies only. The values shown are medians in the respective bin.}
 \label{fig:SHM}
 \end{figure}


We start with inspecting the SHM relation at redshift $z\!=\!0$ for both the \cone\ and \ctwo\ data sets. The results (for central galaxies)
can be viewed in \Fig{fig:SHM} alongside the best-fit model of \citet{Rodriguez-Puebla17}.\footnote{The data (including the error estimates for the halo masses) of \citet{Rodriguez-Puebla17} is for redshift $z\!=\!0.0$ and has been kindly provided by Aldo Rodriguez-Puebla.} Even though that model encapsulates data for both central and satellite galaxies, we only compare against centrals -- as suggested by Rodriguez-Puebla \citep[private communication, but see also][]{Rodriguez-Puebla12}. To generate this plot the data has been binned logarithmically in halo mass for the $x$-axis and the $y$-axis shows the median $M_{*}/M_{\rm halo}$ of all central galaxies in that bin. For most of the models the additional constraints of the \ctwo\ calibration lead to no appreciable difference. The stellar-to-halo mass ratio is essentially determined purely by the SMF, with the difference that here it only applies to central galaxies. We note that the stellar-to-halo mass ratio from the \DLB\ model agrees significantly better with the observational results when the additional \ctwo\ constraints are used, while this relation remains more or less unaffected for the other models.

We have also performed the test where we added the stellar mass of all satellites to the stellar mass of the central galaxy as the halo mass of the central is `inclusive' (i.e. contains all the subhalo masses). This gives rise to a clear effect at the high-mass end of the SHM: the model-to-model variation is marginally reduced for halo masses $M_{\rm halo}>10^{13}\Msun$ and above the observational data by about a factor of two for $M_{\rm halo}>10^{14}\Msun$. It indicates the importance of how to count stellar and halo mass in theoretical models when comparing to observations. Rather than showing the respective plots here we present them in \Fig{fig:SHM+MstarSatellites} in the \App{app:SHM}.

\subsection{SHM peak value $\left(M_{*}/M_{\rm halo}\right)^{\rm peak}$}
 \begin{figure}
   \includegraphics[width=\columnwidth]{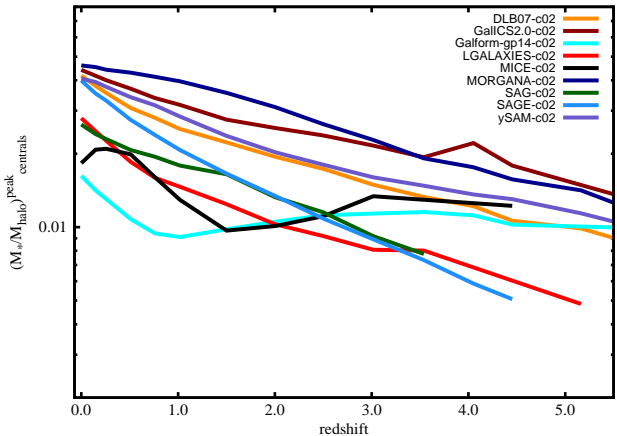}
   \caption{Redshift evolution of the peak value of the SHM relation for \ctwo\ models.}
 \label{fig:peakMsMh}
 \end{figure}

In \Fig{fig:peakMsMh} we show the redshift evolution of the peak value $\left(M_{*}/M_{\rm halo}\right)^{\rm peak}$ of the SHM relation. The value is obtained by spline-interpolation using four times as many bins as shown in \Fig{fig:SHM}, but smoothing the curve to reduce the noise; further, only bins with at least 50 galaxies are considered.\footnote{The actual curves can be viewed in \Fig{fig:SHMzred-model}} We observe a general trend for all models in the sense that the star formation efficiency declines with increasing redshift -- as reported before by, for instance, \citet{Moster13} and \citet{Hudson15}. However, there appears to be only little (if any) evolution for \galform\ and \mice. This agrees with the findings of \citet{Mitchell16} who reported a very strong dependence of $\left(M_{*}/M_{\rm halo}\right)^{\rm peak}$ on model parameters, especially for the \galform\ model also used in their work.

\subsection{SHM peak position $\left(M_{\rm halo}\right)^{\rm peak}$}
 \begin{figure}
   \includegraphics[width=\columnwidth]{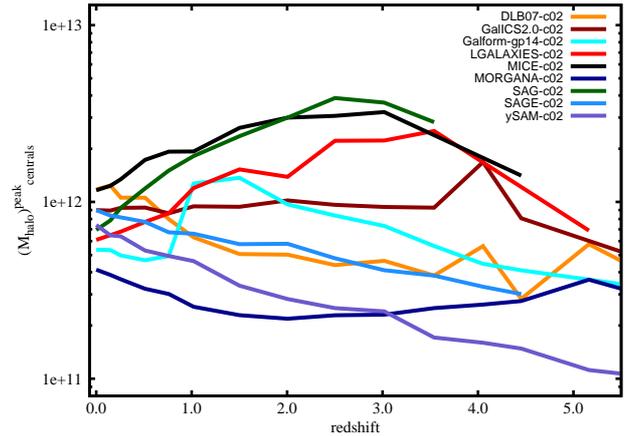}
   \caption{Redshift evolution of the peak position of the SHM relation for central galaxies in the \ctwo\ models.}
 \label{fig:peakMh}
 \end{figure}

The value of $M_{\rm halo}$ where the SHM relation peaks -- referred to as $\left(M_{\rm halo}\right)^{\rm peak}$ here -- coincides with the knee of the SMF: it is the halo mass for which star formation is most effective and least influenced by either stellar or AGN feedback \citep{Moster10,Yang12}. Assuming a simple relation with the typical mass of collapsed objects $M_{\bigstar}$\footnote{Usually defined as the mass of a 1-$\sigma$ peak in the density field at a given redshift, and not to be confused with $M_{*}$.} and its evolution within a hierarchical structure formation scenario we na\"{\i}vely expect $\left(M_{\rm halo}\right)^{\rm peak}$ to drop with redshift. While a range of models (\sage, \galform, \ysam, \DLB) show such a trend, at least marginally, \sag, \lgalaxy, and \mice\ actually have a rising $\left(M_{\rm halo}\right)^{\rm peak}$ value until redshift $z\sim3$ -- noting that these two SAM models are the ones applying an automated calibration procedure and \mice\ is the HOD model. The remaining models \morgana\ and \galics\ favour no evolution. As outlined before, there is no clear consensus yet in the literature as to whether this value is evolving \citep{Behroozi13,Moster13,Leauthaud12} or not \citep{Hudson15,McCracken15}. It is clear from our analysis that this quantity is model-dependent.

\subsection{SHM relation scatter}
 \begin{figure}
   \includegraphics[width=\columnwidth]{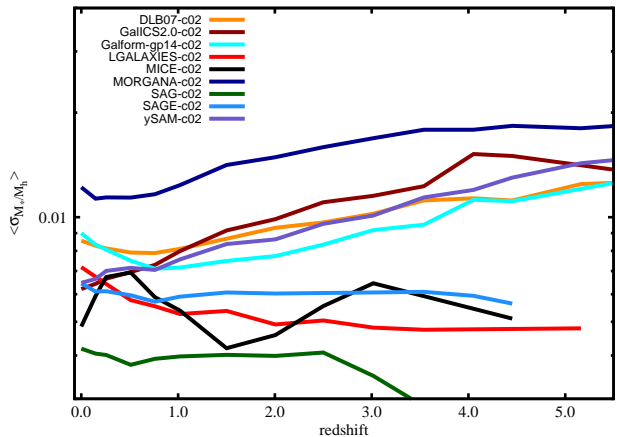}
   \caption{Redshift evolution of the mean of the scatter of the SHM relation for central galaxies in the \ctwo\ models.}
 \label{fig:peaksigmaMsMh}
 \end{figure}

The relation between stellar and halo mass is related to the star formation efficiency, and the evolution of galaxies is thought to be dominated by the growth of halo mass \citep{White78}. But the correlation between $M_{*}$ and $M_{\rm halo}$ is not tight -- as suggested by observations and models \citep[both SAMs and hydrodynamical simulations, as shown in][]{gp15}. That means that the halo mass alone is not sufficient to predict the stellar mass of the galaxy residing in it. The situation is actually far more complex for satellite galaxies whose halo has lost dark matter due to tidal stripping or are already completely disrupted leaving us with an orphan galaxy. Because of this we will again restrict our analysis to central galaxies.

Putting aside the origin of the scatter in the SHM relation, we acknowledge that this scatter could be both halo mass and redshift dependent -- as shown and investigated in great detail in \citet{Matthee17}. For that reason it might be best to investigate the evolution of the scatter in appropriately chosen halo mass bins. However, for the work presented here we refrain from it and simply define the mean scatter $\langle \sigma_{M_{*}/M_{\rm halo}}\rangle$ as the arithmetic mean of the scatter in each of the bins used for \Fig{fig:SHM}; and that scatter corresponds to half of the 25--75 percentiles of the distribution of $M_{*}/M_{\rm halo}$ values in that bin. For the relation of the scatter with halo mass -- at least for redshift $z=0$ -- we refer the reader to \Fig{fig:SHMsigma-model} where the medians and 25--75 percentiles are shown for each model. Again, only bins with at least 50 galaxies in it will be considered.

We show the redshift evolution of $\langle \sigma_{M_{*}/M_{\rm halo}}\rangle$ in \Fig{fig:peaksigmaMsMh}. We find that the evolution -- if any -- is very mild, but the normalisation is highly model dependent. This is agreement with the findings of, for instance, \cite{gp15} where the scatter in the SHM relation and its evolution was also investigated for \galform\ and \lgalaxy: the larger dispersion seen for the former model comes from treating feedback differently in disks and bulges as opposed to only depending of the halo mass for the latter model. 

\section{Conclusions}\label{sec:conclusions}
This work continues the efforts presented in \citet{Knebe15} of comparing a plethora of galaxy formation models applied to the same cosmological simulation.  Here we have compared nine galaxy formation models run on a larger simulation than previously, with a box of co-moving width $125$\hMpc, and with a dark-matter particle mass of $1.24\times10^9$\hMsun. We also use the same halo catalogues \citep[as identified with \rockstar,][]{Behroozi12} and merger trees \citep[generated with \textsc{ConsistentTrees},][]{Behroozi12b}. In this work more unifying assumptions have also been made and models have been calibrated to a common set of observational data. However, some `memory' of any previous calibration which served as the starting point for this work is retained; this is especially true for the models that are tuned manually. This may particularly affect the `prescriptions' and `predictions' of the models. All the data used here are publicly available.\footnote{The resulting galaxy catalogues (ca. 40GB of data) are stored on a data server to which access will be granted upon request to the leading author.}

The nine models summarised in \Tab{tab:models} have been calibrated in two ways: i) Just to the stellar mass function at $z\!=\!0$ and ii) To the SMF, the star formation rate function, the cold gas mass, and the black hole--bulge mass relation at $z\sim0$ together with the SMF at higher redshift $z\!=\!2$, i.e. the `CARNage calibration' set (described in \Sec{sec:observations}). 

When calibrating all our models just to the SMF at $z\!=\!0$ -- as presented in \Sec{sec:calibration01} -- the scatter is significantly reduced compared to models run with parameters fixed by other datasets. This indicates that the main conclusion of our earlier paper \citep{Knebe15}, i.e. that the scatter was driven by the lack of re-calibration, was correct. We re-confirm that galaxy formation models need to be recalibrated to the specific simulation, halo finder and merger tree being considered and, in general, cannot simply be re-run using parameters obtained from a different underlying simulation. At the high-mass end, where there are few objects, some models overproduce the number of massive galaxies. When calibrating simultaneously to all five constraints -- as studied in \Sec{sec:calibration02} -- the scatter in the SMF at $z\!=\!0$ naturally grows. Even so, the scatter is still less than that obtained when using models with standard parameters tuned to other simulation datasets.

The observed shape of the star formation rate function at $z=0$ is reproduced by models calibrated just to the SMF at $z\!=\!0$. When this star formation rate function is added as a constraint, the scatter is reduced. This same trend is also seen for the cold gas mass. The observed BHBM relation is reproduced by the models even before this is used as part of the calibration, but several codes included this in previous incarnations of the models and hence the starting parameters used for the calibrations presented here introduce a `memory' of this into the catalogues.

The most difficult constraint to match proved to be the SMF at $z\!=\!2$. Most models struggle to obtain an acceptable fit and little improvement is seen whether this constraint is included or not. Even though this is a well-known problem \citep{Fontanot09,Weinmann12,Mitchell14,Henriques15,Somerville15,Hirschmann16} we will study this phenomenon in more detail for the models presented here in a companion paper (paper~II, Asquith et al., in preparation).  But note that the models that best respond to the addition of the high-redshift observational data during the calibration are the two that apply an automated tuning procedure (i.e. \lgalaxy\ and \sag) and the HOD model \mice.

Viewing, for instance, \Fig{fig:CGMF-c01} (CGF for \cone) or \ref{fig:SFRzred} (cSFRD for \cone\ and \ctwo) that show a noticeable model-to-model variation for properties not used during the calibration, one might be inclined to question the predictive power of galaxy formation models. But before jumping to such conclusions, one needs to remember that these models are `tools' that help to explore and eventually understand galaxy formation. And, as mentioned before, calibration itself is a tool too, and not a goal. Whenever a certain property is not matched satisfactorily, the model eventually allows for deeper insight to be gained regarding the driving astrophysics of said property. 

In one way or another, all SAM and HOD models capture our present-day knowledge about galaxy formation, which is certainly not yet complete. These techniques are sufficiently rapid to be able to explore different physical prescriptions and change model parameters to probe the poorly understood aspects of galaxy formation. This is actually where the strength of them lies: while the other approach of studying galaxy formation by means of hydrodynamical simulation is also based upon similar assumptions -- at least regarding star formation and feedback -- it is considerably slower than SAM and HOD models and hence parameter-space exploration is rather prohibitive for it. Furthermore, theoretical models might also focus on different aspects of galaxies; while one model may aim at providing reasonable gas fractions, another model may have its strength in reproducing the star formation rate function -- with or without calibrating to it. And the model-to-model variation eventually is a reflection of model design and implementation of the actual physical phenomena into these tools.

We have also studied the stellar-to-halo mass relation and its evolution with redshift in \Sec{sec:SHM}. The SHM relation represents the conversion efficiency from baryonic matter into stars within dark haloes and hence gives great insight into galaxy formation. However, there is still controversy as to whether the SHM relation evolves with redshift or remains constant. We know from observations (and models) that the star formation rate peaks at about redshift $z\!=\!2$--$3$. But how exactly does this relate to the evolution of haloes? Focusing only on central galaxies, we find that for the majority of the models used here there is a clear trend for the position and value of the peak in the SHM relation to decline with redshift, albeit with prominent variations in the normalisation of this evolutionary trend. The situation is less clear for the scatter in the SHM relation: some models predict it to be marginally declining whereas other favour an increase.

In summary, we have presented a range of galaxy formation models calibrated to the same set of observations. These observations were chosen to be as complementary as possible and designed to test different aspects of the galaxy formation process. This choice places the various processes that govern galaxy formation into tension and the general success of the models demonstrates their robustness. The data from these galaxy formation models are available to the community. This paper provides a solid ground upon which build future explorations of the physical processes that govern the evolution of galaxies.

\section*{Acknowledgments}
The authors would like to express special thanks to the Carnegie Observatories for their support and hospitality during the workshop `Cosmic CARNage' where all the calibration issues were discussed and the roadmap laid out for the work presented here. This work further builds up upon the results obtained during the three week workshop `nIFTy Cosmology' hosted by the Instituto de Fisica Teorica (IFT-UAM/CSIC in Madrid) and we also like to thank it for its hospitality and support, via the Centro de Excelencia Severo Ochoa Program under Grant No. SEV-2012-0249. We further acknowledge the financial support of the 2014 University of Western Australia Research Collaboration Award for `Fast Approximate Synthetic Universes for the SKA', the ARC Centre of Excellence for All Sky Astrophysics (CAASTRO) grant number CE110001020, and the two ARC Discovery Projects DP130100117 and DP140100198. We also recognize support from the Universidad Autonoma de Madrid (UAM) for the infrastructure of the `nIFTy Cosmology' workshop.

We like to express our special thanks to Rachel Somerville, Gabriella De Lucia, and Pierluigi Monaco for stimulating discussions; they all provided highly valuable comments on draft versions of this paper that helped to substantially improve it and bring it to its present shape. We further thank the anonymous referee whose comments and suggestions helped to improve the paper.

AK is supported by the {\it Ministerio de Econom\'ia y Competitividad} and the {\it Fondo Europeo de Desarrollo Regional} (MINECO/FEDER, UE) in Spain through grant AYA2015-63810-P. He also acknowledges support from the {\it Australian Research Council} (ARC) grant DP140100198. 
FJC acknowledges support from the Spanish Ministerio de Econom\'ia y Competitividad project ESP2015-66861.
SAC acknowledges grants from Consejo Nacional de Investigaciones Cient\'ificas y T\'ecnicas (PIP-112-201301-00387), Agencia Nacional de Promoci\'on Cient\'ifica y Tecnol\'ogica (PICT-2013-0317), and Universidad Nacional de La Plata (UNLP 11-G124), Argentina.
DJC acknowledges receipt of a QEII Fellowship from the Australian Government.
PJE is supported by the SSimPL programme and the Sydney Institute for Astronomy (SIfA), DP130100117.
FF acknowledges financial contribution from the grants PRIN INAF 2010 `From the dawn of galaxy formation' and PRIN INAF 2014 `Glittering kaleidoscopes in the sky: the multifaceted nature and role of Galaxy Clusters' (1.05.01.94.02).
VGP acknowledges support from a European Research Council Starting Grant (DEGAS-259586) and from the University of Portsmouth's Dennis Sciama Fellowship. This work used the DiRAC Data Centric system at Durham University, operated by the Institute for Computational Cosmology on behalf of the STFC DiRAC HPC Facility (www.dirac.ac.uk). This equipment was funded by BIS National E-infrastructure capital grant ST/K00042X/1, STFC capital grant ST/H008519/1, and STFC DiRAC Operations grant ST/K003267/1 and Durham University. DiRAC is part of the National E-Infrastructure.    
The work of BH was supported by Advanced Grant 246797 \textsc{GALFORMOD} from the European Research Council.
MH acknowledges financial support from the European Research Council via an Advanced Grant under grant agreement no. 321323 NEOGAL.
PM has been supported by a FRA2012 grant of the University of Trieste, PRIN2010-2011 (J91J12000450001) from MIUR, and Consorzio per la Fisica di Trieste.
NDP was supported by BASAL PFB-06 CATA, and Fondecyt 1150300.  Part of the calculations presented here were run using the Geryon cluster at the Center for Astro-Engineering at U. Catolica, which received funding from QUIMAL 130008 and Fondequip AIC-57.
CP acknowledges support of the Australian Research Council (ARC) through Future Fellowship FT130100041 and Discovery Project DP140100198. WC and CP acknowledge support of ARC DP130100117.
AP acknowledges support from beca FI and 2009-SGR-1398 from Generalitat de Catalunya, project AYA2012-39620 and AYA2015-71825 from MICINN, and from a European Research Council Starting Grant (LENA-678282). 
RA is funded by the Science and Technology Funding Council (STFC) through a studentship.
RSS thanks the Downsbrough family for their generous support.
PAT acknowledges support from the Science and Technology Facilities Council (grant number ST/L000652/1).
SKY acknowledges support from the Korean National Research Foundation (NRF-2017R1A2A1A05001116). This study was performed under the umbrella of the joint collaboration between Yonsei University Observatory and the Korean Astronomy and Space Science Institute. The supercomputing time for the numerical simulations was kindly provided by KISTI (KSC-2014-G2-003). 

The authors contributed to this paper in the following ways: AK, FRP, VGP and PAT formed the core team with AK, FRP, and VGP writing the paper. AK \& FRP organized week \#2 of the 'nIFTy Cosmology' workshop where this work was initiated. AB organized the follow-up workshop 'Cosmic CARNage' where all the discussions about the common calibration took place and out of which this paper emerged. JO supplied the simulation and halo catalogue for the work presented here. The following authors performed the SAM or HOD modelling using their codes, in particular FJC, AC, SAC, DC, ARHS, FF, VGP, BH, JL, PM, CVM, and SY actively ran their models with the assistance of JH and CS. WC, DC, PJE, CP, and JO assisted with the analysis and data format issues. All authors proof-read and commented on the paper.

This research has made use of NASA's Astrophysics Data System (ADS) and the arXiv preprint server.

\bibliographystyle{mnras}
\bibliography{archive.bib}

\begin{thebibliography}{}
\makeatletter
\relax
\def\mn@urlcharsother{\let\do\@makeother \do\$\do\&\do\#\do\^\do\_\do\%\do\~}
\def\mn@doi{\begingroup\mn@urlcharsother \@ifnextchar [ {\mn@doi@}
  {\mn@doi@[]}}
\def\mn@doi@[#1]#2{\def\@tempa{#1}\ifx\@tempa\@empty \href
  {http://dx.doi.org/#2} {doi:#2}\else \href {http://dx.doi.org/#2} {#1}\fi
  \endgroup}
\def\mn@eprint#1#2{\mn@eprint@#1:#2::\@nil}
\def\mn@eprint@arXiv#1{\href {http://arxiv.org/abs/#1} {{\tt arXiv:#1}}}
\def\mn@eprint@dblp#1{\href {http://dblp.uni-trier.de/rec/bibtex/#1.xml}
  {dblp:#1}}
\def\mn@eprint@#1:#2:#3:#4\@nil{\def\@tempa {#1}\def\@tempb {#2}\def\@tempc
  {#3}\ifx \@tempc \@empty \let \@tempc \@tempb \let \@tempb \@tempa \fi \ifx
  \@tempb \@empty \def\@tempb {arXiv}\fi \@ifundefined
  {mn@eprint@\@tempb}{\@tempb:\@tempc}{\expandafter \expandafter \csname
  mn@eprint@\@tempb\endcsname \expandafter{\@tempc}}}

\bibitem[\protect\citeauthoryear{{Avila} et~al.,}{{Avila}
  et~al.}{2014}]{Avila14}
{Avila} S.,  et~al., 2014, \mn@doi [\mnras] {10.1093/mnras/stu799}, \href
  {http://adsabs.harvard.edu/abs/2014MNRAS.441.3488A} {441, 3488}

\bibitem[\protect\citeauthoryear{{Baldry}, {Glazebrook}  \& {Driver}}{{Baldry}
  et~al.}{2008}]{Baldry08}
{Baldry} I.~K.,  {Glazebrook} K.,   {Driver} S.~P.,  2008, \mn@doi [\mnras]
  {10.1111/j.1365-2966.2008.13348.x}, \href
  {http://adsabs.harvard.edu/abs/2008MNRAS.388..945B} {388, 945}

\bibitem[\protect\citeauthoryear{{Baldry} et~al.,}{{Baldry}
  et~al.}{2012}]{baldry_etal12}
{Baldry} I.~K.,  et~al., 2012, \mn@doi [\mnras]
  {10.1111/j.1365-2966.2012.20340.x}, \href
  {http://adsabs.harvard.edu/abs/2012MNRAS.421..621B} {421, 621}

\bibitem[\protect\citeauthoryear{{Baugh}}{{Baugh}}{2006}]{baugh_review_2006}
{Baugh} C.~M.,  2006, Rep. Prog. Phys., 69, 3101

\bibitem[\protect\citeauthoryear{{Behroozi}, {Wechsler}, {Wu}, {Busha},
  {Klypin}  \& {Primack}}{{Behroozi} et~al.}{2011}]{Behroozi12b}
{Behroozi} P.~S.,  {Wechsler} R.~H.,  {Wu} H.-Y.,  {Busha} M.~T.,  {Klypin}
  A.~A.,   {Primack} J.~R.,  2011, preprint, \href
  {http://adsabs.harvard.edu/abs/2011arXiv1110.4370B} {} (\mn@eprint {arXiv}
  {1110.4370})

\bibitem[\protect\citeauthoryear{{Behroozi}, {Wechsler}  \& {Wu}}{{Behroozi}
  et~al.}{2013a}]{Behroozi12}
{Behroozi} P.~S.,  {Wechsler} R.~H.,   {Wu} H.-Y.,  2013a, \mn@doi [\apj]
  {10.1088/0004-637X/762/2/109}, \href
  {http://adsabs.harvard.edu/abs/2013ApJ...762..109B} {762, 109}

\bibitem[\protect\citeauthoryear{{Behroozi}, {Wechsler}  \&
  {Conroy}}{{Behroozi} et~al.}{2013b}]{Behroozi13}
{Behroozi} P.~S.,  {Wechsler} R.~H.,   {Conroy} C.,  2013b, \mn@doi [\apj]
  {10.1088/0004-637X/770/1/57}, \href
  {http://adsabs.harvard.edu/abs/2013ApJ...770...57B} {770, 57}

\bibitem[\protect\citeauthoryear{{Benson}}{{Benson}}{2014}]{Benson14}
{Benson} A.~J.,  2014, \mn@doi [\mnras] {10.1093/mnras/stu1630}, \href
  {http://adsabs.harvard.edu/abs/2014MNRAS.444.2599B} {444, 2599}

\bibitem[\protect\citeauthoryear{{Benson}, {Borgani}, {De Lucia},
  {Boylan-Kolchin}  \& {Monaco}}{{Benson} et~al.}{2012}]{Benson12b}
{Benson} A.~J.,  {Borgani} S.,  {De Lucia} G.,  {Boylan-Kolchin} M.,   {Monaco}
  P.,  2012, \mn@doi [\mnras] {10.1111/j.1365-2966.2011.20002.x}, \href
  {http://adsabs.harvard.edu/abs/2012MNRAS.419.3590B} {419, 3590}

\bibitem[\protect\citeauthoryear{{Boselli}, {Cortese}, {Boquien}, {Boissier},
  {Catinella}, {Lagos}  \& {Saintonge}}{{Boselli} et~al.}{2014}]{Boselli14}
{Boselli} A.,  {Cortese} L.,  {Boquien} M.,  {Boissier} S.,  {Catinella} B.,
  {Lagos} C.,   {Saintonge} A.,  2014, \mn@doi [\aap]
  {10.1051/0004-6361/201322312}, \href
  {http://adsabs.harvard.edu/abs/2014A%26A...564A..66B} {564, A66}

\bibitem[\protect\citeauthoryear{{Bower}, {Benson}, {Malbon}, {Helly}, {Frenk},
  {Baugh}, {Cole}  \& {Lacey}}{{Bower} et~al.}{2006a}]{Bower06}
{Bower} R.~G.,  {Benson} A.~J.,  {Malbon} R.,  {Helly} J.~C.,  {Frenk} C.~S.,
  {Baugh} C.~M.,  {Cole} S.,   {Lacey} C.~G.,  2006a, \mn@doi [\mnras]
  {10.1111/j.1365-2966.2006.10519.x}, \href
  {http://ukads.nottingham.ac.uk/abs/2006MNRAS.370..645B} {370, 645}

\bibitem[\protect\citeauthoryear{{Bower}, {Benson}, {Malbon}, {Helly}, {Frenk},
  {Baugh}, {Cole}  \& {Lacey}}{{Bower} et~al.}{2006b}]{bower_agn_2006}
{Bower} R.~G.,  {Benson} A.~J.,  {Malbon} R.,  {Helly} J.~C.,  {Frenk} C.~S.,
  {Baugh} C.~M.,  {Cole} S.,   {Lacey} C.~G.,  2006b, MNRAS, 370, 645

\bibitem[\protect\citeauthoryear{{Bower}, {Vernon}, {Goldstein}, {Benson},
  {Lacey}, {Baugh}, {Cole}  \& {Frenk}}{{Bower} et~al.}{2010}]{Bower10}
{Bower} R.~G.,  {Vernon} I.,  {Goldstein} M.,  {Benson} A.~J.,  {Lacey} C.~G.,
  {Baugh} C.~M.,  {Cole} S.,   {Frenk} C.~S.,  2010, \mn@doi [\mnras]
  {10.1111/j.1365-2966.2010.16991.x}, \href
  {http://adsabs.harvard.edu/abs/2010MNRAS.407.2017B} {407, 2017}

\bibitem[\protect\citeauthoryear{{Carretero}, {Castander}, {Gazta{\~n}aga},
  {Crocce}  \& {Fosalba}}{{Carretero} et~al.}{2015}]{Carretero14}
{Carretero} J.,  {Castander} F.~J.,  {Gazta{\~n}aga} E.,  {Crocce} M.,
  {Fosalba} P.,  2015, \mn@doi [\mnras] {10.1093/mnras/stu2402}, \href
  {http://adsabs.harvard.edu/abs/2015MNRAS.447..646C} {447, 646}

\bibitem[\protect\citeauthoryear{{Cattaneo} et~al.,}{{Cattaneo}
  et~al.}{2017}]{Cattaneo17}
{Cattaneo} A.,  et~al., 2017, \mn@doi [\mnras] {10.1093/mnras/stx1597}, \href
  {http://adsabs.harvard.edu/abs/2017MNRAS.471.1401C} {471, 1401}

\bibitem[\protect\citeauthoryear{{Cheung} et~al.,}{{Cheung}
  et~al.}{2016}]{Cheung16}
{Cheung} E.,  et~al., 2016, \mn@doi [\nat] {10.1038/nature18006}, \href
  {http://adsabs.harvard.edu/abs/2016Natur.533..504C} {533, 504}

\bibitem[\protect\citeauthoryear{{Croton} et~al.,}{{Croton}
  et~al.}{2006}]{Croton06}
{Croton} D.~J.,  et~al., 2006, \mn@doi [\mnras]
  {10.1111/j.1365-2966.2005.09675.x}, \href
  {http://ukads.nottingham.ac.uk/abs/2006MNRAS.365...11C} {365, 11}

\bibitem[\protect\citeauthoryear{{Croton} et~al.,}{{Croton}
  et~al.}{2016}]{Croton16}
{Croton} D.~J.,  et~al., 2016, \mn@doi [\apjs] {10.3847/0067-0049/222/2/22},
  \href {http://adsabs.harvard.edu/abs/2016ApJS..222...22C} {222, 22}

\bibitem[\protect\citeauthoryear{{Dav{\'e}}}{{Dav{\'e}}}{2009}]{Dave09}
{Dav{\'e}} R.,  2009, in {Jogee} S.,  {Marinova} I.,  {Hao} L.,   {Blanc}
  G.~A.,  eds,  Astronomical Society of the Pacific Conference Series Vol. 419,
  Galaxy Evolution: Emerging Insights and Future Challenges. p.~347 (\mn@eprint
  {arXiv} {0901.3149})

\bibitem[\protect\citeauthoryear{{De Lucia} \& {Blaizot}}{{De Lucia} \&
  {Blaizot}}{2007}]{delucia_sam_2007}
{De Lucia} G.,  {Blaizot} J.,  2007, MNRAS, 375, 2

\bibitem[\protect\citeauthoryear{{Dom{\'{\i}}nguez S{\'a}nchez}
  et~al.,}{{Dom{\'{\i}}nguez S{\'a}nchez} et~al.}{2011}]{Dominguez-Sanchez11}
{Dom{\'{\i}}nguez S{\'a}nchez} H.,  et~al., 2011, \mn@doi [\mnras]
  {10.1111/j.1365-2966.2011.19263.x}, \href
  {http://adsabs.harvard.edu/abs/2011MNRAS.417..900D} {417, 900}

\bibitem[\protect\citeauthoryear{{Fontanot}, {De Lucia}, {Monaco}, {Somerville}
   \& {Santini}}{{Fontanot} et~al.}{2009}]{Fontanot09}
{Fontanot} F.,  {De Lucia} G.,  {Monaco} P.,  {Somerville} R.~S.,   {Santini}
  P.,  2009, \mn@doi [\mnras] {10.1111/j.1365-2966.2009.15058.x}, \href
  {http://adsabs.harvard.edu/abs/2009MNRAS.397.1776F} {397, 1776}

\bibitem[\protect\citeauthoryear{{Frenk} \& {White}}{{Frenk} \&
  {White}}{2012}]{Frenk12}
{Frenk} C.~S.,  {White} S.~D.~M.,  2012, \mn@doi [Annalen der Physik]
  {10.1002/andp.201200212}, \href
  {http://adsabs.harvard.edu/abs/2012AnP...524..507F} {524, 507}

\bibitem[\protect\citeauthoryear{{Gargiulo} et~al.,}{{Gargiulo}
  et~al.}{2015}]{gargiulo_2014}
{Gargiulo} I.~D.,  et~al., 2015, \mn@doi [\mnras] {10.1093/mnras/stu2272},
  \href {http://adsabs.harvard.edu/abs/2015MNRAS.446.3820G} {446, 3820}

\bibitem[\protect\citeauthoryear{{Gonzalez-Perez}, {Lacey}, {Baugh}, {Lagos},
  {Helly}, {Campbell}  \& {Mitchell}}{{Gonzalez-Perez} et~al.}{2014}]{gp14}
{Gonzalez-Perez} V.,  {Lacey} C.~G.,  {Baugh} C.~M.,  {Lagos} C.~D.~P.,
  {Helly} J.,  {Campbell} D.~J.~R.,   {Mitchell} P.~D.,  2014, \mn@doi [\mnras]
  {10.1093/mnras/stt2410}, \href
  {http://adsabs.harvard.edu/abs/2014MNRAS.439..264G} {439, 264}

\bibitem[\protect\citeauthoryear{{Gruppioni} et~al.,}{{Gruppioni}
  et~al.}{2015}]{Gruppioni15}
{Gruppioni} C.,  et~al., 2015, \mn@doi [\mnras] {10.1093/mnras/stv1204}, \href
  {http://adsabs.harvard.edu/abs/2015MNRAS.451.3419G} {451, 3419}

\bibitem[\protect\citeauthoryear{{Guo} et~al.,}{{Guo} et~al.}{2011}]{Guo11}
{Guo} Q.,  et~al., 2011, \mn@doi [\mnras] {10.1111/j.1365-2966.2010.18114.x},
  \href {http://adsabs.harvard.edu/abs/2011MNRAS.413..101G} {413, 101}

\bibitem[\protect\citeauthoryear{{Guo}, {White}, {Angulo}, {Henriques},
  {Lemson}, {Boylan-Kolchin}, {Thomas}  \& {Short}}{{Guo} et~al.}{2013}]{Guo13}
{Guo} Q.,  {White} S.,  {Angulo} R.~E.,  {Henriques} B.,  {Lemson} G.,
  {Boylan-Kolchin} M.,  {Thomas} P.,   {Short} C.,  2013, \mn@doi [\mnras]
  {10.1093/mnras/sts115}, \href
  {http://adsabs.harvard.edu/abs/2013MNRAS.428.1351G} {428, 1351}

\bibitem[\protect\citeauthoryear{{Guo} et~al.,}{{Guo} et~al.}{2015}]{gp15}
{Guo} Q.,  et~al., 2015, preprint, \href
  {http://adsabs.harvard.edu/abs/2015arXiv151200015G} {} (\mn@eprint {arXiv}
  {1512.00015})

\bibitem[\protect\citeauthoryear{{Henriques}, {Thomas}, {Oliver}  \&
  {Roseboom}}{{Henriques} et~al.}{2009a}]{henriques_mcmc_2009}
{Henriques} B.~M.~B.,  {Thomas} P.~A.,  {Oliver} S.,   {Roseboom} I.,  2009a,
  MNRAS, 396, 535

\bibitem[\protect\citeauthoryear{{Henriques}, {Thomas}, {Oliver}  \&
  {Roseboom}}{{Henriques} et~al.}{2009b}]{Henriques09}
{Henriques} B.~M.~B.,  {Thomas} P.~A.,  {Oliver} S.,   {Roseboom} I.,  2009b,
  \mn@doi [\mnras] {10.1111/j.1365-2966.2009.14730.x}, \href
  {http://ukads.nottingham.ac.uk/abs/2009MNRAS.396..535H} {396, 535}

\bibitem[\protect\citeauthoryear{{Henriques}, {White}, {Thomas}, {Angulo},
  {Guo}, {Lemson}  \& {Springel}}{{Henriques} et~al.}{2013}]{Henriques13}
{Henriques} B.~M.~B.,  {White} S.~D.~M.,  {Thomas} P.~A.,  {Angulo} R.~E.,
  {Guo} Q.,  {Lemson} G.,   {Springel} V.,  2013, \mn@doi [\mnras]
  {10.1093/mnras/stt415}, \href
  {http://adsabs.harvard.edu/abs/2013MNRAS.431.3373H} {431, 3373}

\bibitem[\protect\citeauthoryear{{Henriques}, {White}, {Thomas}, {Angulo},
  {Guo}, {Lemson}, {Springel}  \& {Overzier}}{{Henriques}
  et~al.}{2014}]{Henriques15}
{Henriques} B.,  {White} S.,  {Thomas} P.,  {Angulo} R.,  {Guo} Q.,  {Lemson}
  G.,  {Springel} V.,   {Overzier} R.,  2014, preprint, \href
  {http://adsabs.harvard.edu/abs/2014arXiv1410.0365H} {} (\mn@eprint {arXiv}
  {1410.0365})

\bibitem[\protect\citeauthoryear{{Hirschmann}, {De Lucia}  \&
  {Fontanot}}{{Hirschmann} et~al.}{2016}]{Hirschmann16}
{Hirschmann} M.,  {De Lucia} G.,   {Fontanot} F.,  2016, \mn@doi [\mnras]
  {10.1093/mnras/stw1318}, \href
  {http://adsabs.harvard.edu/abs/2016MNRAS.461.1760H} {461, 1760}

\bibitem[\protect\citeauthoryear{{Hudson} et~al.,}{{Hudson}
  et~al.}{2015}]{Hudson15}
{Hudson} M.~J.,  et~al., 2015, \mn@doi [\mnras] {10.1093/mnras/stu2367}, \href
  {http://adsabs.harvard.edu/abs/2015MNRAS.447..298H} {447, 298}

\bibitem[\protect\citeauthoryear{{Ilbert} et~al.,}{{Ilbert}
  et~al.}{2013}]{ilbert_etal13}
{Ilbert} O.,  et~al., 2013, \mn@doi [\aap] {10.1051/0004-6361/201321100}, \href
  {http://adsabs.harvard.edu/abs/2013A%26A...556A..55I} {556, A55}

\bibitem[\protect\citeauthoryear{{Kauffmann}, {White}  \&
  {Guiderdoni}}{{Kauffmann} et~al.}{1993}]{Kauffmann93}
{Kauffmann} G.,  {White} S.~D.~M.,   {Guiderdoni} B.,  1993, \mnras, \href
  {http://adsabs.harvard.edu/abs/1993MNRAS.264..201K} {264, 201}

\bibitem[\protect\citeauthoryear{{Knebe} et~al.,}{{Knebe}
  et~al.}{2015}]{Knebe15}
{Knebe} A.,  et~al., 2015, \mn@doi [\mnras] {10.1093/mnras/stv1149}, \href
  {http://adsabs.harvard.edu/abs/2015MNRAS.451.4029K} {451, 4029}

\bibitem[\protect\citeauthoryear{{Kormendy} \& {Ho}}{{Kormendy} \&
  {Ho}}{2013}]{Kormendy13}
{Kormendy} J.,  {Ho} L.~C.,  2013, \mn@doi [\araa]
  {10.1146/annurev-astro-082708-101811}, \href
  {http://adsabs.harvard.edu/abs/2013ARA%26A..51..511K} {51, 511}

\bibitem[\protect\citeauthoryear{{Lacey} et~al.,}{{Lacey}
  et~al.}{2016}]{Lacey16}
{Lacey} C.~G.,  et~al., 2016, \mn@doi [\mnras] {10.1093/mnras/stw1888}, \href
  {http://adsabs.harvard.edu/abs/2016MNRAS.462.3854L} {462, 3854}

\bibitem[\protect\citeauthoryear{{Lagos} et~al.,}{{Lagos}
  et~al.}{2016}]{Lagos16}
{Lagos} C.~d.~P.,  et~al., 2016, \mn@doi [\mnras] {10.1093/mnras/stw717}, \href
  {http://adsabs.harvard.edu/abs/2016MNRAS.459.2632D} {459, 2632}

\bibitem[\protect\citeauthoryear{{Lara-L{\'o}pez} et~al.,}{{Lara-L{\'o}pez}
  et~al.}{2010}]{Lara-Lopez10}
{Lara-L{\'o}pez} M.~A.,  et~al., 2010, \mn@doi [\aap]
  {10.1051/0004-6361/201014803}, \href
  {http://adsabs.harvard.edu/abs/2010A%26A...521L..53L} {521, L53}

\bibitem[\protect\citeauthoryear{{Leauthaud} et~al.,}{{Leauthaud}
  et~al.}{2012}]{Leauthaud12}
{Leauthaud} A.,  et~al., 2012, \mn@doi [\apj] {10.1088/0004-637X/744/2/159},
  \href {http://adsabs.harvard.edu/abs/2012ApJ...744..159L} {744, 159}

\bibitem[\protect\citeauthoryear{{Lee} \& {Yi}}{{Lee} \& {Yi}}{2013}]{lee13}
{Lee} J.,  {Yi} S.~K.,  2013, \mn@doi [\apj] {10.1088/0004-637X/766/1/38},
  \href {http://adsabs.harvard.edu/abs/2013ApJ...766...38L} {766, 38}

\bibitem[\protect\citeauthoryear{{Lee} et~al.,}{{Lee} et~al.}{2014}]{Lee14}
{Lee} J.,  et~al., 2014, preprint, \href
  {http://adsabs.harvard.edu/abs/2014arXiv1410.1241L} {} (\mn@eprint {arXiv}
  {1410.1241})

\bibitem[\protect\citeauthoryear{Li \& White}{Li \&
  White}{2009}]{li_distribution_2009}
Li C.,  White S. D.~M.,  2009, Monthly Notices of the Royal Astronomical
  Society, 398, 2177

\bibitem[\protect\citeauthoryear{{Lilly}, {Le Fevre}, {Hammer}  \&
  {Crampton}}{{Lilly} et~al.}{1996}]{Lilly96}
{Lilly} S.~J.,  {Le Fevre} O.,  {Hammer} F.,   {Crampton} D.,  1996, \mn@doi
  [\apjl] {10.1086/309975}, \href
  {http://adsabs.harvard.edu/abs/1996ApJ...460L...1L} {460, L1}

\bibitem[\protect\citeauthoryear{{Lu} et~al.,}{{Lu} et~al.}{2014}]{Lu14}
{Lu} Y.,  et~al., 2014, \mn@doi [\apj] {10.1088/0004-637X/795/2/123}, \href
  {http://adsabs.harvard.edu/abs/2014ApJ...795..123L} {795, 123}

\bibitem[\protect\citeauthoryear{{Madau} \& {Dickinson}}{{Madau} \&
  {Dickinson}}{2014}]{Madau14}
{Madau} P.,  {Dickinson} M.,  2014, \mn@doi [\araa]
  {10.1146/annurev-astro-081811-125615}, \href
  {http://adsabs.harvard.edu/abs/2014ARA%26A..52..415M} {52, 415}

\bibitem[\protect\citeauthoryear{{Madau}, {Ferguson}, {Dickinson},
  {Giavalisco}, {Steidel}  \& {Fruchter}}{{Madau} et~al.}{1996}]{Madau96}
{Madau} P.,  {Ferguson} H.~C.,  {Dickinson} M.~E.,  {Giavalisco} M.,  {Steidel}
  C.~C.,   {Fruchter} A.,  1996, \mn@doi [\mnras] {10.1093/mnras/283.4.1388},
  \href {http://adsabs.harvard.edu/abs/1996MNRAS.283.1388M} {283, 1388}

\bibitem[\protect\citeauthoryear{{Mannucci}, {Cresci}, {Maiolino}, {Marconi}
  \& {Gnerucci}}{{Mannucci} et~al.}{2010}]{Mannucci10}
{Mannucci} F.,  {Cresci} G.,  {Maiolino} R.,  {Marconi} A.,   {Gnerucci} A.,
  2010, \mn@doi [\mnras] {10.1111/j.1365-2966.2010.17291.x}, \href
  {http://adsabs.harvard.edu/abs/2010MNRAS.408.2115M} {408, 2115}

\bibitem[\protect\citeauthoryear{{Matthee}, {Schaye}, {Crain}, {Schaller},
  {Bower}  \& {Theuns}}{{Matthee} et~al.}{2017}]{Matthee17}
{Matthee} J.,  {Schaye} J.,  {Crain} R.~A.,  {Schaller} M.,  {Bower} R.,
  {Theuns} T.,  2017, \mn@doi [\mnras] {10.1093/mnras/stw2884}, \href
  {http://adsabs.harvard.edu/abs/2017MNRAS.465.2381M} {465, 2381}

\bibitem[\protect\citeauthoryear{{McConnell} \& {Ma}}{{McConnell} \&
  {Ma}}{2013}]{McConnell13}
{McConnell} N.~J.,  {Ma} C.-P.,  2013, \mn@doi [\apj]
  {10.1088/0004-637X/764/2/184}, \href
  {http://adsabs.harvard.edu/abs/2013ApJ...764..184M} {764, 184}

\bibitem[\protect\citeauthoryear{{McCracken} et~al.,}{{McCracken}
  et~al.}{2015}]{McCracken15}
{McCracken} H.~J.,  et~al., 2015, \mn@doi [\mnras] {10.1093/mnras/stv305},
  \href {http://adsabs.harvard.edu/abs/2015MNRAS.449..901M} {449, 901}

\bibitem[\protect\citeauthoryear{{Mitchell}, {Lacey}, {Cole}  \&
  {Baugh}}{{Mitchell} et~al.}{2014}]{Mitchell14}
{Mitchell} P.~D.,  {Lacey} C.~G.,  {Cole} S.,   {Baugh} C.~M.,  2014, \mn@doi
  [\mnras] {10.1093/mnras/stu1639}, \href
  {http://adsabs.harvard.edu/abs/2014MNRAS.444.2637M} {444, 2637}

\bibitem[\protect\citeauthoryear{{Mitchell}, {Lacey}, {Baugh}  \&
  {Cole}}{{Mitchell} et~al.}{2016}]{Mitchell16}
{Mitchell} P.~D.,  {Lacey} C.~G.,  {Baugh} C.~M.,   {Cole} S.,  2016, \mn@doi
  [\mnras] {10.1093/mnras/stv2741}, \href
  {http://adsabs.harvard.edu/abs/2016MNRAS.456.1459M} {456, 1459}

\bibitem[\protect\citeauthoryear{{Monaco}, {Fontanot}  \& {Taffoni}}{{Monaco}
  et~al.}{2007}]{Monaco07}
{Monaco} P.,  {Fontanot} F.,   {Taffoni} G.,  2007, \mn@doi [\mnras]
  {10.1111/j.1365-2966.2006.11253.x}, \href
  {http://ukads.nottingham.ac.uk/abs/2007MNRAS.375.1189M} {375, 1189}

\bibitem[\protect\citeauthoryear{{Mortlock} et~al.,}{{Mortlock}
  et~al.}{2015}]{Mortlock15}
{Mortlock} A.,  et~al., 2015, \mn@doi [\mnras] {10.1093/mnras/stu2403}, \href
  {http://adsabs.harvard.edu/abs/2015MNRAS.447....2M} {447, 2}

\bibitem[\protect\citeauthoryear{{Moster}, {Somerville}, {Maulbetsch}, {van den
  Bosch}, {Macci{\`o}}, {Naab}  \& {Oser}}{{Moster} et~al.}{2010}]{Moster10}
{Moster} B.~P.,  {Somerville} R.~S.,  {Maulbetsch} C.,  {van den Bosch} F.~C.,
  {Macci{\`o}} A.~V.,  {Naab} T.,   {Oser} L.,  2010, \mn@doi [\apj]
  {10.1088/0004-637X/710/2/903}, \href
  {http://adsabs.harvard.edu/abs/2010ApJ...710..903M} {710, 903}

\bibitem[\protect\citeauthoryear{{Moster}, {Naab}  \& {White}}{{Moster}
  et~al.}{2013}]{Moster13}
{Moster} B.~P.,  {Naab} T.,   {White} S.~D.~M.,  2013, \mn@doi [\mnras]
  {10.1093/mnras/sts261}, \href
  {http://adsabs.harvard.edu/abs/2013MNRAS.428.3121M} {428, 3121}

\bibitem[\protect\citeauthoryear{{Mutch}, {Poole}  \& {Croton}}{{Mutch}
  et~al.}{2013}]{Mutch13a}
{Mutch} S.~J.,  {Poole} G.~B.,   {Croton} D.~J.,  2013, \mn@doi [\mnras]
  {10.1093/mnras/sts182}, \href
  {http://adsabs.harvard.edu/abs/2013MNRAS.428.2001M} {428, 2001}

\bibitem[\protect\citeauthoryear{{Muzzin} et~al.,}{{Muzzin}
  et~al.}{2013}]{Muzzin13}
{Muzzin} A.,  et~al., 2013, \mn@doi [\apj] {10.1088/0004-637X/777/1/18}, \href
  {http://adsabs.harvard.edu/abs/2013ApJ...777...18M} {777, 18}

\bibitem[\protect\citeauthoryear{{Nagamine}, {Cen}, {Hernquist}, {Ostriker}  \&
  {Springel}}{{Nagamine} et~al.}{2004}]{Nagamine04}
{Nagamine} K.,  {Cen} R.,  {Hernquist} L.,  {Ostriker} J.~P.,   {Springel} V.,
  2004, \mn@doi [\apj] {10.1086/421379}, \href
  {http://adsabs.harvard.edu/abs/2004ApJ...610...45N} {610, 45}

\bibitem[\protect\citeauthoryear{{Panter}, {Jimenez}, {Heavens}  \&
  {Charlot}}{{Panter} et~al.}{2007}]{Panter07}
{Panter} B.,  {Jimenez} R.,  {Heavens} A.~F.,   {Charlot} S.,  2007, \mn@doi
  [\mnras] {10.1111/j.1365-2966.2007.11909.x}, \href
  {http://adsabs.harvard.edu/abs/2007MNRAS.378.1550P} {378, 1550}

\bibitem[\protect\citeauthoryear{{Peeples}, {Werk}, {Tumlinson}, {Oppenheimer},
  {Prochaska}, {Katz}  \& {Weinberg}}{{Peeples} et~al.}{2014}]{Peeples14}
{Peeples} M.~S.,  {Werk} J.~K.,  {Tumlinson} J.,  {Oppenheimer} B.~D.,
  {Prochaska} J.~X.,  {Katz} N.,   {Weinberg} D.~H.,  2014, \mn@doi [\apj]
  {10.1088/0004-637X/786/1/54}, \href
  {http://adsabs.harvard.edu/abs/2014ApJ...786...54P} {786, 54}

\bibitem[\protect\citeauthoryear{{Planck Collaboration} et~al.,}{{Planck
  Collaboration} et~al.}{2014}]{Planck2013}
{Planck Collaboration} et~al., 2014, \mn@doi [\aap]
  {10.1051/0004-6361/201321591}, \href
  {http://adsabs.harvard.edu/abs/2014A%26A...571A..16P} {571, A16}

\bibitem[\protect\citeauthoryear{{Pujol} et~al.,}{{Pujol}
  et~al.}{2017}]{Pujol17}
{Pujol} A.,  et~al., 2017, \mn@doi [\mnras] {10.1093/mnras/stx913}, \href
  {http://adsabs.harvard.edu/abs/2017MNRAS.469..749P} {469, 749}

\bibitem[\protect\citeauthoryear{{Rodrigues}, {Vernon}  \& {Bower}}{{Rodrigues}
  et~al.}{2017}]{Rodrigues17}
{Rodrigues} L.~F.~S.,  {Vernon} I.,   {Bower} R.~G.,  2017, \mn@doi [\mnras]
  {10.1093/mnras/stw3269}, \href
  {http://adsabs.harvard.edu/abs/2017MNRAS.466.2418R} {466, 2418}

\bibitem[\protect\citeauthoryear{{Rodr{\'{\i}}guez-Puebla}, {Drory}  \&
  {Avila-Reese}}{{Rodr{\'{\i}}guez-Puebla} et~al.}{2012}]{Rodriguez-Puebla12}
{Rodr{\'{\i}}guez-Puebla} A.,  {Drory} N.,   {Avila-Reese} V.,  2012, \mn@doi
  [\apj] {10.1088/0004-637X/756/1/2}, \href
  {http://adsabs.harvard.edu/abs/2012ApJ...756....2R} {756, 2}

\bibitem[\protect\citeauthoryear{{Rodr{\'{\i}}guez-Puebla}, {Primack},
  {Avila-Reese}  \& {Faber}}{{Rodr{\'{\i}}guez-Puebla}
  et~al.}{2017}]{Rodriguez-Puebla17}
{Rodr{\'{\i}}guez-Puebla} A.,  {Primack} J.~R.,  {Avila-Reese} V.,   {Faber}
  S.~M.,  2017, \mn@doi [\mnras] {10.1093/mnras/stx1172}, \href
  {http://adsabs.harvard.edu/abs/2017MNRAS.470..651R} {470, 651}

\bibitem[\protect\citeauthoryear{{Ruiz} et~al.,}{{Ruiz}
  et~al.}{2015}]{ruiz2014}
{Ruiz} A.~N.,  et~al., 2015, \mn@doi [\apj] {10.1088/0004-637X/801/2/139},
  \href {http://adsabs.harvard.edu/abs/2015ApJ...801..139R} {801, 139}

\bibitem[\protect\citeauthoryear{{Salim}, {Lee}, {Dav{\'e}}  \&
  {Dickinson}}{{Salim} et~al.}{2015}]{Salim15}
{Salim} S.,  {Lee} J.~C.,  {Dav{\'e}} R.,   {Dickinson} M.,  2015, \mn@doi
  [\apj] {10.1088/0004-637X/808/1/25}, \href
  {http://adsabs.harvard.edu/abs/2015ApJ...808...25S} {808, 25}

\bibitem[\protect\citeauthoryear{{Silk} \& {Mamon}}{{Silk} \&
  {Mamon}}{2012}]{Silk12}
{Silk} J.,  {Mamon} G.~A.,  2012, \mn@doi [Research in Astronomy and
  Astrophysics] {10.1088/1674-4527/12/8/004}, \href
  {http://adsabs.harvard.edu/abs/2012RAA....12..917S} {12, 917}

\bibitem[\protect\citeauthoryear{{Somerville} \& {Dav{\'e}}}{{Somerville} \&
  {Dav{\'e}}}{2015}]{Somerville15}
{Somerville} R.~S.,  {Dav{\'e}} R.,  2015, \mn@doi [\araa]
  {10.1146/annurev-astro-082812-140951}, \href
  {http://adsabs.harvard.edu/abs/2015ARA%26A..53...51S} {53, 51}

\bibitem[\protect\citeauthoryear{{Somerville} \& {Primack}}{{Somerville} \&
  {Primack}}{1999}]{Somerville99}
{Somerville} R.~S.,  {Primack} J.~R.,  1999, \mn@doi [\mnras]
  {10.1046/j.1365-8711.1999.03032.x}, \href
  {http://adsabs.harvard.edu/abs/1999MNRAS.310.1087S} {310, 1087}

\bibitem[\protect\citeauthoryear{{Somerville}, {Primack}  \&
  {Faber}}{{Somerville} et~al.}{2001}]{Somerville01}
{Somerville} R.~S.,  {Primack} J.~R.,   {Faber} S.~M.,  2001, \mn@doi [\mnras]
  {10.1046/j.1365-8711.2001.03975.x}, \href
  {http://adsabs.harvard.edu/abs/2001MNRAS.320..504S} {320, 504}

\bibitem[\protect\citeauthoryear{{Springel}}{{Springel}}{2005}]{Springel05}
{Springel} V.,  2005, \mn@doi [\mnras] {10.1111/j.1365-2966.2005.09655.x},
  \href {http://adsabs.harvard.edu/abs/2005MNRAS.364.1105S} {364, 1105}

\bibitem[\protect\citeauthoryear{{Tomczak}, {Quadri}, {Tran}, {Labb{\'e}},
  {Straatman}, {Papovich}, {Glazebrook}  \& {et al.}}{{Tomczak}
  et~al.}{2014}]{Tomczak14}
{Tomczak} A.~R.,  {Quadri} R.~F.,  {Tran} K.-V.~H.,  {Labb{\'e}} I.,
  {Straatman} C.~M.~S.,  {Papovich} C.,  {Glazebrook} K.,   {et al.} 2014,
  \mn@doi [\apj] {10.1088/0004-637X/783/2/85}, \href
  {http://adsabs.harvard.edu/abs/2014ApJ...783...85T} {783, 85}

\bibitem[\protect\citeauthoryear{{Wang}, {De Lucia}  \& {Weinmann}}{{Wang}
  et~al.}{2013}]{Wang13}
{Wang} L.,  {De Lucia} G.,   {Weinmann} S.~M.,  2013, \mn@doi [\mnras]
  {10.1093/mnras/stt188}, \href
  {http://adsabs.harvard.edu/abs/2013MNRAS.431..600W} {431, 600}

\bibitem[\protect\citeauthoryear{{Weinmann}, {Pasquali}, {Oppenheimer},
  {Finlator}, {Mendel}, {Crain}  \& {Macci{\`o}}}{{Weinmann}
  et~al.}{2012}]{Weinmann12}
{Weinmann} S.~M.,  {Pasquali} A.,  {Oppenheimer} B.~D.,  {Finlator} K.,
  {Mendel} J.~T.,  {Crain} R.~A.,   {Macci{\`o}} A.~V.,  2012, \mn@doi [\mnras]
  {10.1111/j.1365-2966.2012.21931.x}, \href
  {http://adsabs.harvard.edu/abs/2012MNRAS.426.2797W} {426, 2797}

\bibitem[\protect\citeauthoryear{{White} \& {Rees}}{{White} \&
  {Rees}}{1978}]{White78}
{White} S.~D.~M.,  {Rees} M.~J.,  1978, \mnras, \href
  {http://adsabs.harvard.edu/abs/1978MNRAS.183..341W} {183, 341}

\bibitem[\protect\citeauthoryear{{White}, {Somerville}  \& {Ferguson}}{{White}
  et~al.}{2015}]{White15}
{White} C.~E.,  {Somerville} R.~S.,   {Ferguson} H.~C.,  2015, \mn@doi [\apj]
  {10.1088/0004-637X/799/2/201}, \href
  {http://adsabs.harvard.edu/abs/2015ApJ...799..201W} {799, 201}

\bibitem[\protect\citeauthoryear{{Wilkins}, {Trentham}  \& {Hopkins}}{{Wilkins}
  et~al.}{2008}]{Wilkins08}
{Wilkins} S.~M.,  {Trentham} N.,   {Hopkins} A.~M.,  2008, \mn@doi [\mnras]
  {10.1111/j.1365-2966.2008.12885.x}, \href
  {http://adsabs.harvard.edu/abs/2008MNRAS.385..687W} {385, 687}

\bibitem[\protect\citeauthoryear{{Yang}, {Mo}, {van den Bosch}, {Zhang}  \&
  {Han}}{{Yang} et~al.}{2012}]{Yang12}
{Yang} X.,  {Mo} H.~J.,  {van den Bosch} F.~C.,  {Zhang} Y.,   {Han} J.,  2012,
  \mn@doi [\apj] {10.1088/0004-637X/752/1/41}, \href
  {http://adsabs.harvard.edu/abs/2012ApJ...752...41Y} {752, 41}

\bibitem[\protect\citeauthoryear{{van Daalen}, {Henriques}, {Angulo}  \&
  {White}}{{van Daalen} et~al.}{2015}]{vanDaalen16}
{van Daalen} M.~P.,  {Henriques} B.~M.~B.,  {Angulo} R.~E.,   {White} S.~D.~M.,
   2015, preprint, \href {http://adsabs.harvard.edu/abs/2015arXiv151200008V} {}
  (\mn@eprint {arXiv} {1512.00008})

\makeatother
\end{thebibliography}

\appendix

\section{The Observational Data} \label{app:observations}
Aiming at generating model galaxies with properties that can reasonably reproduce the observed statistical properties of observed galaxies, one of the objectives of the Cosmic CARNage workshop\footnote{\url{http://users.obs.carnegiescience.edu/abenson/CARnage.html}} -- the successor of the nIFTy Cosmology workshop where the whole comparison was initiated -- was to discuss appropriate observations to use for a common calibration of the participating galaxy formation models. We summarize the outcome of this debate here and list the chosen calibration data sets in \Tab{tab:constraints}.\footnote{While the last column in \Tab{tab:constraints} gives the references from which the data has been obtained, we also provide the link to a page with the actual observational data files used throughout this study: \url{http://popia.ft.uam.es/public/CARNageSet.zip}} The set of observations is designed to constrain parameters from a wide range of modelled physical processes, yet is observationally well established. The decision was to use the stellar mass function (SMF, both at $z\!=\!0$ and $z\!=\!2$), the star formation rate function (SFRF), the black hole--bulge mass relation (BHBM), and the fraction of mass in cold gas (CGMF). This set of observational constraints probes several different aspects of the galaxy formation models that are all inter-related yet nevertheless sufficiently independent of each other.

However, we also like to state that calibration is a `tool' -- not a `goal' -- for galaxy formation models. What we present here as a data set is tailored to be useful for the purpose of this project, i.e. model comparison.

\subsection{Stellar Mass Function} \label{app:SMF}
The literature contains a great number of (local) measurements of the galaxy stellar mass function deduced using empirically determined mass-to-light ratios. Stellar masses determined this way therefore rely on several implicit assumptions regarding the stellar initial mass function, star formation histories and the integrated effects of dust attenuation. Hence these estimates can suffer from large systematic uncertainties.

It was a matter of some debate amongst workshop participants whether it was better to compare the models to stellar mass functions or to luminosity functions.  In principle the latter is more straightforward as the models have known star-formation histories and no conversion is required for the observational data.  However, the reliance on an accurate dust model outweigh this advantage. Hence we decided to compare a quantity directly measurable in the models: the mass in stars, and treat the differences between the observational predictions as an estimate of the systematic error.

The observed stellar mass function used here is a compilation of the data presented in \citet{baldry_etal12,li_distribution_2009,Baldry08} for redshift $z\!=\!0$ and \citet{Tomczak14,Muzzin13,ilbert_etal13,Dominguez-Sanchez11} for $z\!=\!2$. The different data sets at each redshift are formally incompatible with each other within the error bars, suggesting that there are systematic errors between them.  For that reason, the maximum and minimum observational estimates (taking into account the error bars) are used as a measure of the systematic uncertainty between observations; the precise details of the procedure can be found in Appendic~C of \citet{Henriques13}. 

\subsection{Star Formation Rate Function} \label{app:SFRF}
We use the star formation rate function as presented in Table~1 of \citet{Gruppioni15} for the redshift interval $z\in[0.0,0.3]$. These data comes from a flux-limited sample of galaxies observed with the \textit{Herschel} satellite giving the total (IR+UV) instantaneous star formation rates. These data was compared against model galaxies at redshift $z\!=\!0.15$.

\subsection{Cold Gas Fractions} \label{app:CGF}
For the mass fraction of HI+H$_2$ the decision was to use the \citet{Boselli14} data, which is based on a volume limited sample, within the range $\log_{10}{M_{*}/\Msun}\in [{9.15},{10.52}]$ in stellar mass. For the data used here we combined the information for the two methods used for the X factors; see Table~4 in \citet{Boselli14}. We further agreed not to use a cut for separating active from passive galaxies during the calibration -- given that such a cut can be model dependant. Therefore, while the \citet{Peeples14} data was also discussed, it eventually was not adopted as it only contains star forming (active) galaxies.

\subsection{Black Hole--Bulge Mass Relation} \label{app:BHBM}
The black hole--bulge mass relation used for calibration is a compilation of both the data presented in \citet{McConnell13} and \citet{Kormendy13}.

\section{Un-Calibrated Catalogues} \label{app:uc}
 \begin{figure}
   \includegraphics[width=\columnwidth]{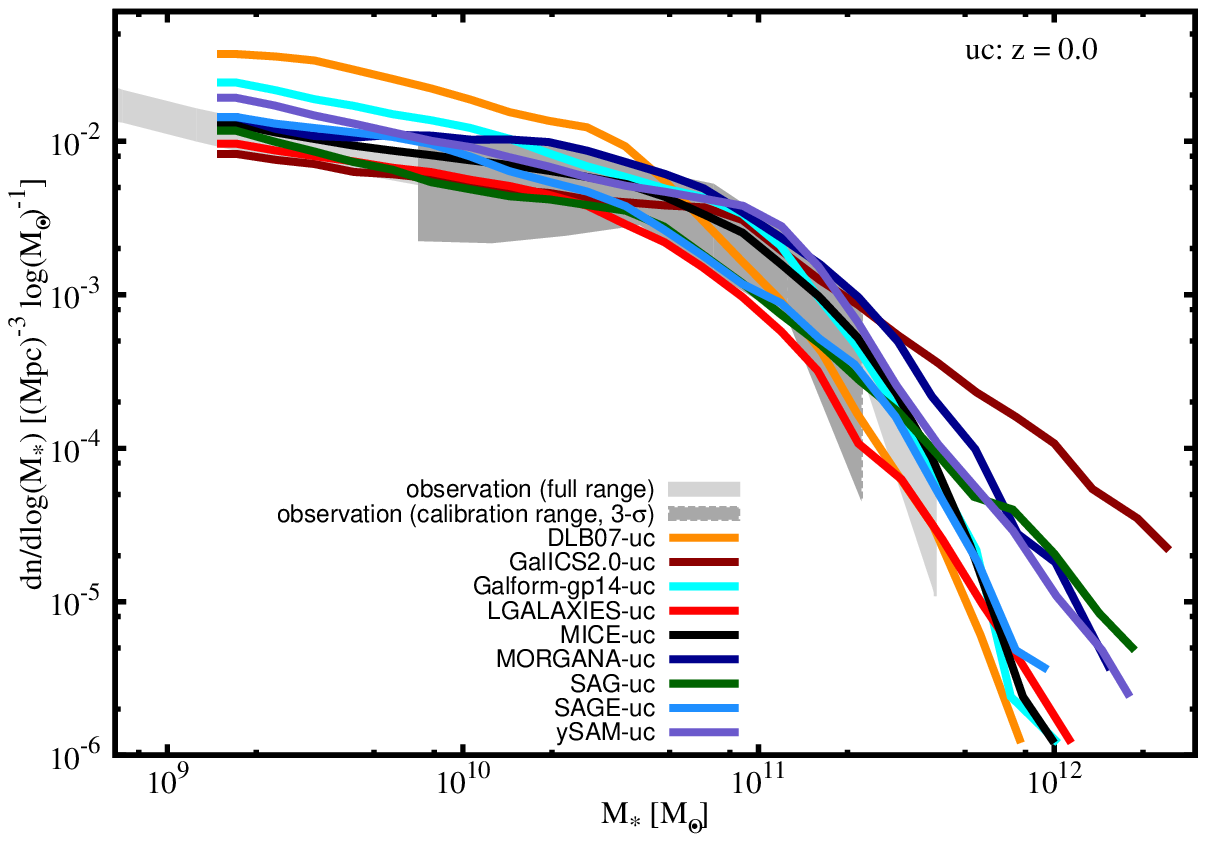}\\
   \caption{Same as \Fig{fig:SMFz0-c01}, but for galaxy catalogues not re-calibrated.}
 \label{fig:SMFz0-uc}
 \end{figure}

While we only studied commonly calibrated models in the main part of this paper, we show here in \Fig{fig:SMFz0-uc} the stellar mass function at redshift $z\!=\!0$ for each model when used without any re-calibration and how it compares to the CARNage calibration SMF. This SMF could be compared to the upper panel of Fig.2 in \citet{Knebe15}. There is a difference though: in \citet{Knebe15} the simulation featured a much smaller volume of (62.5$^3$)\hMpc$^3$ whereas we are using here (125$^3$)\hMpc$^3$. Further, merger trees in \citet{Knebe15} were constructed with \textsc{MergerTree} whereas we are here using \textsc{ConsistenTrees}. As has been shown in \citet{Avila14} this will have an impact on the quality of the trees and \citet{Lee14} further show that (and how) this impacts upon SAMs.

\section{Stellar-to-Halo Mass Relation} \label{app:SHM}
\subsection{Redshift Evolution} \label{app:SHMzred}
 \begin{figure*}
   \includegraphics[width=0.75\columnwidth]{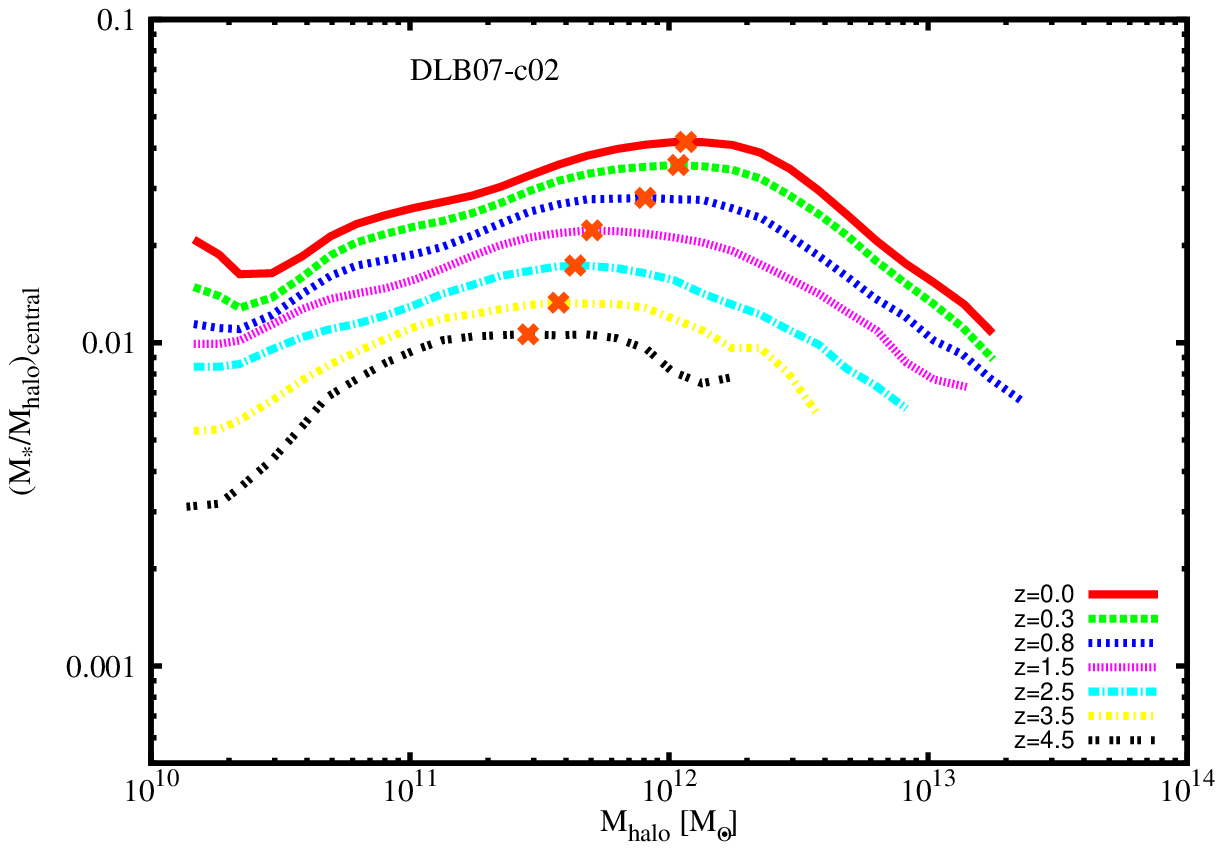}
   \includegraphics[width=0.75\columnwidth]{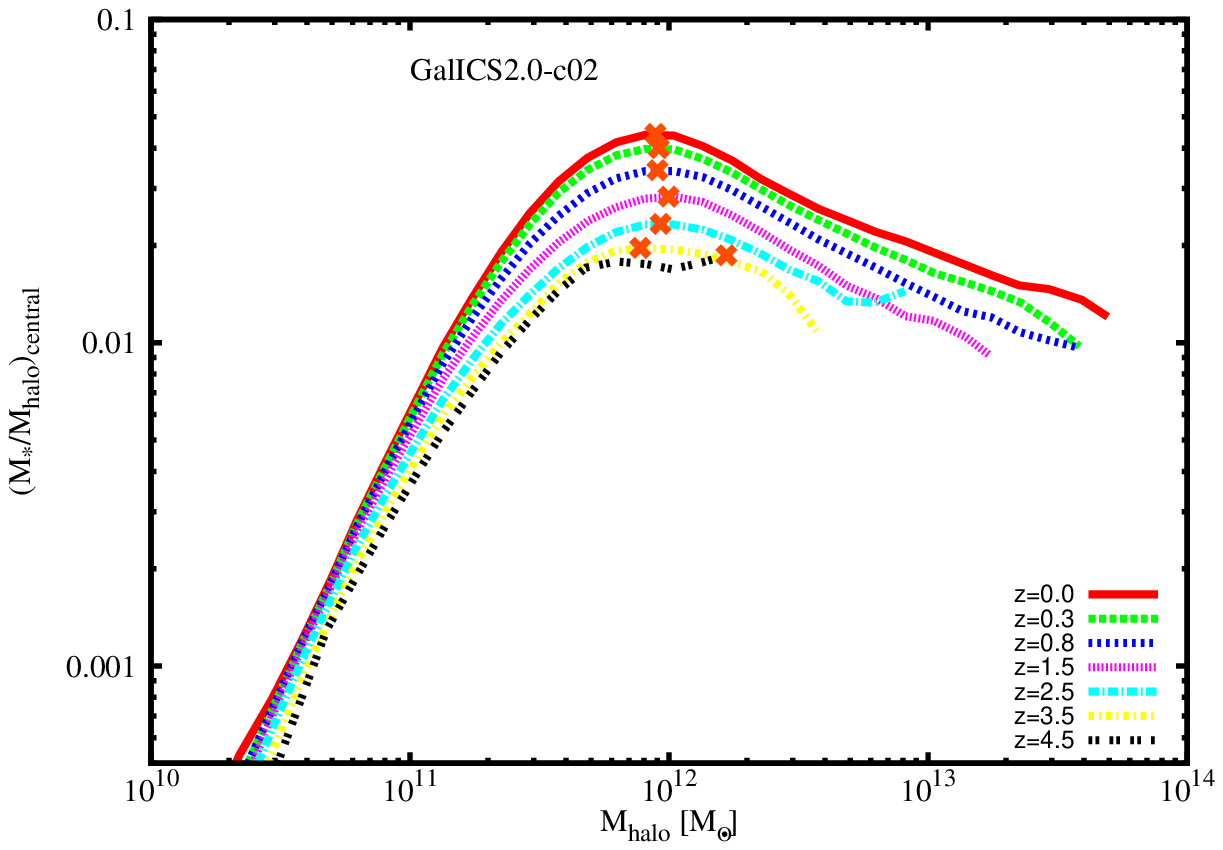}
   \includegraphics[width=0.75\columnwidth]{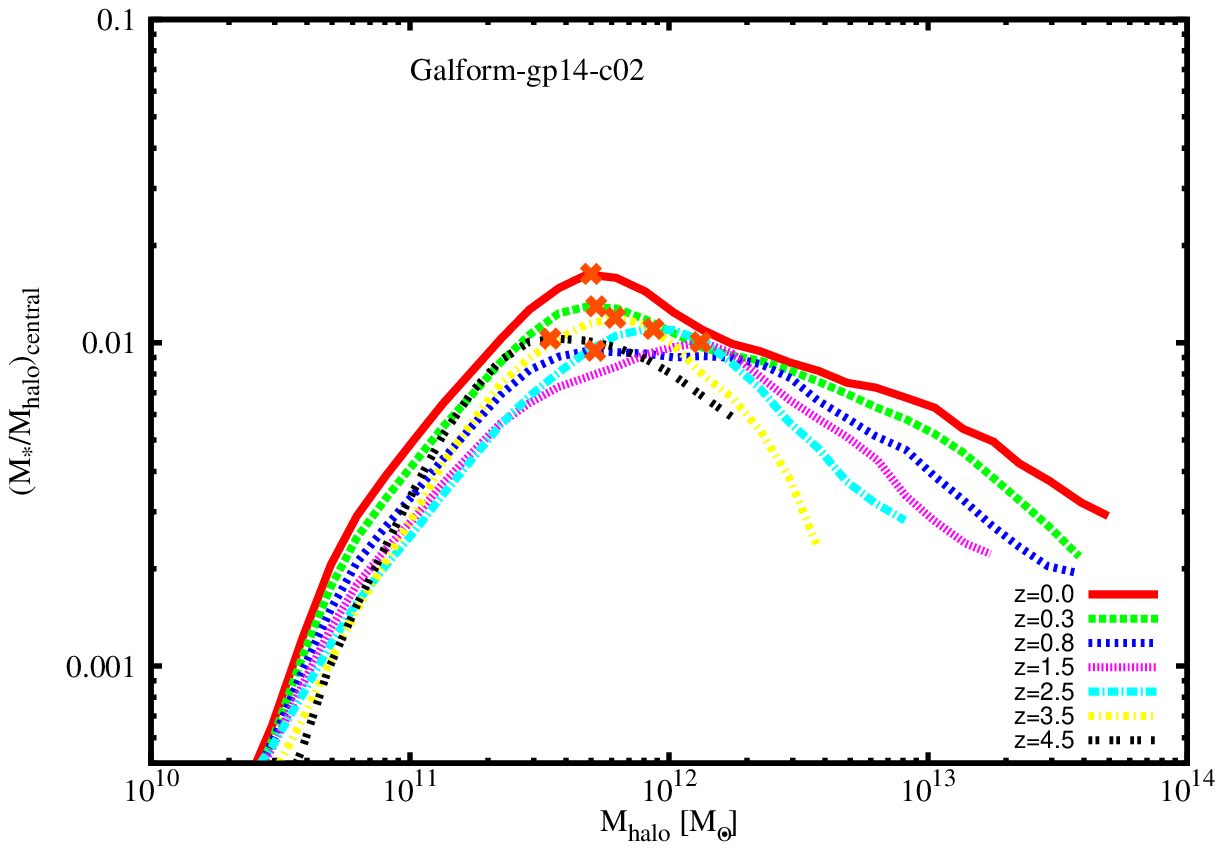}
   \includegraphics[width=0.75\columnwidth]{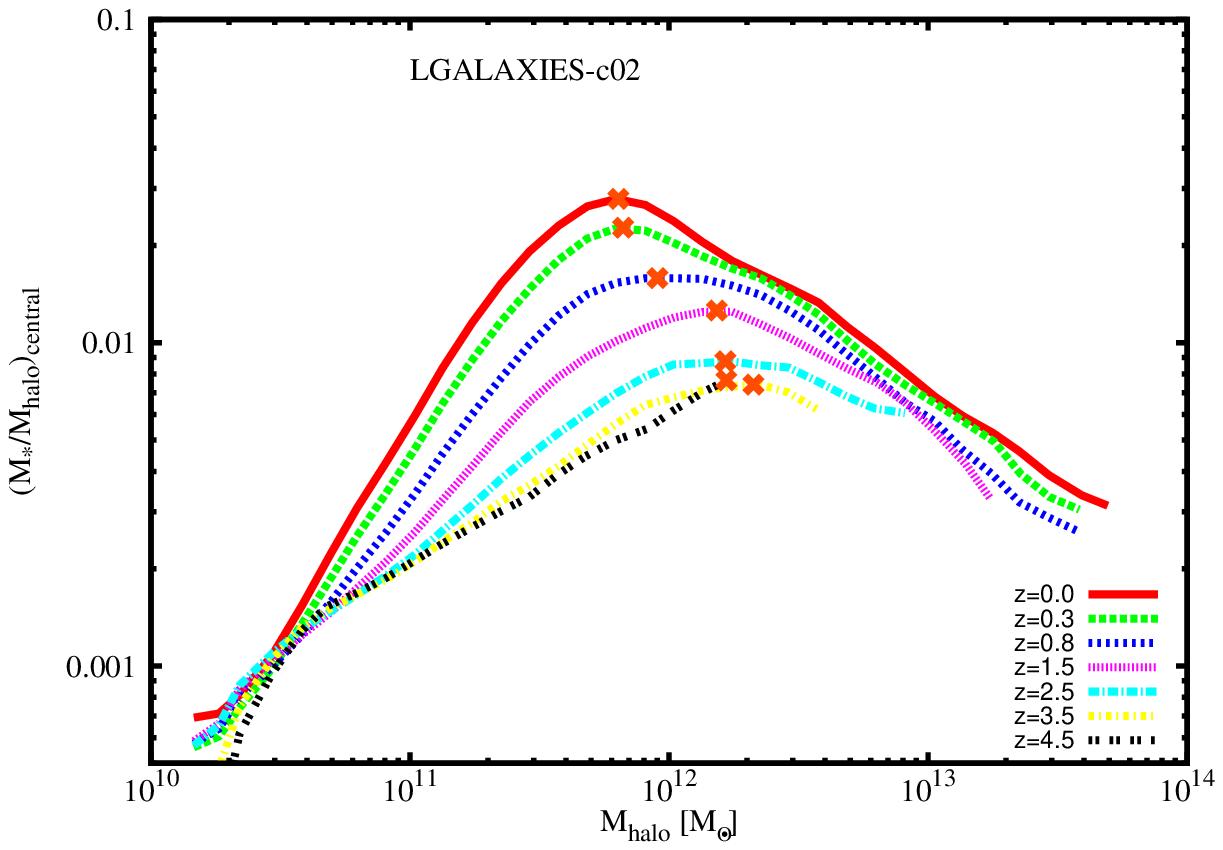}
   \includegraphics[width=0.75\columnwidth]{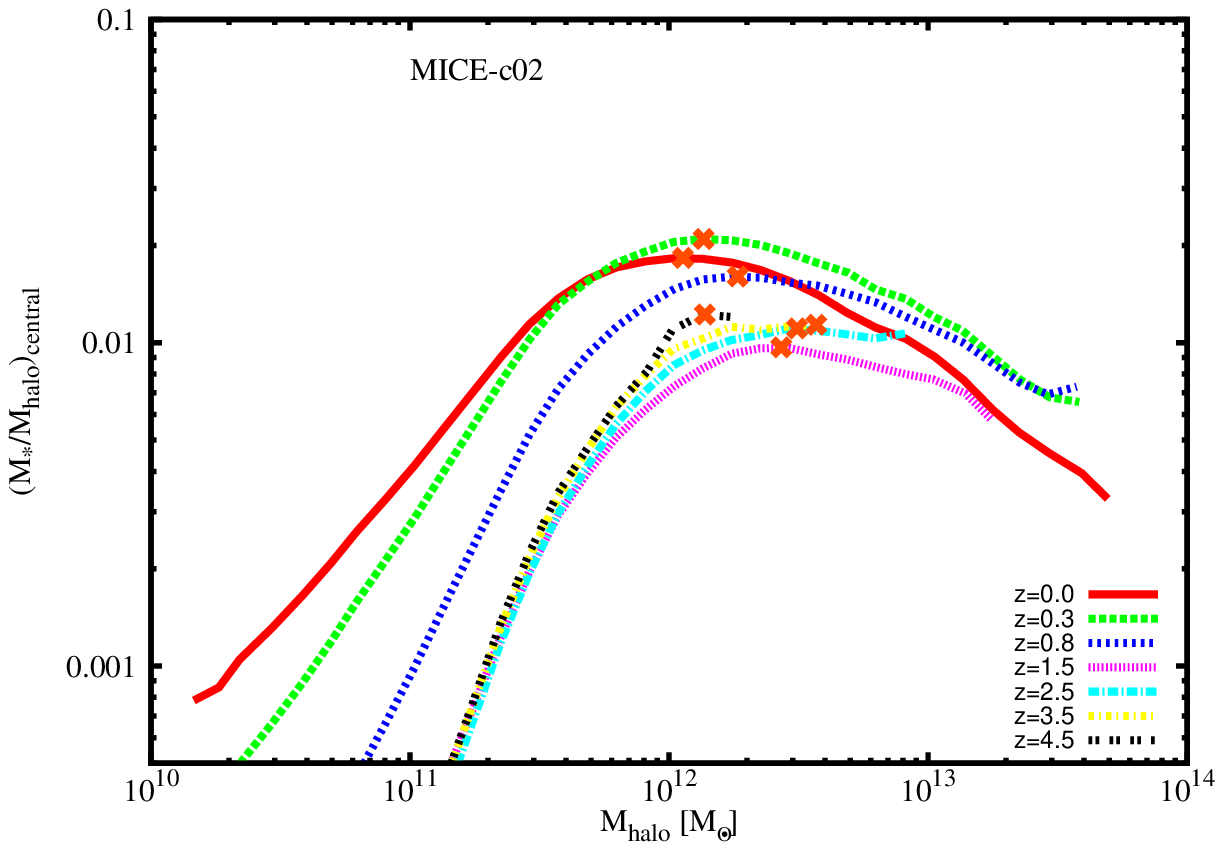}
   \includegraphics[width=0.75\columnwidth]{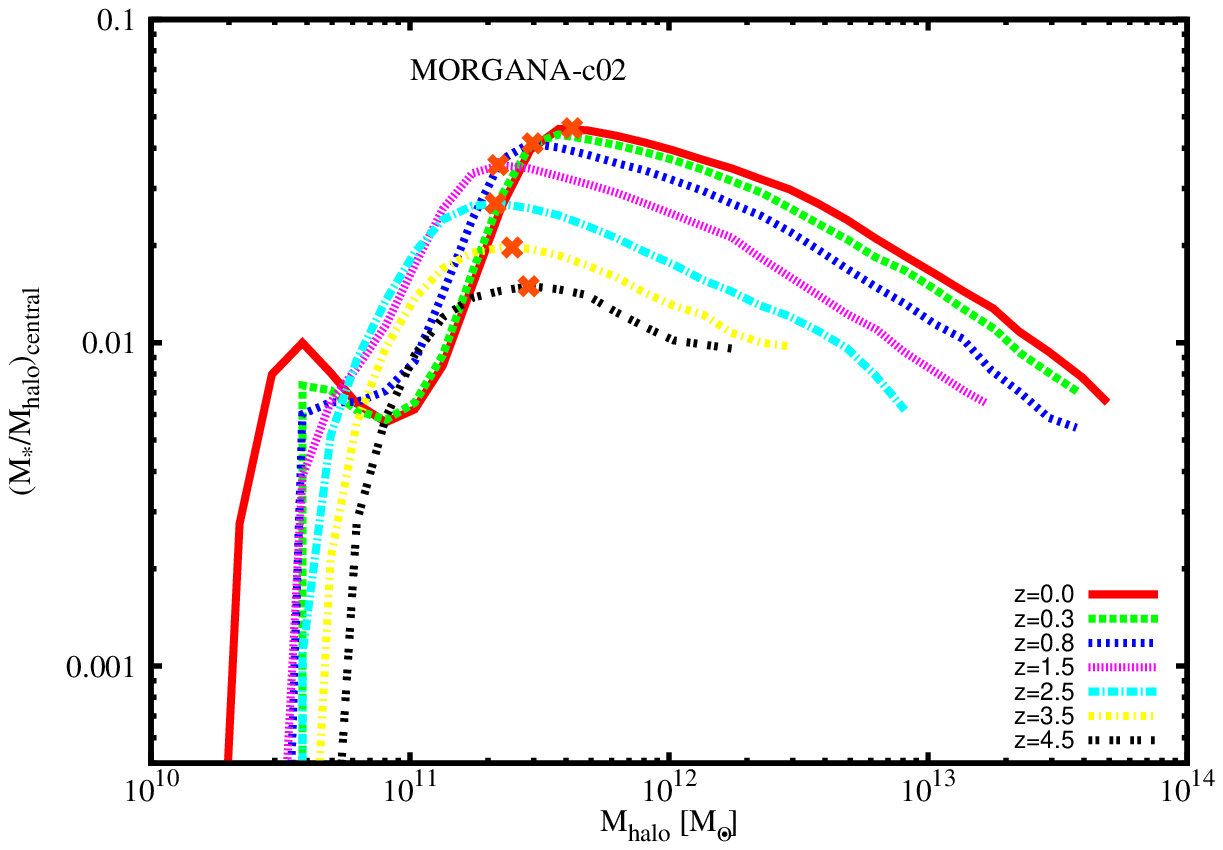}
   \includegraphics[width=0.75\columnwidth]{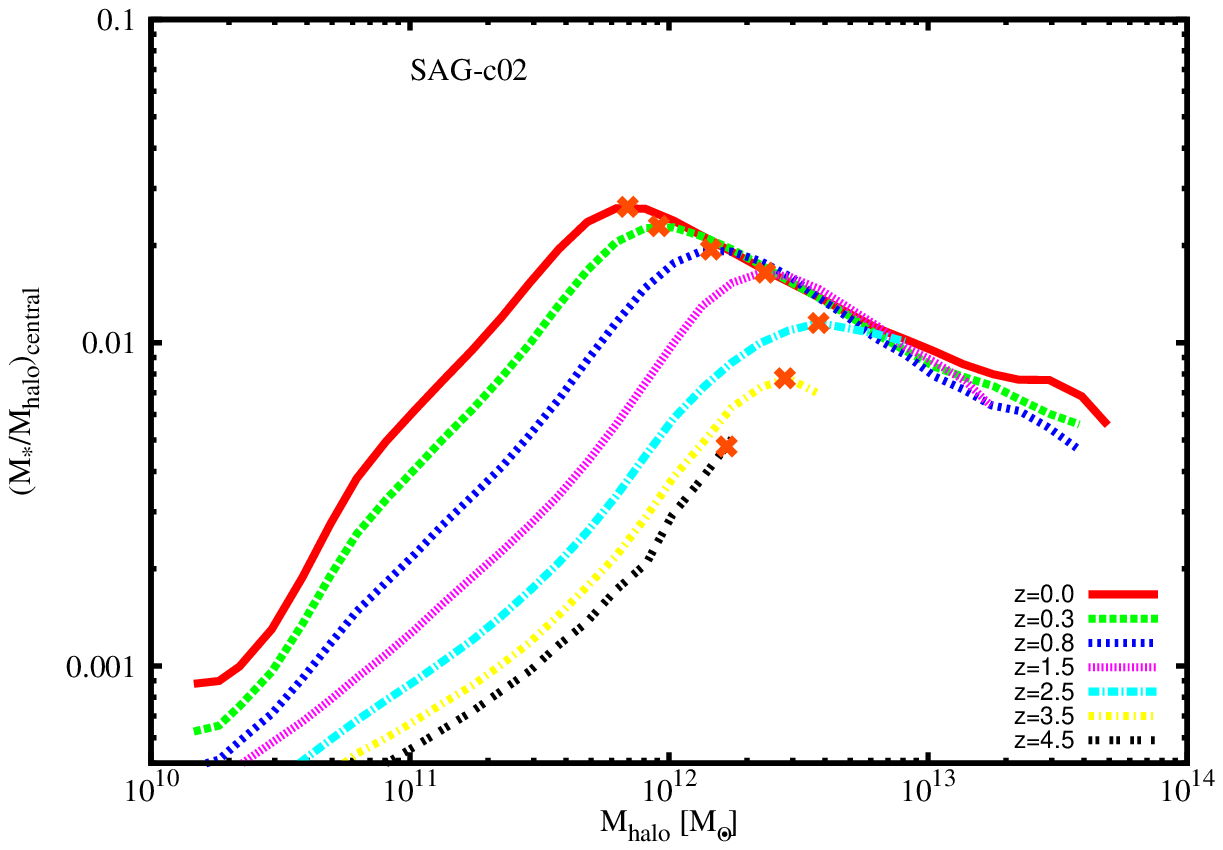}
   \includegraphics[width=0.75\columnwidth]{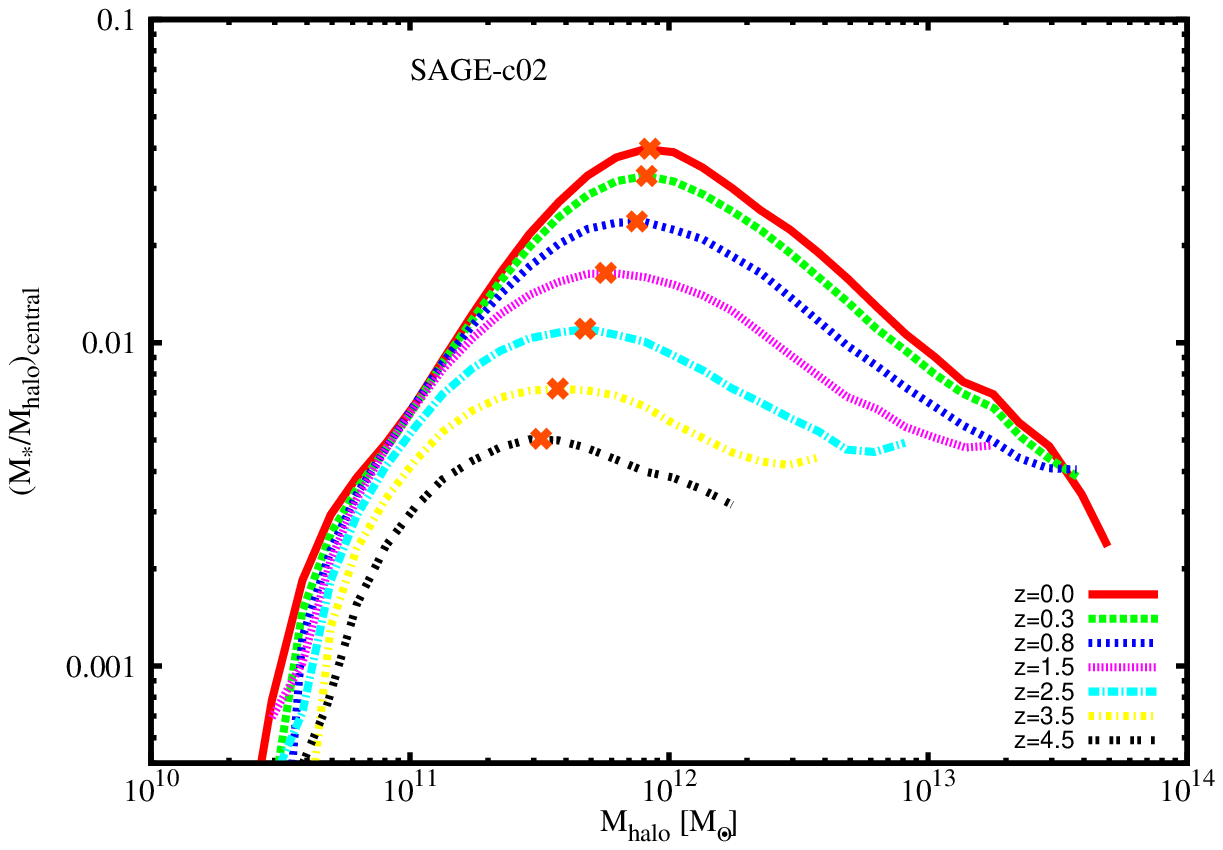}
   \includegraphics[width=0.75\columnwidth]{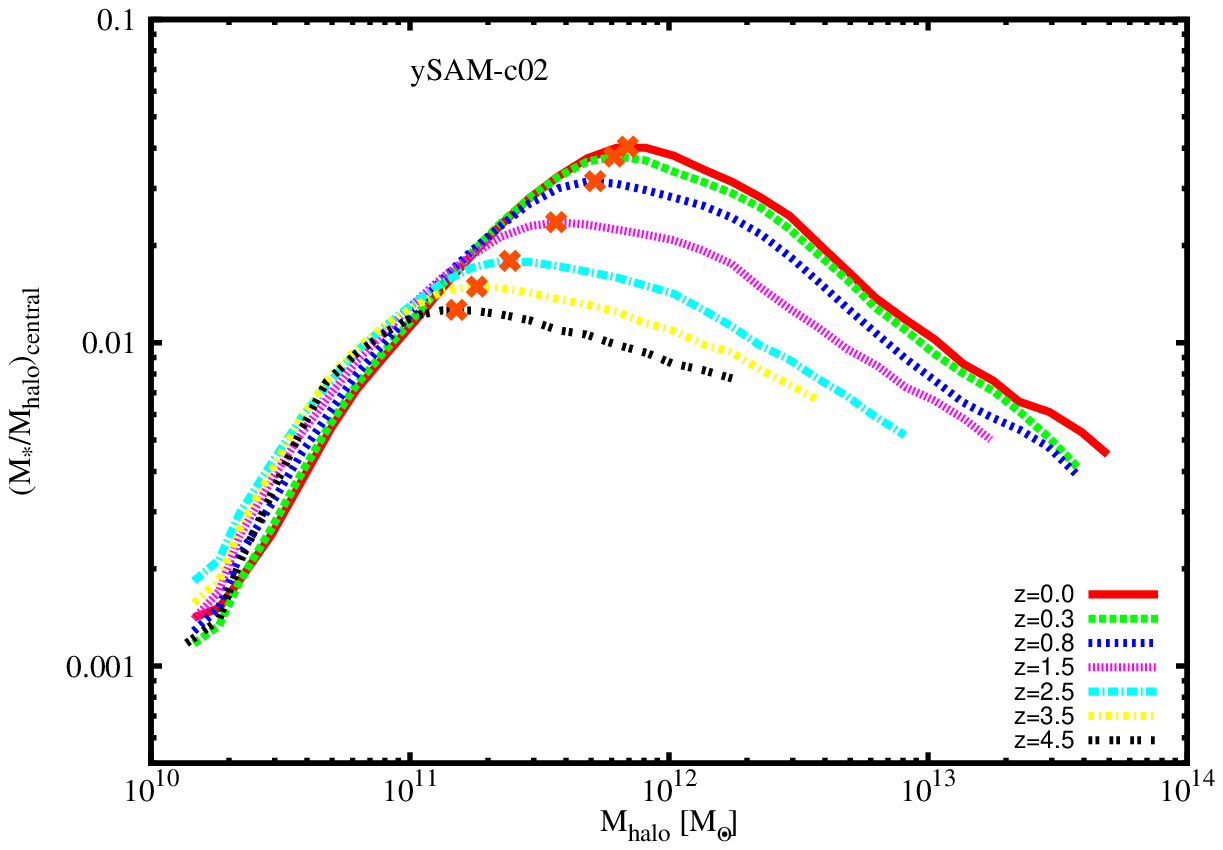}
   \caption{Evolution of the SHM relation for all \ctwo\ models. For clarity only every second available redshift is shown. The grey cross marks the peak (position) used in \Sec{sec:SHM}.}
 \label{fig:SHMzred-model}
 \end{figure*}
 
 In \Fig{fig:SHM} we have presented the SHM relations for all \ctwo\ models at redshift $z\!=\!0$, but later on studied the redshift evolution in Figs.~\ref{fig:peakMsMh} through~\ref{fig:peaksigmaMsMh}. Here we now show in \Fig{fig:SHMzred-model} for each model individually the SHM relations at all considered redshifts that directly entered into the calculation of the peak value and position.

\subsection{Individual Variance} \label{app:SHMsigma}
 \begin{figure*}
   \includegraphics[width=0.75\columnwidth]{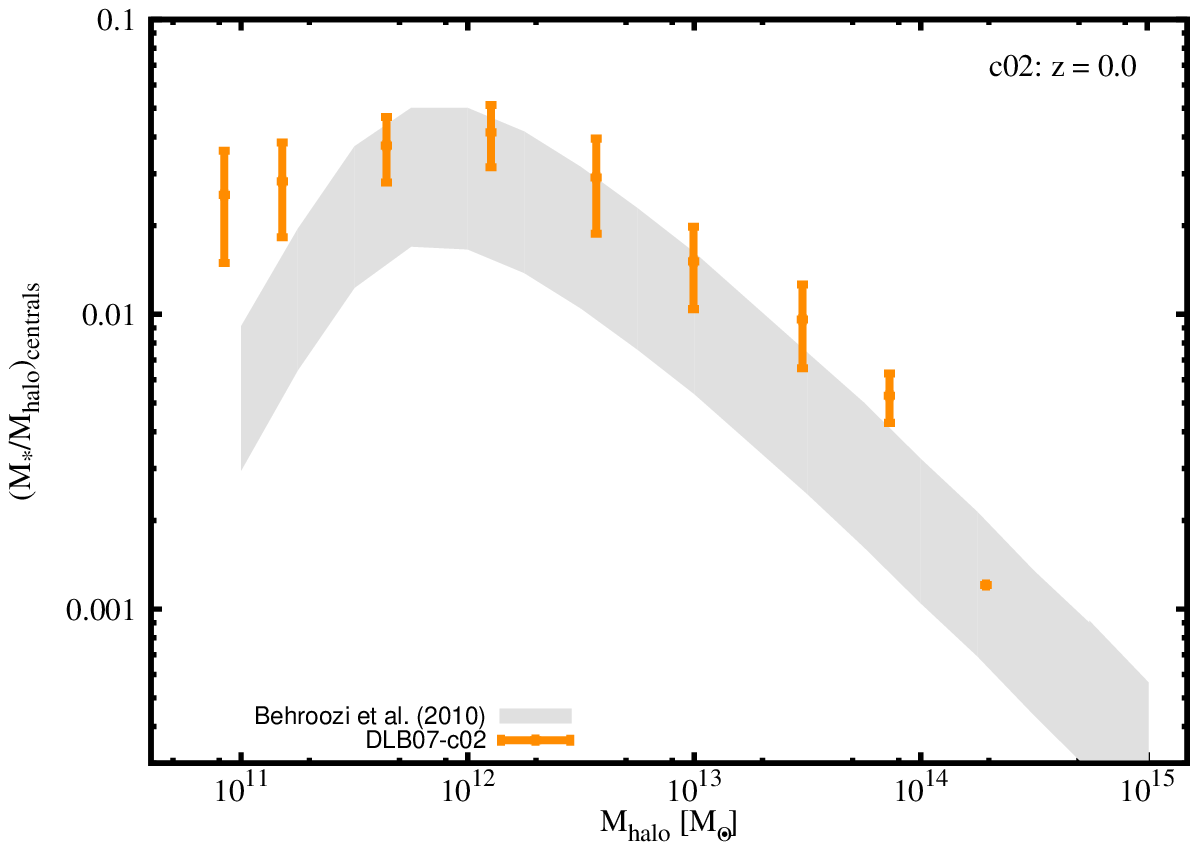}
   \includegraphics[width=0.75\columnwidth]{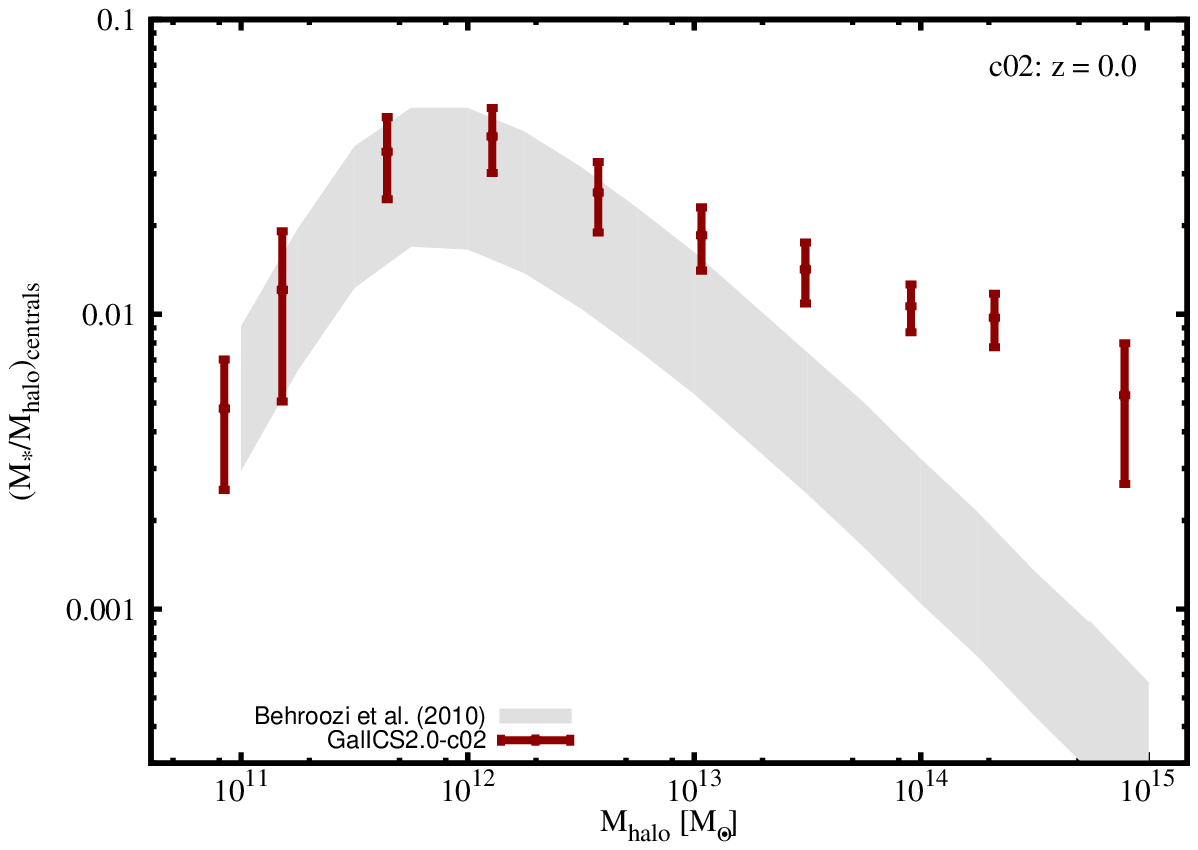}
   \includegraphics[width=0.75\columnwidth]{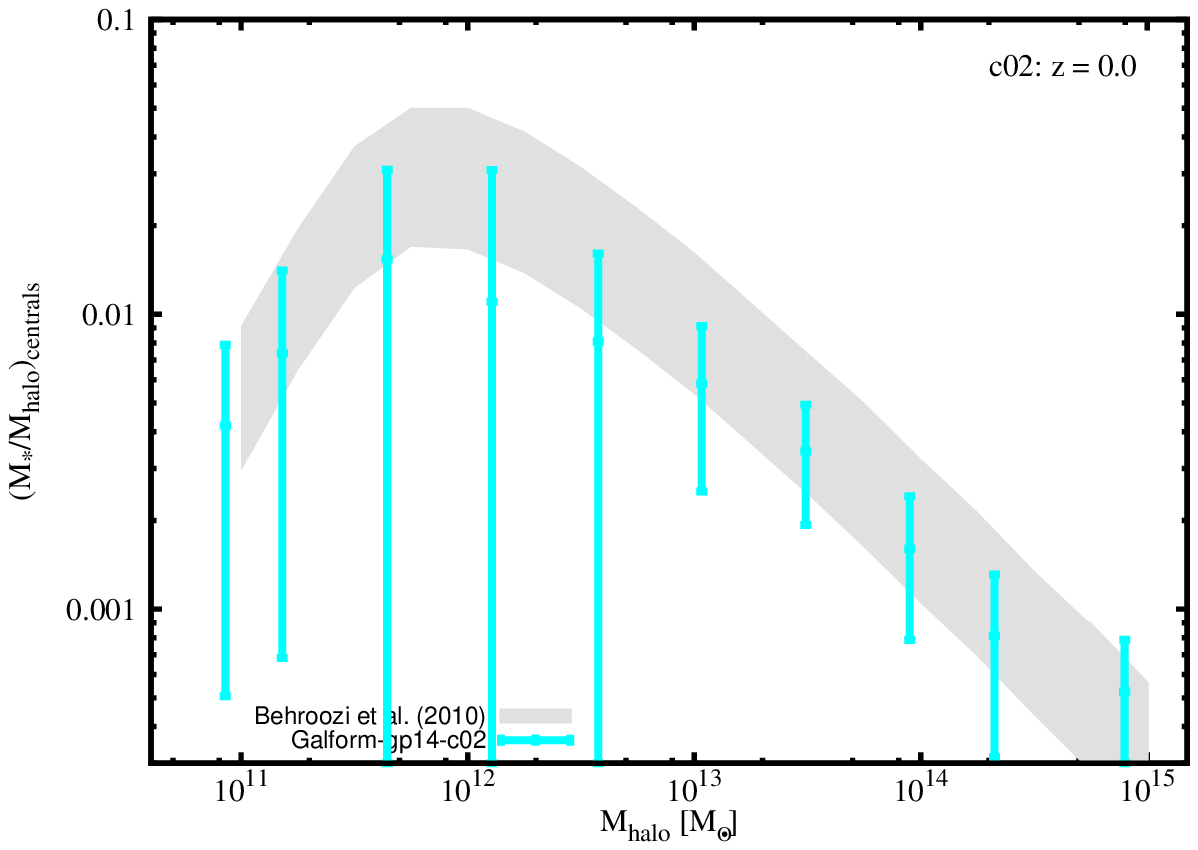}
   \includegraphics[width=0.75\columnwidth]{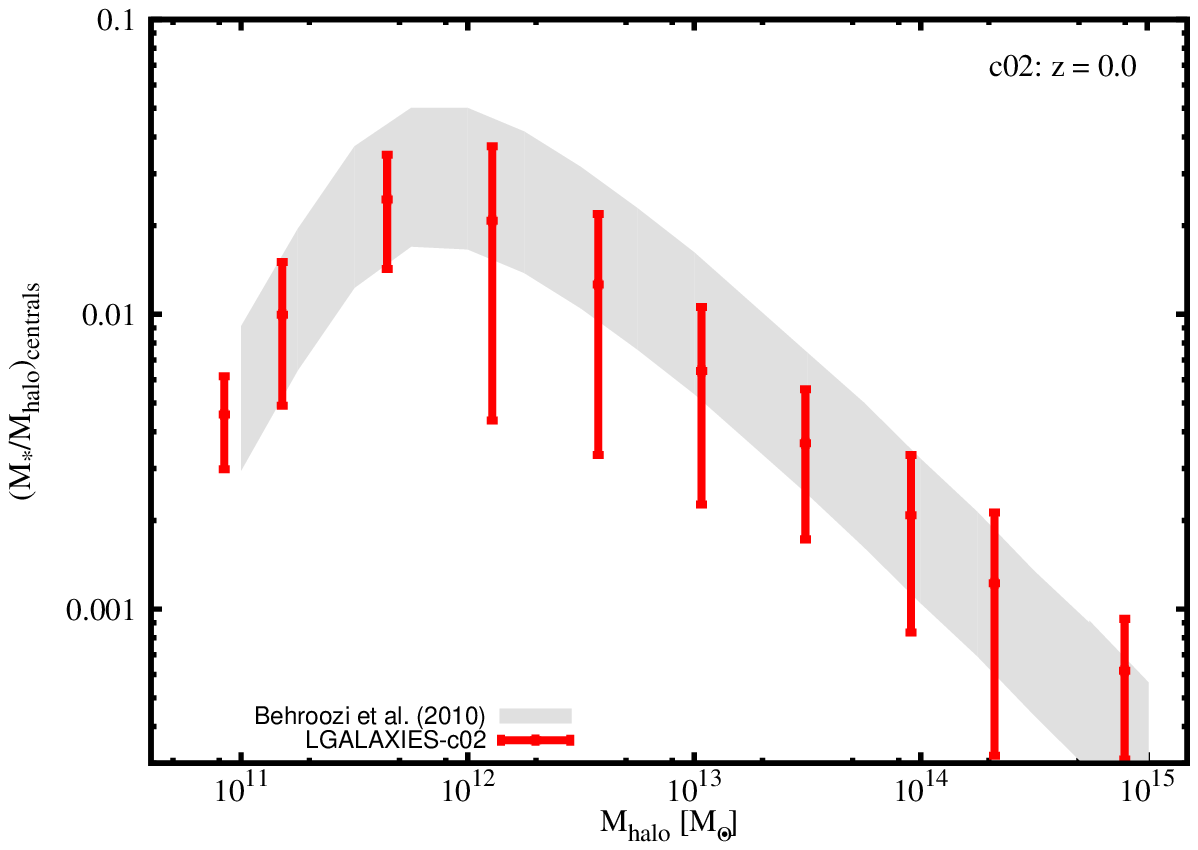}
   \includegraphics[width=0.75\columnwidth]{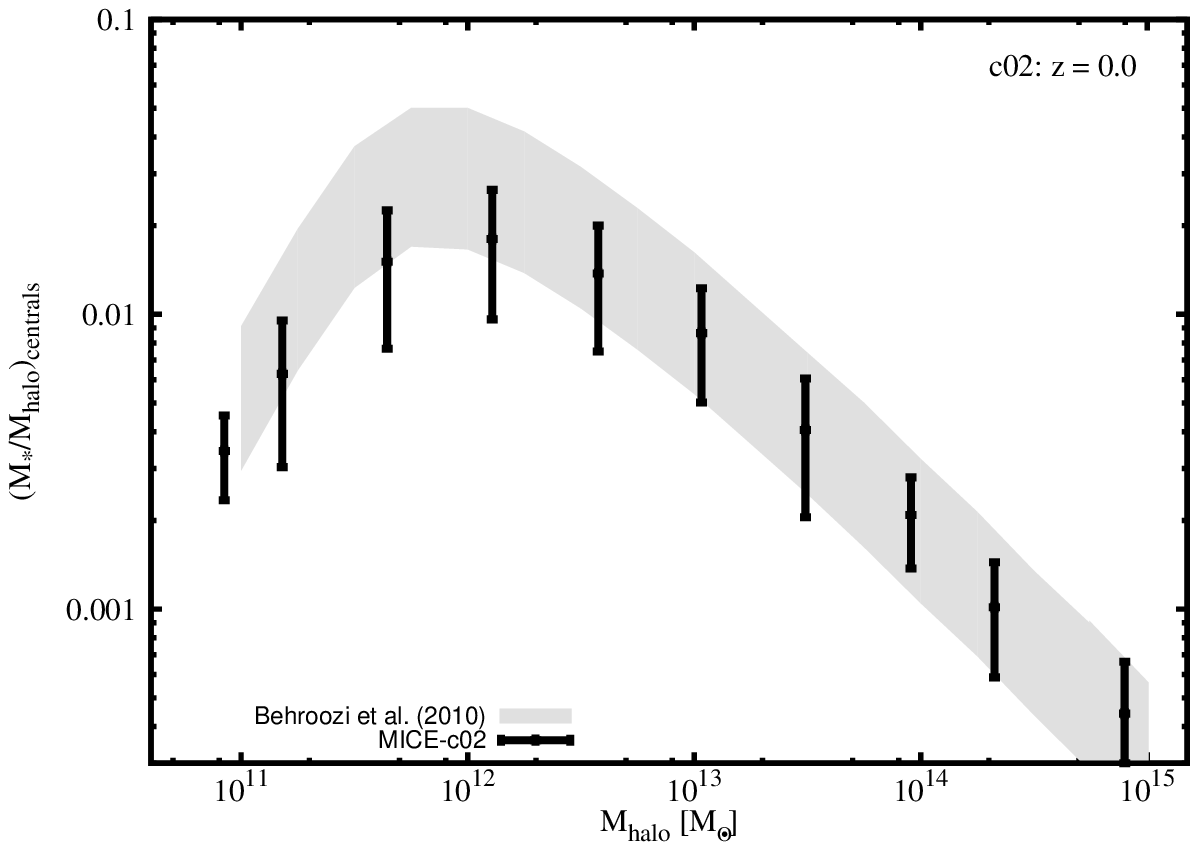}
   \includegraphics[width=0.75\columnwidth]{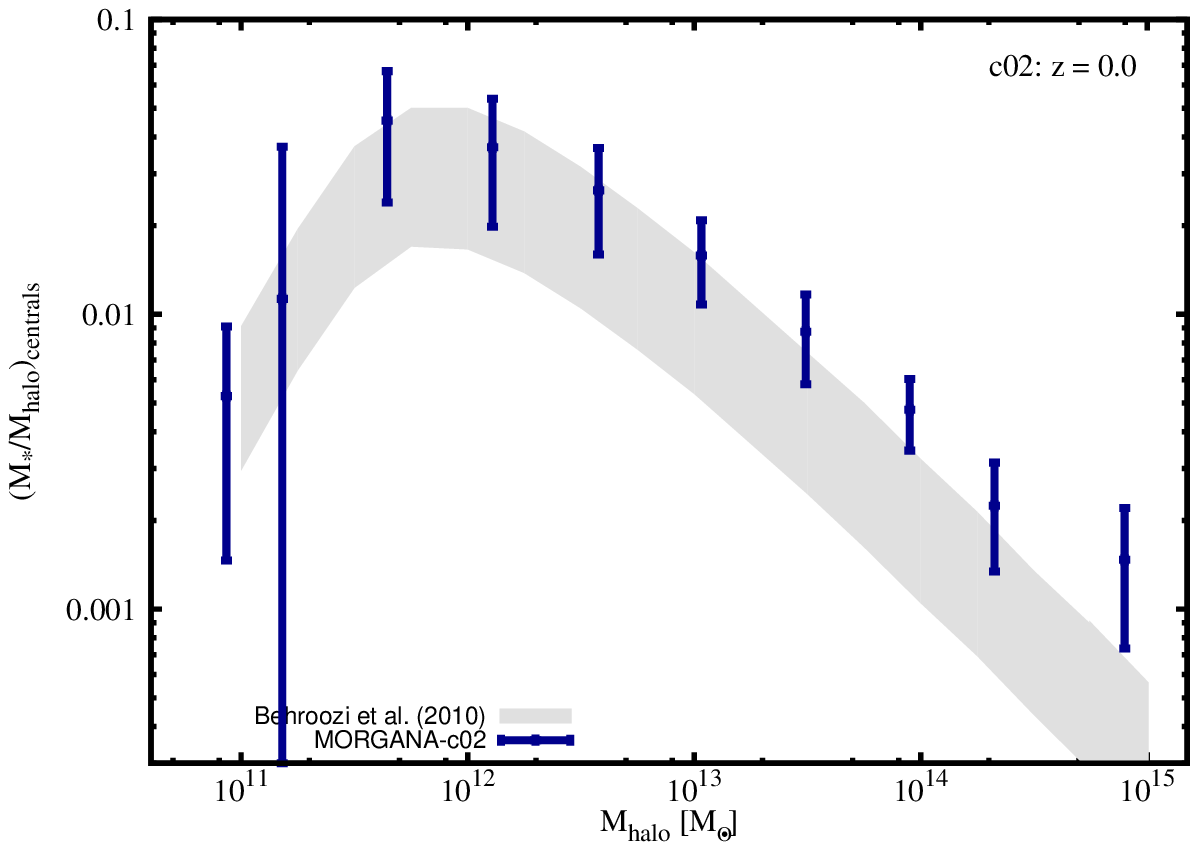}
   \includegraphics[width=0.75\columnwidth]{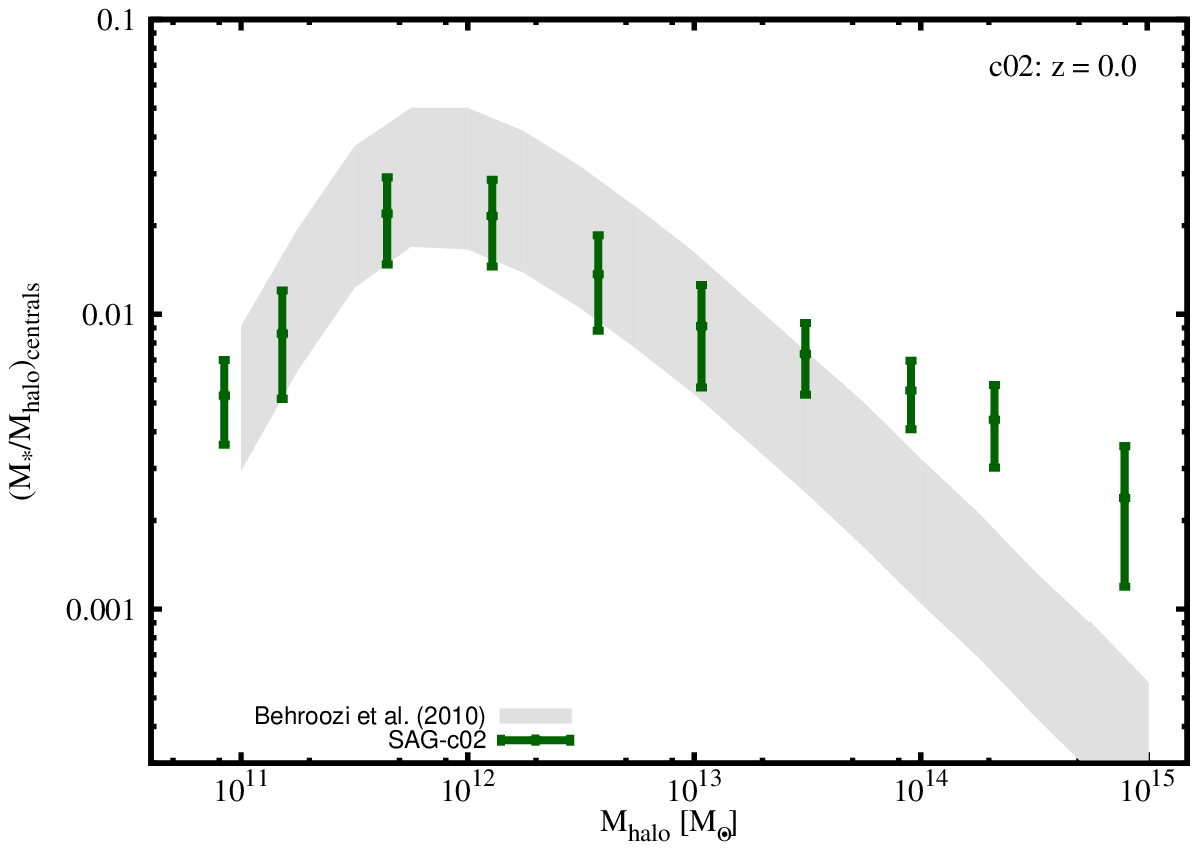}
   \includegraphics[width=0.75\columnwidth]{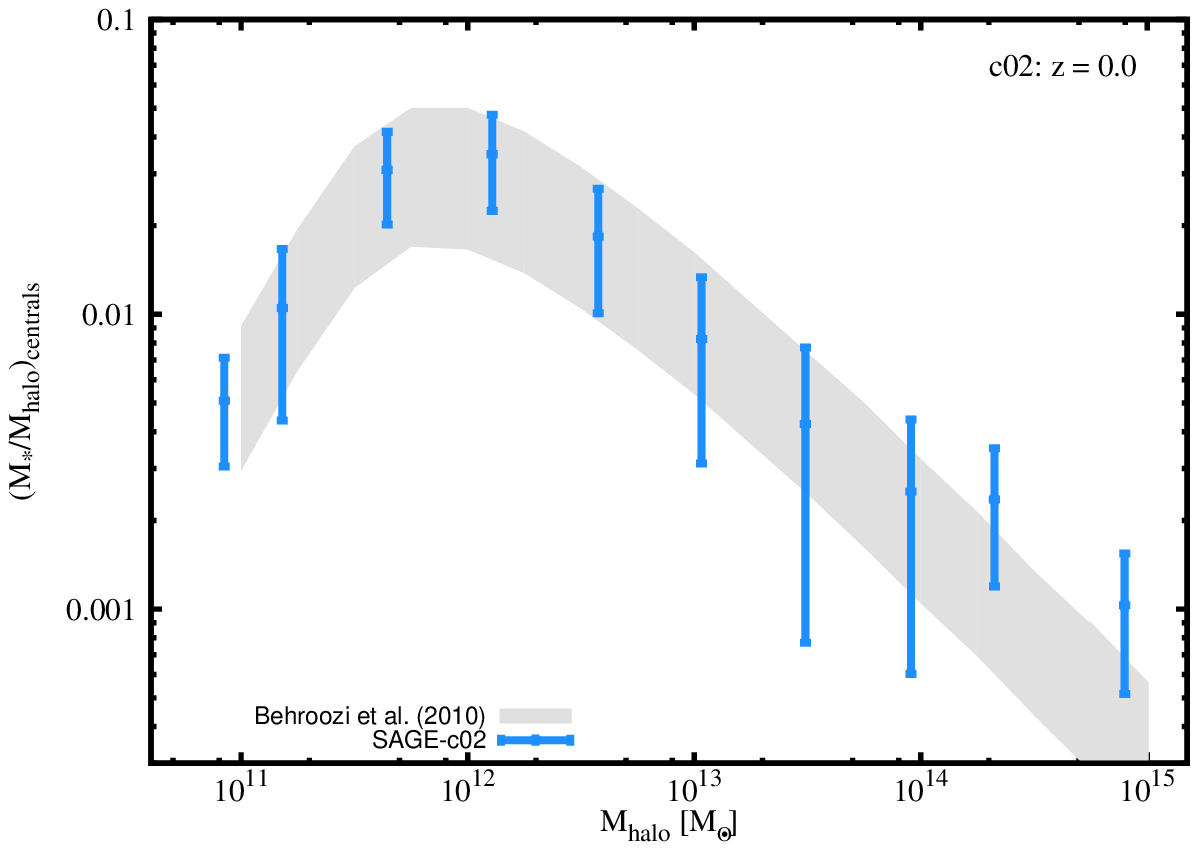}
   \includegraphics[width=0.75\columnwidth]{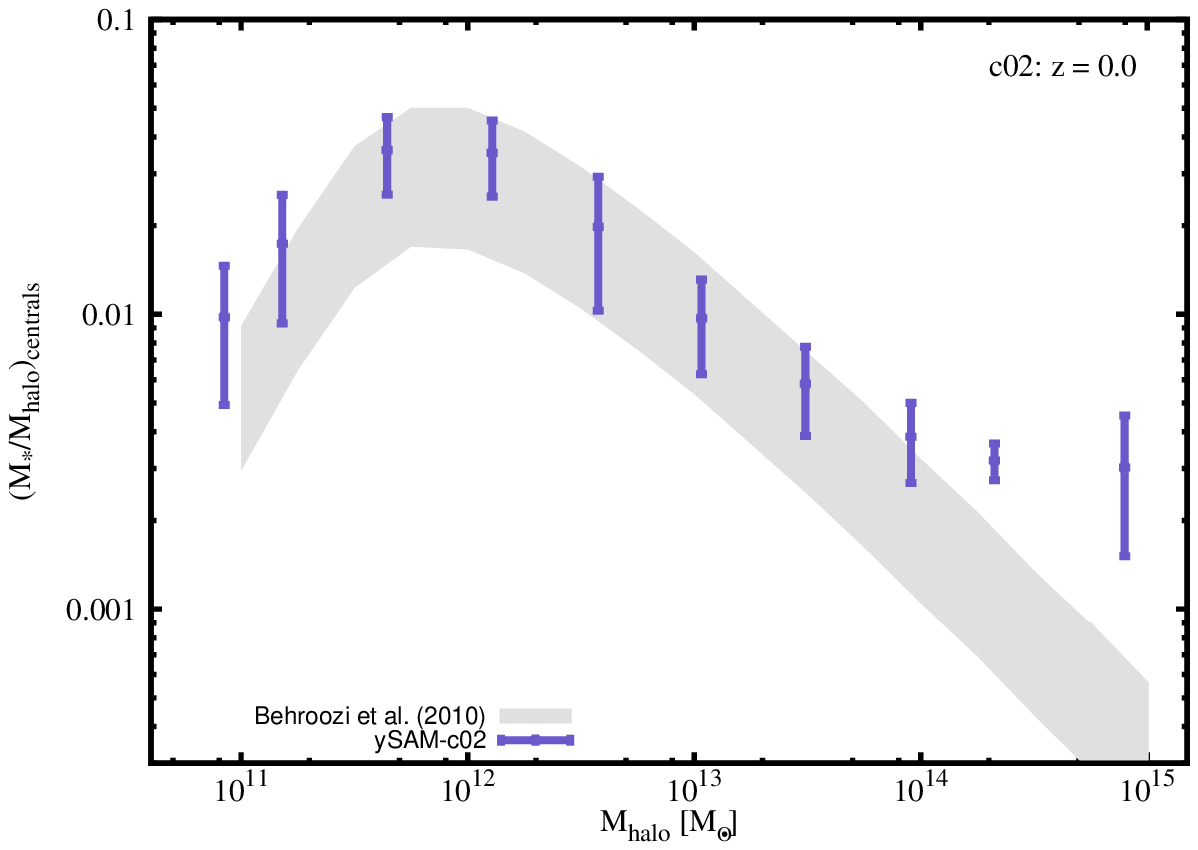}
   \caption{The SHM relation for all models at redshift $z\!=\!0$ including error bars defined as 25--75 percentiles of the distribution in each bin.}
 \label{fig:SHMsigma-model}
 \end{figure*}

In \Fig{fig:peaksigmaMsMh} we have shown the mean scatter of the SHM relation defined as the 25--75 percentiles of the distribution in each $M_{\rm halo}$ bin. In \Fig{fig:SHMsigma-model} we now give an example of the scatter of each model at redshift $z\!=\!0$.

\subsection{Addition of Satellite Stellar Mass} \label{app:SHM+MstarSatellites}
\begin{figure}
   \includegraphics[width=\columnwidth]{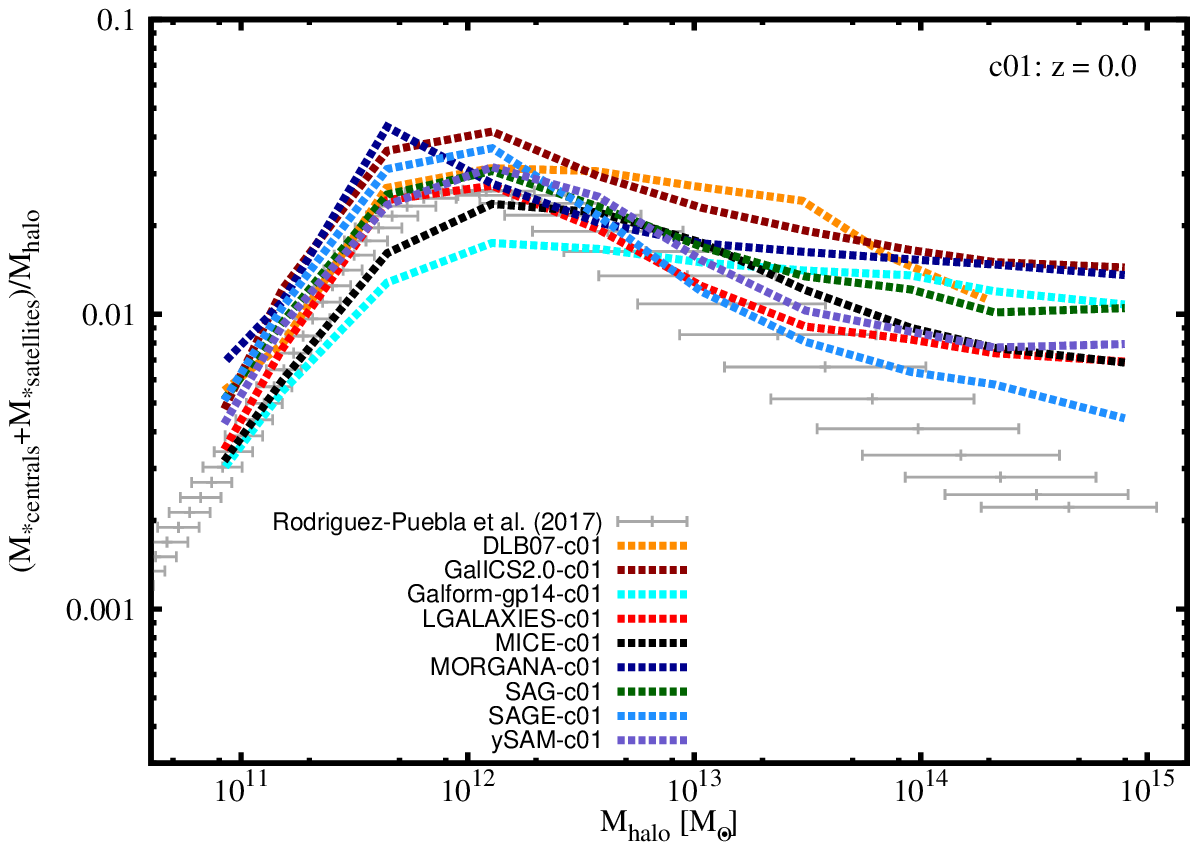}
   \includegraphics[width=\columnwidth]{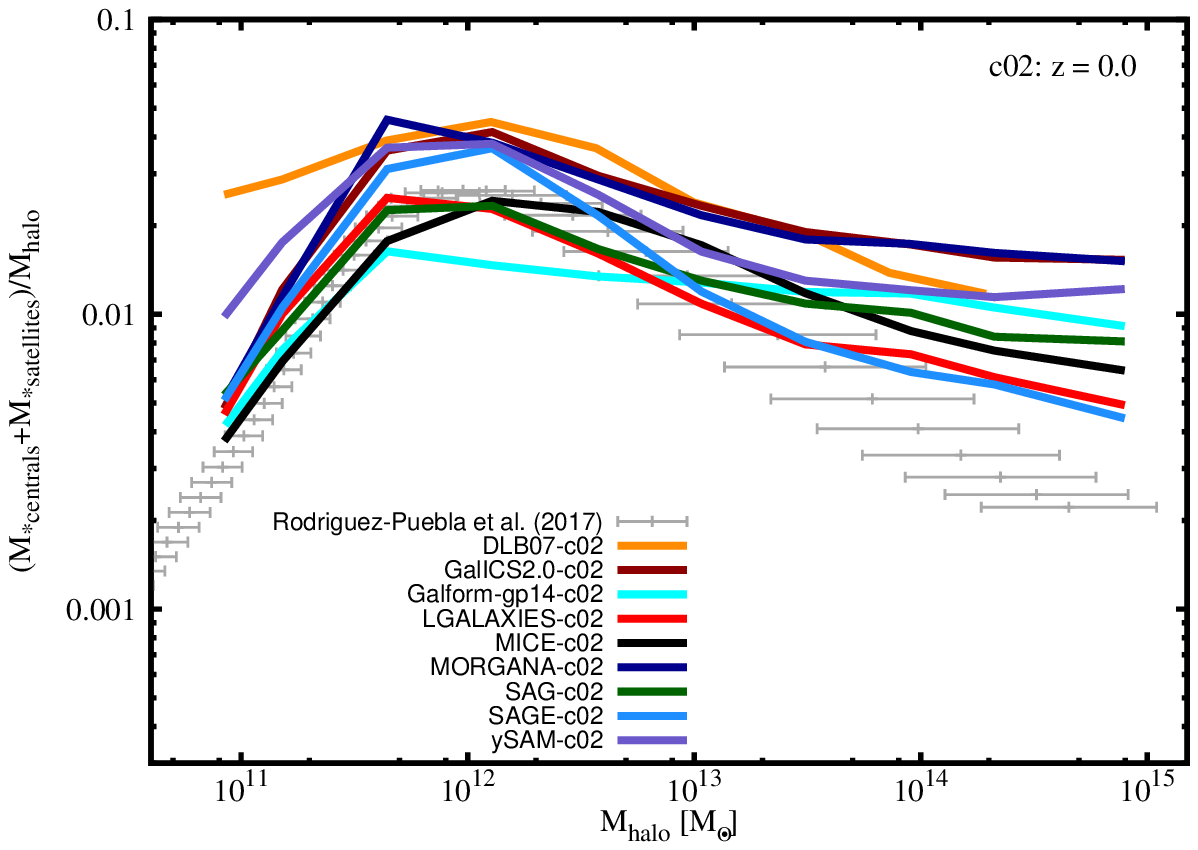}
   \caption{Stellar to halo mass ratio as a function of halo mass for central galaxies only, but with the stellar mass of the satellite galaxies added to it. The values shown are medians in the respective bin.}
 \label{fig:SHM+MstarSatellites}
 \end{figure}

As mentioned in \Sec{sec:SHMrelation} we have performed the test of adding the stellar mass of satellite galaxies to the stellar mass of the central when calculating the SHM ratio as the halo mass entering this ratio also contains the masses of the subhaloes. The resulting changes to the original \Fig{fig:SHM} can be viewed here in \Fig{fig:SHM+MstarSatellites} where the upper panel shows calibration \cone\ and the lower panel \ctwo. We notice a substantial effect, especially at the high-$M_*$ end, i.e. for haloes that host a pronounced number of subhaloes and satellites, respectively. This test highlights the importance of how to count stellar and halo mass in theoretical models when comparing to observations, but it does not change our conclusions.

\subsection{Addition of Cold Gas Mass} \label{app:SHM+Mgas}
\begin{figure}
   \includegraphics[width=\columnwidth]{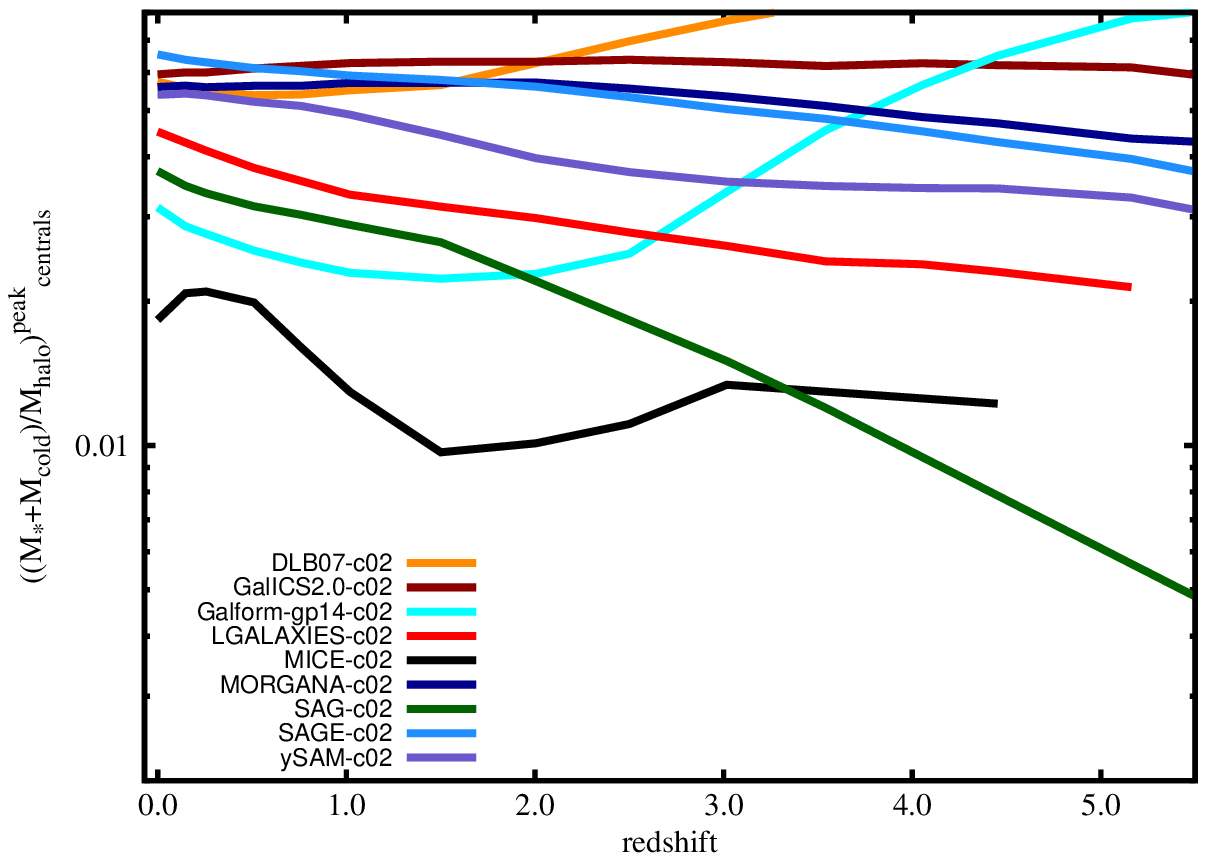}
   \includegraphics[width=\columnwidth]{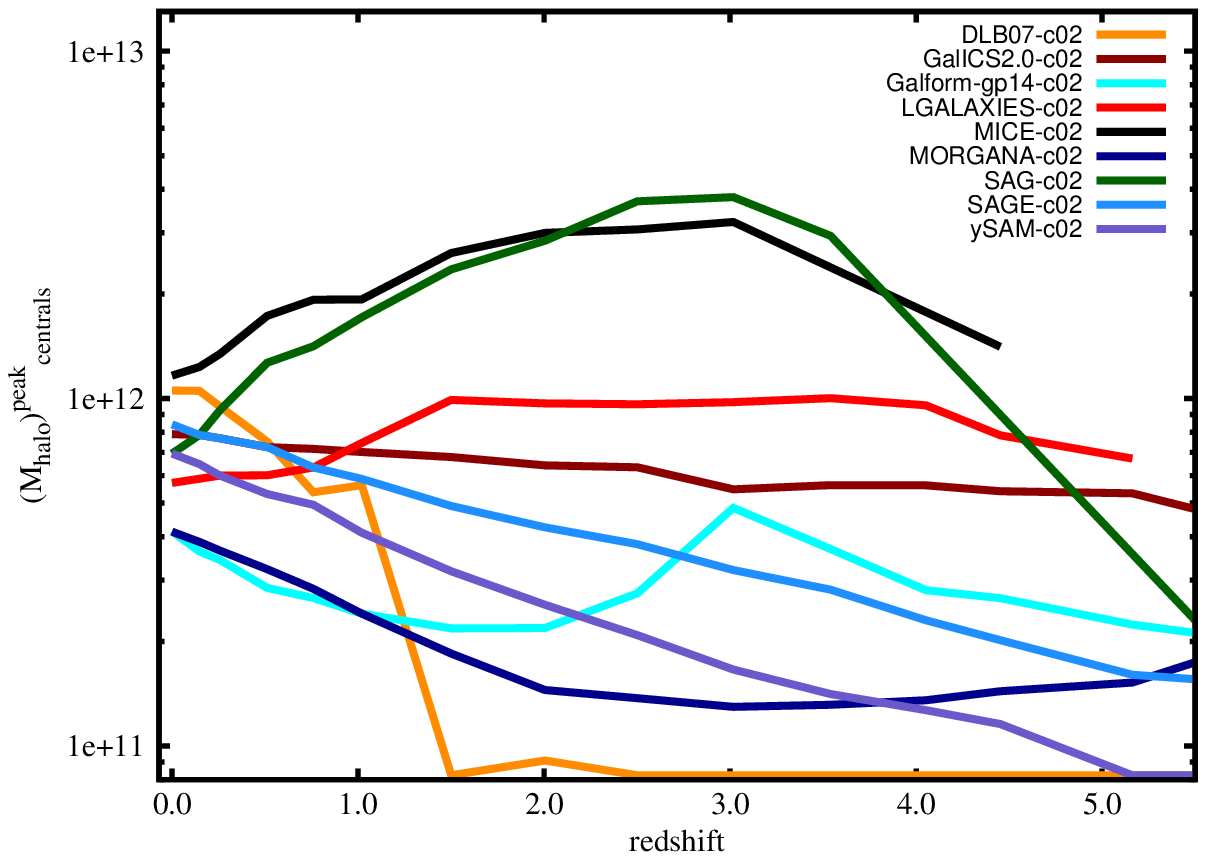}
   \includegraphics[width=\columnwidth]{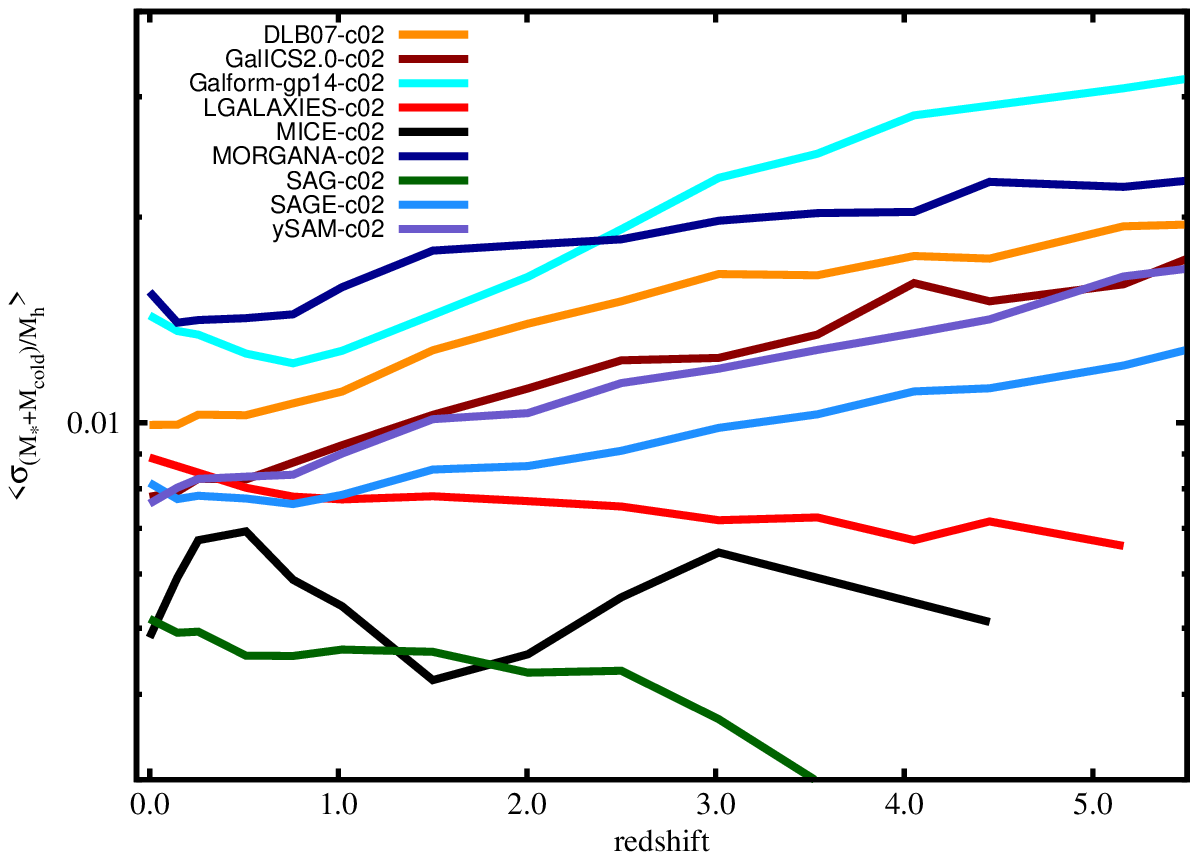}
   \caption{Redshift evolution of the peak value (upper panel), the peak position (middle panel),  the mean of the scatter (lower panel) of the SHM relation for central galaxies when also adding the cold gas mass to the stellar mass of the central in the \ctwo\ models.}
 \label{fig:SHMevolution+Mgas}
 \end{figure}

We have also performed the test of adding the cold gas mass to the stellar mass of the central galaxy when calculating the SHM ratio (which should then rather be called the baryon-to-halo mass relation, but we continue calling it SHM). The results can be viewed in \Fig{fig:SHMevolution+Mgas} which shows the anologies to Figs.~\ref{fig:peakMsMh}-\ref{fig:peaksigmaMsMh}. Even though some models do show rather distinct changes, the trends are nevertheless preserved and our conclusions not affected, respectively. 

\bsp

\label{lastpage}

\end{document}